\documentclass[12pt]{iopart}
\usepackage{graphicx}
\usepackage{hyperref}
\usepackage{subcaption} 
\expandafter\let\csname equation*\endcsname\relax
\expandafter\let\csname endequation*\endcsname\relax
\usepackage{amsmath, amssymb,amsfonts}
\usepackage{multirow}
\usepackage{color}
\usepackage{array}
\usepackage{makecell}
\usepackage{acronym}
\usepackage{xspace}
\usepackage{comment}
\usepackage{booktabs}
\usepackage{pdflscape}
\usepackage{geometry}
\usepackage{orcidlink}
\usepackage{siunitx}

\def\GW{{\rm gw}}

\begin{document}
\title[ ]{Decadal upgrade strategy for KAGRA toward post-O5 gravitational-wave astronomy}

\author{%
T.~Akutsu\,\orcidlink{0000-0003-0733-7530}$^{1,2}$, 
M.~Ando$^{3,4}$, 
M.~Aoumi$^{5}$, 
A.~Araya\,\orcidlink{0000-0002-6884-2875}$^{6}$, 
Y.~Aso\,\orcidlink{0000-0002-1902-6695}$^{1,7}$, 
L.~Baiotti\,\orcidlink{0000-0003-0458-4288}$^{8}$, 
R.~Bajpai\,\orcidlink{0000-0003-0495-5720}$^{9}$, 
K.~Cannon\,\orcidlink{0000-0003-4068-6572}$^{4}$, 
A.~H.-Y.~Chen$^{10}$, 
D.~Chen\,\orcidlink{0000-0003-1433-0716}$^{11}$
% ...
\\
{(The KAGRA Collaboration)}%
}%

\medskip

\address{$^{1}$Gravitational Wave Science Project, National Astronomical Observatory of Japan, 2-21-1 Osawa, Mitaka City, Tokyo 181-8588, Japan}
\address{$^{2}$Advanced Technology Center, National Astronomical Observatory of Japan, 2-21-1 Osawa, Mitaka City, Tokyo 181-8588, Japan}
\address{$^{3}$Department of Physics, The University of Tokyo, 7-3-1 Hongo, Bunkyo-ku, Tokyo 113-0033, Japan}
\address{$^{4}$Research Center for the Early Universe (RESCEU), The University of Tokyo, 7-3-1 Hongo, Bunkyo-ku, Tokyo 113-0033, Japan}
\address{$^{5}$Institute for Cosmic Ray Research, KAGRA Observatory, The University of Tokyo, 238 Higashi-Mozumi, Kamioka-cho, Hida City, Gifu 506-1205, Japan}
\address{$^{6}$Earthquake Research Institute, The University of Tokyo, 1-1-1 Yayoi, Bunkyo-ku, Tokyo 113-0032, Japan}
\address{$^{7}$The Graduate University for Advanced Studies (SOKENDAI), 2-21-1 Osawa, Mitaka City, Tokyo 181-8588, Japan}
\address{$^{8}$International College, Osaka University, 1-1 Machikaneyama-cho, Toyonaka City, Osaka 560-0043, Japan}
\address{$^{9}$Accelerator Laboratory, High Energy Accelerator Research Organization (KEK), 1-1 Oho, Tsukuba City, Ibaraki 305-0801, Japan}
\address{$^{10}$Institute of Physics, National Yang Ming Chiao Tung University, 101 Univ. Street, Hsinchu, Taiwan}
\address{$^{11}$Kamioka Branch, National Astronomical Observatory of Japan, 238 Higashi-Mozumi, Kamioka-cho, Hida City, Gifu 506-1205, Japan}
% ...
\eads{\mailto{kentaro.komori@phys.s.u-tokyo.ac.jp}, \mailto{soichiro@icrr.u-tokyo.ac.jp}}
\title[ ]{ }
\date{\today}
\acrodef{GW}[GW]{gravitational wave}
\acrodef{LVK}[LVK]{LIGO Scientific, Virgo and KAGRA Collaboration}
\acrodef{BH}[BH]{black hole}
\acrodef{NS}[NS]{neutron star}
\acrodef{CBC}[CBC]{compact binary coalescence}
\acrodef{BBH}[BBH]{binary black hole}
\acrodef{BNS}[BNS]{binary neutron star}
\acrodef{NSBH}[NSBH]{neutron star--black hole binary}
\acrodef{SNR}[SNR]{signal-to-noise ratio}
\acrodef{EoS}[EoS]{equation of state}

\newcommand{\lvk}{\ac{LVK}\xspace}
\newcommand{\snr}{\ac{SNR}\xspace}
\newcommand{\snrs}{\acp{SNR}\xspace}
\newcommand{\gw}{\ac{GW}\xspace}
\newcommand{\gws}{\acp{GW}\xspace}
\renewcommand{\ns}{\ac{NS}\xspace}
\newcommand{\nss}{\acp{NS}\xspace}
\newcommand{\bh}{\ac{BH}\xspace}
\newcommand{\bhs}{\acp{BH}\xspace}
\renewcommand{\nss}{\acp{NS}\xspace}
\renewcommand{\bhs}{\acp{BH}\xspace}
\newcommand{\bns}{\ac{BNS}\xspace}
\newcommand{\bnss}{\acp{BNS}\xspace}
\newcommand{\nsbh}{\ac{NSBH}\xspace}
\newcommand{\bbh}{\ac{BBH}\xspace}
\newcommand{\bbhs}{\acp{BBH}\xspace}
\newcommand{\cbc}{\ac{CBC}\xspace}
\newcommand{\cbcs}{\acp{CBC}\xspace}
\newcommand{\eos}{\ac{EoS}\xspace}
\newcommand{\eoss}{\acp{EoS}\xspace}

\newcommand{\calA}{\mathcal{A}}
\newcommand{\uc}{\mathrm{c}}
\newcommand{\ud}{\mathrm{d}}
\newcommand{\calE}{\mathcal{E}}
\newcommand{\uf}{\mathrm{f}}
\newcommand{\calF}{\mathcal{F}}
\newcommand{\calI}{\mathcal{I}}
\newcommand{\calL}{\mathcal{L}}
\newcommand{\calM}{\mathcal{M}}
\newcommand{\un}{\mathrm{n}}
\newcommand{\calO}{\mathcal{O}}
\newcommand{\up}{\mathrm{p}}
\newcommand{\ur}{\mathrm{r}}
\newcommand{\calR}{\mathcal{R}}
\newcommand{\ut}{\mathrm{t}}

\newcommand{\thr}{\mathrm{th}}
\newcommand{\dl}{d_\mathrm{L}}
\newcommand{\fin}{\mathrm{fin}}
\newcommand{\ini}{\mathrm{ini}}
\newcommand{\re}{\mathrm{ref}}
\newcommand{\prim}{\mathrm{prim}}
\newcommand{\hyp}{\mathchar`-}
\newcommand{\obs}{\mathrm{obs}}
\newcommand{\Fp}{F_+}
\newcommand{\Fc}{F_\times}
\newcommand{\dd}[1]{\mathrm{d}{#1}\xspace}

\acrodef{SGWB}[SGWB]{stochastic gravitational wave background}


\tableofcontents
\title[ ]{ }

\section{Abstract\label{sec:abstract}}
The KAGRA Collaboration has investigated a ten-year upgrade strategy for the KAGRA gravitational wave detector, considering a total of 14 upgrade options that vary in mirror mass, quantum noise reduction techniques, and the quality of cryogenic suspensions. We evaluated the scientific potential of these configurations with a focus on key targets such as parameter estimation of compact binary coalescences, binary neutron star post-merger signals, and continuous gravitational waves. Rather than aiming to improve all science cases uniformly, we prioritized those most sensitive to the detector configuration. Technical feasibility was assessed based on required hardware developments, associated R\&D efforts, cost, and risk. Our study finds that a high-frequency upgrade plan that enhances sensitivity over a broad frequency range above $\sim 200$~Hz offers the best balance between scientific return and technical feasibility. Such an upgrade would enable sky localization of binary neutron star mergers at 100 Mpc to better than 0.5 deg$^2$ in a LIGO-Virgo-KAGRA network, and improve the measurement precision of tidal deformability parameter by approximately 10\% at median, compared to a network without KAGRA.

\section{Introduction\label{sec:introduction}}
Since the first observation of \gws in 2015, LIGO and Virgo have steadily improved their sensitivities, and more than 250 \gw events have now been reported since then. As their post-O5 upgrades, LIGO and Virgo are considering broadband upgrades such as A\# and Virgo\_nEXT, and further detections of \gws from binary mergers composed of stellar-mass \bhs and \nss are expected. Additionally, the LIGO-India project is progressing and it is expected to start observations by the end of the present decade. By the second half of 2030s or early 2040s, the third generation large-scale laser interferometers such as the Einstein Telescope and Cosmic Explorer are expected to begin operations.

As \gw detection becomes routine, KAGRA must strategically consider how to upgrade its capability and performance after its initial detection of \gws. Furthermore, upgrading KAGRA will require a different strategy compared to LIGO and Virgo, as KAGRA has taken a significantly different approach to reducing coating thermal noise. While Advanced LIGO and Advanced Virgo use large fused silica mirrors at room temperature to increase the beam size, KAGRA cools its sapphire mirrors to cryogenic temperatures. In KAGRA, heat extraction from the laser beam impinging on the test masses is achieved through the sapphire fibers that suspend the test masses. Therefore, when injecting higher laser power into the test masses, thicker and shorter fibers are necessary to extract more heat. While higher laser power is required to reduce quantum noise at high frequencies, the use of thick and short fibers for heat extraction increases suspension thermal noise. As a result, upgrading KAGRA will not be as straightforward as upgrading room-temperature interferometers. Modifying the cryogenic suspension system also requires more time than room-temperature suspensions. Additionally, the limited space in the underground tunnel further restricts changes to the detector layout.

With this in mind, KAGRA established the Future Planning Committee (FPC) in 2018 to explore various upgrade options from both technological and scientific perspectives. In particular, four upgrade proposals were selected based on the criteria of being feasible within five years and a budget of 500 million yen. The technologies and scientific potential of these proposals were summarized in the 2019 FPC White Paper~\cite{FPCWP2019}. The technological aspects were later published as Ref.~\cite{KAGRAplus}, while the scientific aspects were published as Ref.~\cite{KAGRAScience}.

The four proposed upgrade options were as follows:
\begin{itemize}
\item LF: Optimized for detecting intermediate-mass \bbhs ($\sim$ 100 $M_{\odot}$).
\item HF: Optimized for sky localization of \bns mergers.
\item 40kg: Enlarge sapphire mirrors from 23 kg to 40 kg.
\item FDS: The introduction of frequency-dependent squeezing.
\end{itemize}
After evaluating the significance and feasibility of the technologies required for each proposal, HF received the highest overall score.

On the other hand, a comprehensive assessment of all the possible scientific benefits of these four upgrades revealed that there is no upgrade option that can improve all science cases. In other words, in order to select the upgrade candidate solely based on scientific merit, a prioritization of specific science goals is required. However, no conclusion was reached at that time on which science case should be prioritized, leaving the discussion open for the next revision of the White Paper.

The revision of this White Paper was actually not carried out for some time. Instead, with the initiative of the Future Strategy Committee (FSC), which was reorganized from the FPC, the first updated version was written and released in 2024. Reflecting on the difficulty of selecting an upgrade option purely based on scientific merits, the FSC focused primarily on technical aspects in the new KAGRA Instrument Science White Paper 2024~\cite{ISWP2024}. In this revision, it was again confirmed that the high-frequency upgrade received the highest scores in terms of feasibility of relevant technologies.

Parallel to these efforts by the FSC, the Executive Office established the 10 year Task Force to develop a ten-year roadmap for KAGRA, including plans for O5 and O6~\cite{10yrKickOff}. This paper summarizes the discussions led by the 10 year Task Force.

Building on the considerations of the FPC, the 10 year Task Force proposed upgraded sensitivity options based on the latest LIGO's developments and plans, including A\#, as well as the most recent measurements of KAGRA's detector parameters.

\begin{figure}[htbp]
	\centering
    \includegraphics[width=0.95\linewidth]{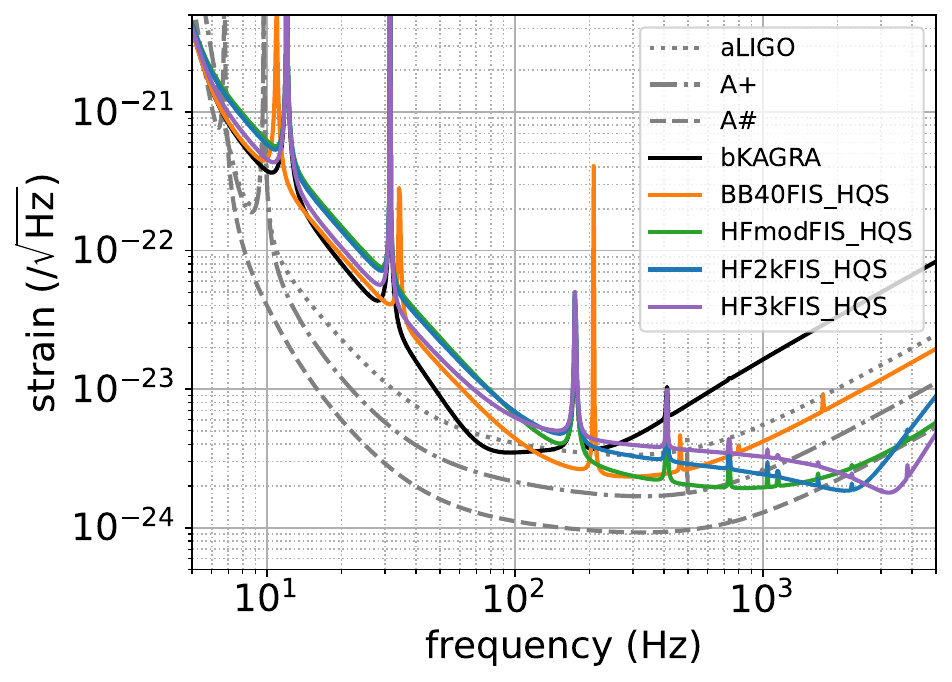}
	\caption{Examples of sensitivity curves for 4 out of the 14 upgrade options considered in this White Paper. These include an option aiming for broadband sensitivity improvement using heavier 40kg sapphire mirrors (BB40), and options that shift the sensitivity band toward higher frequencies (HFmod) or introduce dips around 2~kHz or 3~kHz (HF2k, HF3k) by adjusting the reflectivity of the signal recycling mirror. All of these assume the implementation of frequency-independent squeezing (FIS) and high-quality-factor suspensions (HQS) as originally planned for baseline KAGRA (bKAGRA).} \label{fig:intro_sensitivity}
\end{figure}

A total of 14 upgrade options were considered, based on factors such as the mass of the mirrors, the use of frequency-independent or frequency-dependent squeezing, whether to assume sapphire suspensions with lower mechanical loss than the current ones, and the specific high-frequency range to target through modifications to the signal recycling cavity. Details of these options are summarized in Sec.\ref{sec:hardware}, and the sensitivity curve data can be found in Ref.\cite{10yrTFSensitivity}. Figure~\ref{fig:intro_sensitivity} shows the sensitivity curves for four representative upgrade options, and Table~\ref{tab:acronyms} summarizes the acronyms used for the upgrade plans discussed in this paper.

In the remainder of the White Paper, we first discuss the scientific opportunities enabled by these sensitivity options. Unlike the FPC's approach, which considered all possible science cases, we focus on science topics that we consider particularly important or those that are most impacted by different upgrade options. Finally, we discuss the technologies required to achieve these sensitivities and assess their feasibility.

\begin{table*}[htbp]
\centering
\begin{tabular}{l l}
\hline
BB40 & Broadband configuration using 40 kg test masses \\
HF & High frequency \\
HFmod & Moderate high frequency configuration \\
HF2k & High frequency configuration focused at 2~kHz \\
HF3k & High frequency configuration focused at 3~kHz \\
FIS & Frequency-independent squeezing \\
FDS & Frequency-dependent squeezing \\
HQS & High-quality suspension \\
LB & Larger beam size \\
BC & Better coating \\
\hline
\end{tabular}
\caption{List of acronyms used for the upgrade plans discussed in this paper.}
\label{tab:acronyms}
\end{table*}

\section{Science with KAGRA\label{sec:science}}
In this section, we explore potential science cases for KAGRA in the post-O5 era.
Although KAGRA’s sensitivity is expected to be lower than that of LIGO and Virgo, it plays a valuable complementary role in several areas.
As an additional detector in the global network, KAGRA contributes to improved sky localization of \gw sources.
It enhances the success rate of multi-messenger observations of \gw sources and may enable the next GW170817-like event to be observed as a multi-messenger signal.

The HF configurations of KAGRA extend the network's sensitivity into higher-frequency bands.
This enhancement is particularly beneficial for measuring tidal effects in \bns signals.
Moreover, the HF options enable unique science opportunities for KAGRA, such as detecting post-merger signals from BNSs and continuous signals from rapidly rotating \nss.
On the other hand, limiting the sensitive frequency band may degrade KAGRA’s standalone detection performance by reducing \acp{SNR} of \gw signals.

The results of our study are summarized in Table~\ref{tab:science_summary} and are outlined below:
\begin{itemize}
    \item \textbf{Detection rates:} BB options generally outperform HF options in terms of detection rates of CBC signals. The best configuration is BB40FDS\_HQS\_BC, which predicts $\mathcal{O}(10)$ \bns and $\mathcal{O}(10^2)$ \bbh events annually.
    \item \textbf{Sky localization:} HF configurations generally yield better localization performance than BB options for GW170817-like \bns signals, with the HF-mod variants performing best. In contrast, for GW150914-like \bbh signals, BB configurations yield better localization performance.
    \item \textbf{Tidal deformability:} HF-mod and HF-2k configurations are the most effective for constraining tidal deformabilities of GW170817-like \bns signals, reducing the errors in measuring a tidal parameter, $\tilde{\Lambda}$, by approximately $\sim 10\%$ at median.
    \item \textbf{\bns Post-merger signal:} The HF2k and HF3k configurations are most sensitive to the post-merger signal of \bns, with the maximum event rate estimated at approximately $0.1\,\mathrm{year}^{-1}$.
    \item \textbf{Rapidly rotating \nss:} For detecting sinusoidal signals from rotating neutron stars, HF-mod configurations are preferable due to their enhanced sensitivity in the $200\,\mathrm{Hz}$--$1\,\mathrm{kHz}$ range, where promising sources are expected to emit continuous gravitational waves.
\end{itemize}

\begin{landscape}
\begin{table}
\footnotesize
\centering
\begin{tabular}{c c c c c c c c}
\hline
\multirow{2}{*}{Configuration} & \multicolumn{2}{c}{Annual Detections} & \multicolumn{3}{c}{Error Reduction (\%)} & Annual Detections of & Detectable Pulsars \\
 & BNS & BBH & $\Omega$ (BNS) & $\Omega$ (BBH) & $\tilde{\Lambda}$ & BNS Post-Merger Signals & in ATNF Catalog \\
\hline
BB40FIS\_HQS& $4.2^{+4.7}_{-2.5}$ & $48^{+55}_{-28}$ & 55 & 57 & 3.1 & 4$\times 10^{-5}$ -- 2$\times 10^{-3}$ & 5  \\
BB40FIS\_LB\_HQS& $4.7^{+5.3}_{-2.8}$ & $54^{+61}_{-31}$ & 56 & 58 & 3.1 & 3$\times 10^{-5}$ -- 2$\times 10^{-3}$ & 5  \\
BB23FDS\_HQS& $6.0^{+6.7}_{-3.5}$ & $71^{+80}_{-42}$ & 52 & 56 & 2.3 & 2$\times 10^{-5}$ -- 1$\times 10^{-3}$ & 2  \\
BB40FDS\_HQS& $8.8^{+9.9}_{-5.1}$ & $100^{+110}_{-60}$ & 53 & 58 & 2.5 & 2$\times 10^{-5}$ -- 1$\times 10^{-3}$ & 2  \\
BB40FDS\_LB\_HQS& $10^{+12}_{-6}$ & $120^{+130}_{-70}$ & 53 & 59 & 2.5 & 2$\times 10^{-5}$ -- 1$\times 10^{-3}$ & 2  \\
BB40FDS\_HQS\_BC& $15^{+16}_{-8}$ & $150^{+170}_{-90}$ & 55 & 61 & 2.8 & 2$\times 10^{-5}$ -- 9$\times 10^{-4}$ & 5  \\
HFmodFIS\_HQS& $2.6^{+3.0}_{-1.5}$ & $25^{+28}_{-14}$ & 65 & 52 & 11 & 1$\times 10^{-3}$ -- 6$\times 10^{-2}$ & 10\ \\
HFmodFDS\_HQS& $6.4^{+7.2}_{-3.8}$ & $66^{+75}_{-39}$ & 66 & 56 & 12 & 1$\times 10^{-3}$ -- 6$\times 10^{-2}$ & 10\	 	 	 \\
HF2kFIS\_HQS& $1.5^{+1.7}_{-0.9}$ & $16^{+18}_{-9}$ & 63 & 47 & 11 & 2$\times 10^{-3}$ -- 1$\times 10^{-1}$ & 4\ \\
HF2kFDS\_HQS& $4.0^{+4.6}_{-2.4}$ & $46^{+52}_{-27}$ & 64 & 51 & 12 & 3$\times 10^{-3}$ -- 1$\times 10^{-1}$ & 4\ \\
HF3kFIS\_HQS& $1.4^{+1.6}_{-0.8}$ & $17^{+19}_{-10}$ & 62 & 45 & 7.3 & 2$\times 10^{-3}$ -- 2$\times 10^{-1}$ & 3\ \\
HF3kFDS\_HQS& $2.0^{+2.2}_{-1.1}$ & $26^{+29}_{-15}$ & 59 & 43 & 5.1 & 1$\times 10^{-3}$ -- 1$\times 10^{-1}$ & 2\ \\
HF3k& $0.13^{+0.15}_{-0.08}$ & $2.0^{+2.3}_{-1.2}$ & 43 & 20 & 0.93 & 1$\times 10^{-4}$ -- 3$\times 10^{-2}$ & 0\ \\
HF3kFIS& $0.51^{+0.57}_{-0.3}$ & $5.2^{+5.9}_{-3.1}$ & 60 & 33 & 6.7 & 2$\times 10^{-3}$ -- 2$\times 10^{-1}$ & 3\ \\
\hline
\end{tabular}
\caption{Summary of science case studies with the proposed KAGRA configurations. The first column lists the configuration names. The second and third columns show expected annual detection rates of \bns and \bbh events, respectively (See Sec. \ref{sec:cbc_rates} for details). The fourth and fifth columns show the fractional reduction in sky-localization errors for \bns and \bbh events due to the inclusion of KAGRA in the detector network (See Sec. \ref{sec:cbc_skyloc} for details). These values are calculated as $100 (\Delta \Omega_{\text{w/o KAGRA}} - \Delta \Omega_{\text{w KAGRA}}) / \Delta \Omega_{\text{w/o KAGRA}}$, where $\Delta \Omega_{\text{w KAGRA}}$ and $\Delta \Omega_{\text{w/o KAGRA}}$ denote median sky localization errors with and without KAGRA respectively. The sixth column shows the fractional reduction in the median measurement errors of the tidal deformability parameter, $\tilde{\Lambda}$ (See Sec. \ref{sec:cbc_tides} for details). The seventh column shows the annual detection rates of \bns post-merger signals (See Sec. \ref{sec:cbc_post} for details). The final column shows the number of potentially detectable pulsars in the ATNF catalog (See Sec. \ref{sec:atnf} for details).}
\label{tab:science_summary}
\end{table}
\end{landscape}

\subsection{Compact Binary Coalescence}

\Ac{CBC} refers to the merger of two compact astrophysical objects such as \bhs or \nss that orbit each other and gradually spiral inward due to the emission of \gws.
By the end of the third observing run (O3) of \lvk, 90 \cbc events had been confidently detected \cite{LIGOScientific:2018mvr,LIGOScientific:2020ibl,LIGOScientific:2021usb,KAGRA:2021vkt}, with more than 200 additional events reported during its ongoing fourth observing run (O4).

\subsubsection{Range and Detection Rate} \label{sec:cbc_rates}

\begin{table*}[htbp]
\centering
\begin{tabular}{c c c c c}

\hline
\multirow{2}{*}{Configuration}
& \multirow{2}{*}{BNS Range (Mpc)} & \multicolumn{3}{c}{Annual Detections} \\
      &  & BNS \!& NSBH \!& BBH \\
\hline
BB40FIS\_HQS & 144 & $4.2^{+4.7}_{-2.5}$ & $0.78^{+0.88}_{-0.46}$ & $48^{+55}_{-28}$ \\
BB40FIS\_LB\_HQS & 150 & $4.7^{+5.3}_{-2.8}$ & $0.88^{+0.99}_{-0.51}$ & $54^{+61}_{-31}$ \\
BB23FDS\_HQS & 162 & $6.0^{+6.7}_{-3.5}$ & $1.2^{+1.4}_{-0.7}$ & $71^{+80}_{-42}$ \\
BB40FDS\_HQS & 184 & $8.8^{+9.9}_{-5.1}$ & $1.9^{+2.1}_{-1.1}$ & $100^{+110}_{-60}$ \\
BB40FDS\_LB\_HQS & 194 & $10^{+12}_{-6}$ & $2.1^{+2.4}_{-1.3}$ & $120^{+130}_{-70}$ \\
BB40FDS\_HQS\_BC & 218 & $15^{+16}_{-8}$ & $3.0^{+3.4}_{-1.8}$ & $150^{+170}_{-90}$ \\
HFmodFIS\_HQS & 124 & $2.6^{+3.0}_{-1.5}$ & $0.44^{+0.49}_{-0.25}$ & $25^{+28}_{-14}$ \\
HFmodFDS\_HQS & 166 & $6.4^{+7.2}_{-3.8}$ & $1.2^{+1.4}_{-0.7}$ & $66^{+75}_{-39}$ \\
HF2kFIS\_HQS & 103 & $1.5^{+1.7}_{-0.9}$ & $0.27^{+0.3}_{-0.16}$ & $16^{+18}_{-9}$ \\
HF2kFDS\_HQS & 142 & $4.0^{+4.6}_{-2.4}$ & $0.82^{+0.92}_{-0.48}$ & $46^{+52}_{-27}$ \\
HF3kFIS\_HQS & 99.9 & $1.4^{+1.6}_{-0.8}$ & $0.26^{+0.3}_{-0.15}$ & $17^{+19}_{-10}$ \\
HF3kFDS\_HQS & 111 & $2.0^{+2.2}_{-1.1}$ & $0.42^{+0.47}_{-0.25}$ & $26^{+29}_{-15}$ \\
HF3k & 44.9 & $0.13^{+0.15}_{-0.08}$ & $0.028^{+0.031}_{-0.016}$ & $2.0^{+2.3}_{-1.2}$ \\
HF3kFIS & 71.0 & $0.51^{+0.57}_{-0.3}$ & $0.084^{+0.095}_{-0.049}$ & $5.2^{+5.9}_{-3.1}$ \\
\hline
\end{tabular}
\caption{BNS sensitive ranges and annual detection rates for BNS, NSBH, and BBH sources across various KAGRA configurations. Sensitive ranges are calculated for non-spinning BNS systems with source-frame component masses of $1.4M_\odot$ each. For annual detection rates, median values are presented, with the 5\% and 95\% quantiles indicated as lower and upper subscripts, respectively.}
\label{tab:rate}
\end{table*}

Table \ref{tab:rate} presents \bns sensitive ranges and expected annual detection rates for various KAGRA configurations.
The \bns sensitive ranges are calculated for non-spinning binary systems with source-frame component masses of $1.4M_\odot$ each.
Detection rates are provided for different categories of \gw sources: \bns, \nsbh, and \bbh.
These categories are defined based on the assumption that objects with masses below $3M_\odot$ are classified as \nss, while those with masses above $3M_\odot$ are considered \bhs.
Detection rates are computed as $R \left<V \right>$, where $R$ represents the merger rate per comoving volume, and $\left<V\right>$ represents the sensitive volume \cite{Chen:2017wpg}, averaged over an assumed distribution of masses and spins.
The sensitivity volume is computed with a \snr threshold of $8$ for detection.
Following \cite{Kiendrebeogo:2023hzf}, we adopt the mass and spin distribution derived from fitting the POWER LAW + DIP + BREAK model \cite{Farah:2021qom, KAGRA:2021duu} to \cbcs observed up to the end of O3, and a total merger rate of $240^{+270}_{-140},\mathrm{Gpc^{-3}year^{-1}}$ \cite{KAGRA:2021duu}.
Using $10^6$ samples drawn from this astrophysical model available in \cite{GWTC3PDB}, we compute $R$ and  $\left<V\right>$ for each source category with a Monte-Carlo method.
The IMRPhenomD waveform model \cite{Husa:2015iqa, Khan:2015jqa} is used to compute both the sensitive ranges and the annual detection rates.

As shown in the table, BB options generally outperform HF options.
Furthermore, configurations with $40\,\mathrm{kg}$ mirrors and FDS give more annual detections than those with $23\,\mathrm{kg}$ mirrors and FID respectively.
The best configuration is BB40FDS\_HQS\_BC, which predicts $\mathcal{O}(10)$ \bns and $\mathcal{O}(10^2)$ \bbh events annually at the median.
Among the HF options, the moderate HF configurations tend to demonstrate better performance compared to their counterparts.

\subsubsection{Sky Localization} \label{sec:cbc_skyloc}

We investigate KAGRA's potential to enhance the sky localization of \gw sources across various configurations.
Our analysis assumes a network of the two LIGO detectors located at Hanford and Livingston in the US with the A\# configuration, the Virgo detector with the O5 configuration, and KAGRA.
The noise amplitude spectral densities we employ for LIGO and Virgo are available at \cite{LIGO-T2300041-v1} and \cite{LIGO-T2000012-v2} respectively.
We assume the O5 configuration of Virgo instead of the post-O5 configuration, since estimates on the Virgo's post-O5 sensitivity are not available at the time of writing.

In this study, we consider two types of signals characterized by their source-frame masses, $m^{\mathrm{source}}_1$ and $m^{\mathrm{source}}_2$, and the luminosity distance to the source, $D_{\mathrm{L}}$: \bns signals with $(m^{\mathrm{source}}_1, m^{\mathrm{source}}_2, D_{\mathrm{L}})=(1.4M_\odot,1.4M_\odot,200\,\mathrm{Mpc})$ and \bbh signals with $(m^{\mathrm{source}}_1, m^{\mathrm{source}}_2, D_L)=(30M_\odot,30M_\odot,1\,\mathrm{Gpc})$.
For each signal type, we simulate $10^5$ events with random locations and orientations. 
We then evaluate their sky-localization uncertainties using a Fisher information matrix, calculated with the IMRPhenomD waveform model.
To efficiently compute the derivatives of the waveform, we use the IMRPhenomD model implemented within the \texttt{ripple} library \cite{Edwards:2023sak}, which benefits from auto differentiation of \texttt{jax} \cite{jax2018github}.

The Fisher information matrix approach approximates the posterior distribution as a Gaussian centered around the true parameter values.
This approximation breaks down when the distribution is either non-Gaussian or multimodal.
In this study, we focus on loud sources with median network \snrs of $31$ for \bbhs and $74$ for \bnss, observed by a network of more than three detectors.
Consequently, the sky-position distribution is nearly Gaussian, and we expect our estimates to be accurate.

\begin{figure}[htbp]
	\centering
    \includegraphics[width=0.95\linewidth]{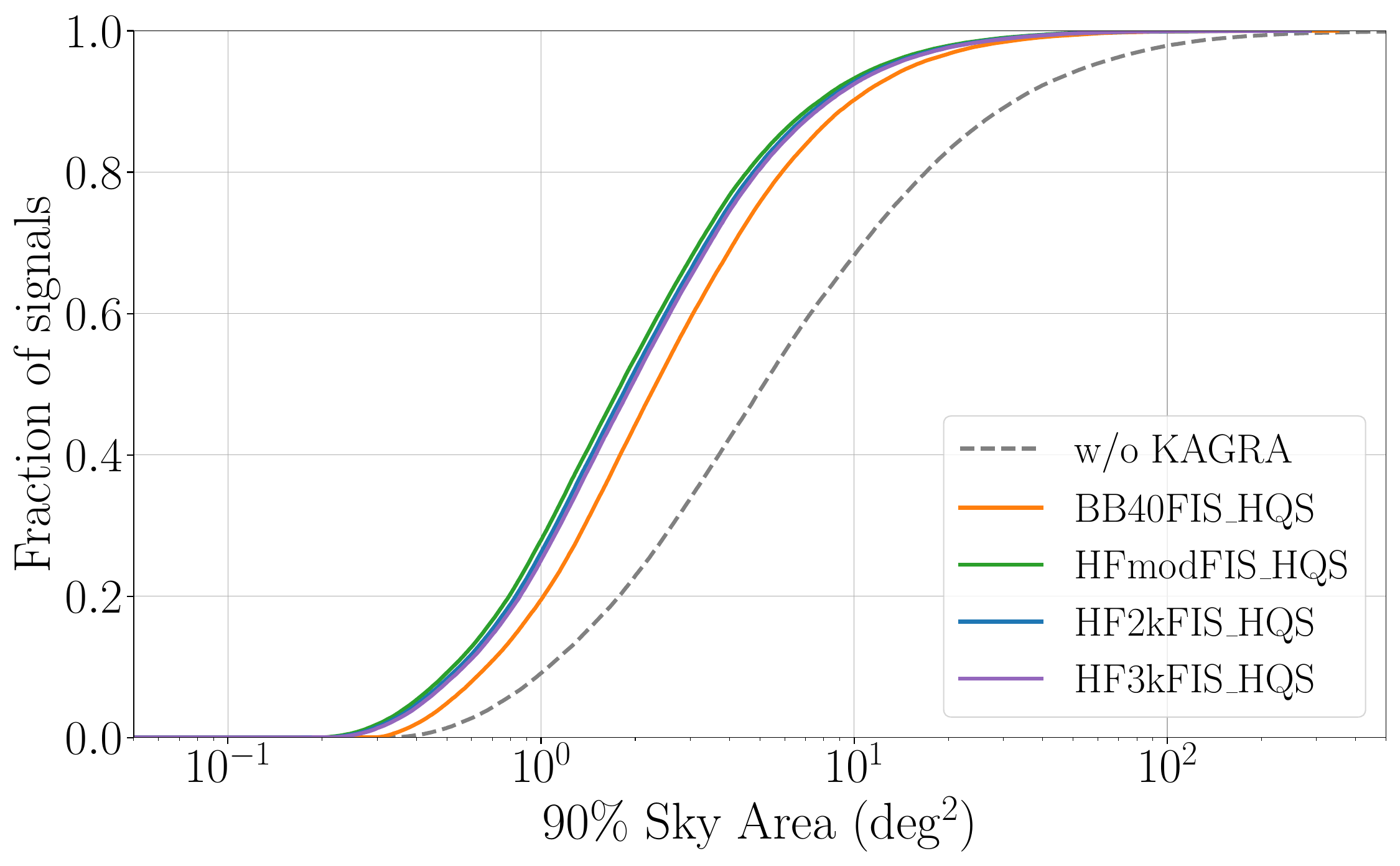}
    \includegraphics[width=0.95\linewidth]{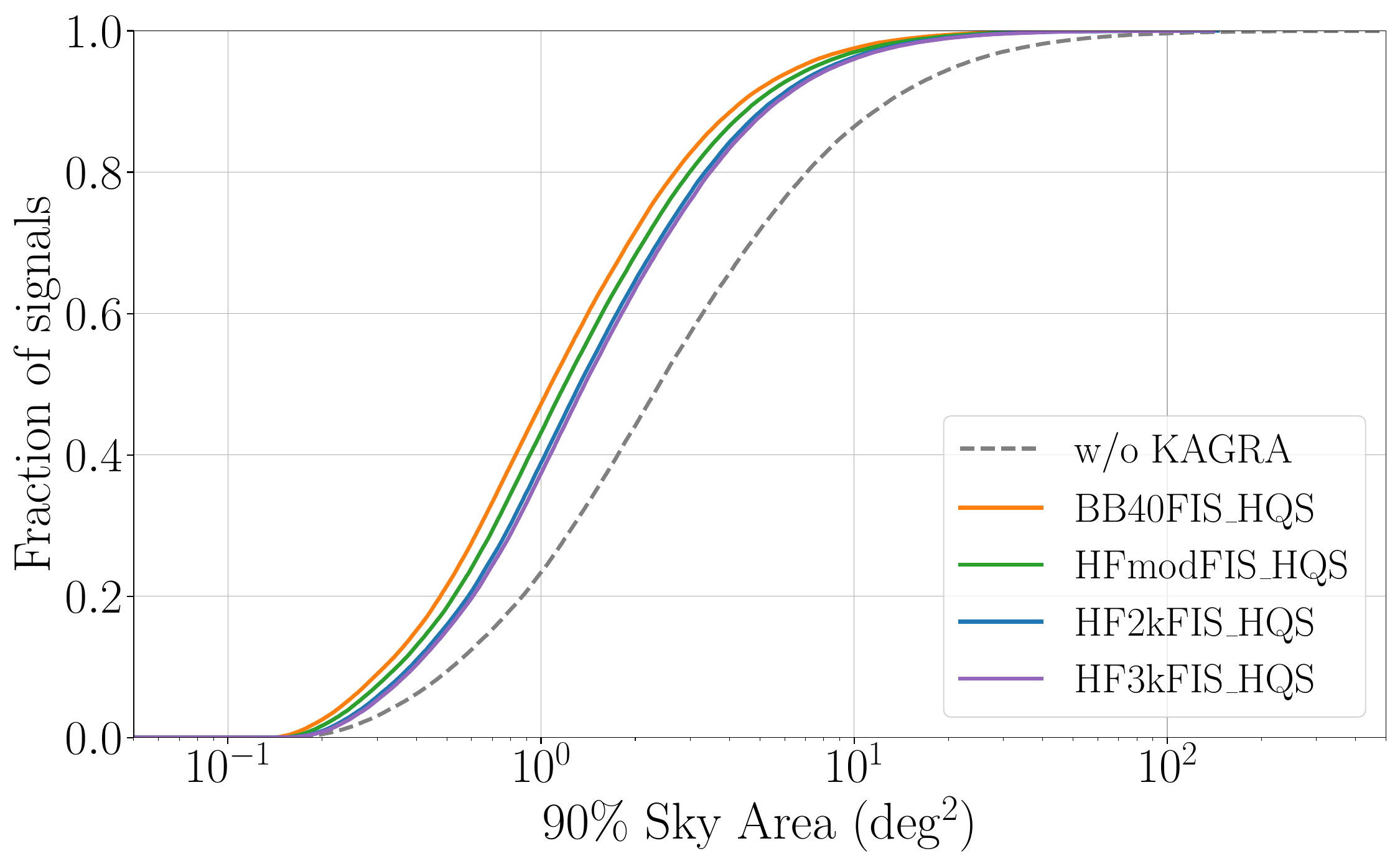}
	\caption{Cumulative distributions of the 90\% credible sky localization areas for the LVK detector network under various representative KAGRA configurations. The upper panel shows the results for binaries with detector-frame masses of $(m_1, m_2)=(1.4M_\odot,1.4M_\odot)$ \cbc at a luminosity distance of $D_{\mathrm{L}}=200\,\mathrm{Mpc}$. The lower panel shows the results for binaries with $(m_1, m_2, D_L)=(30M_\odot,30M_\odot,1\,\mathrm{Gpc})$.} \label{fig:skyloc}
\end{figure}

\begin{table*}[htbp]
\centering
\begin{tabular}{c c c c}
\hline
\multirow{2}{*}{Configuration}
& \multicolumn{2}{c}{Median 90\% sky area ($\mathrm{deg}^2$)} \\
 & $1.4M_\odot$--$1.4M_\odot$ at $200\,\mathrm{Mpc}$ & $30M_\odot$--$30M_\odot$ at $1\,\mathrm{Gpc}$ \\
\hline
Without KAGRA & $5.15$ & $2.39$ & \\
BB40FIS\_HQS & $2.33$ & $1.08$ & \\
BB40FIS\_LB\_HQS & $2.29$ & $1.05$ & \\
BB23FDS\_HQS & $2.48$ & $1.09$ & \\
BB40FDS\_HQS & $2.43$ & $1.05$ & \\
BB40FDS\_LB\_HQS & $2.40$ & $1.02$ & \\
BB40FDS\_HQS\_BC & $2.33$ & $0.96$ & \\
HFmodFIS\_HQS & $1.81$ & $1.19$ & \\
HFmodFDS\_HQS & $1.77$ & $1.09$ & \\
HF2kFIS\_HQS & $1.89$ & $1.33$ & \\
HF2kFDS\_HQS & $1.84$ & $1.22$ & \\
HF3kFIS\_HQS & $1.96$ & $1.38$ & \\
HF3kFDS\_HQS & $2.13$ & $1.41$ & \\
HF3k & $2.92$ & $1.98$ & \\
HF3kFIS & $2.04$ & $1.66$ & \\
\hline
\end{tabular}
\caption{Median 90\% credible sky localization areas for the LVK detector network under various representative KAGRA configurations and different signal types.}
\label{tab:skyloc}
\end{table*}

Figure \ref{fig:skyloc} shows the cumulative distributions of the 90\% sky areas for representative KAGRA configurations. 
For comparison, the results obtained without KAGRA are shown as dashed gray lines.
In either configuration, the inclusion of KAGRA shifts the curve to the left, indicating improved precision in sky localization.
The median 90\% sky area for each configuration is presented in Tab. \ref{tab:skyloc}.

For the \bns signals, the HF options are generally better than the BB options in terms of reducing the sky area.
This is due to the extended bandwidth, which enhances timing accuracy and, consequently, sky localization.
Among the HF configurations, the HF-mod configurations provide the best results.
For \bbh signals, the BB configurations perform better, as these sources merge at lower frequencies.


\subsubsection{Tidal Deformability} \label{sec:cbc_tides}

Coalescing \nss in a \bns system are tidally deformed by their companions, which affects the phase evolution of \gws \cite{Flanagan:2007ix}.
This tidal effect becomes more pronounced in the late stage of the inspiral, and it can be better measured with improvements in sensitivity around the merger frequency, approximately $1\,\mathrm{kHz}$.
By observing the tidal correction to the \gw phase, we can measure the tidal deformability parameters of \nss, $\Lambda_1$ and $\Lambda_2$.
These parameters are dependent on the nuclear \eos, and accurately measuring them has profound implications for nuclear physics.

In this study, we investigate the capabilities of various KAGRA configurations to measure the tidal deformability parameters.
The leading-order tidal correction depends on $\tilde{\Lambda}$ defined as follows,
\begin{equation}
\tilde \Lambda \equiv \frac{16}{13} \frac{(m_1 + 12 m_2) m_1^4 \Lambda_1 + (m_2 + 12 m_1) m^4_2 \Lambda_2}{(m_1 + m_2)^5}.
\end{equation}
Since other tidal parameters are not well constrained, we focus on evaluating the measurement precision of $\tilde{\Lambda}$.

We assess the measurement precisions using a Fisher information matrix, calculated with the IMRPhenomD\_NRTidalv2 waveform model \cite{Dietrich:2019kaq}.
We use the IMRPhenomD\_NRTidalv2 model implemented within the \texttt{ripple} library \cite{Edwards:2023sak} to compute the waveform derivatives efficiently.
We assume a detector network consisting of KAGRA and the two LIGO observatories in Hanford and Livingston with the A\# configuration.
Virgo is not included in this study due to the unavailability of its post-O5 design sensitivity at the time of writing.
As considered in the sky-localization study, our simulations consider non-spinning \bns systems with source-frame masses of $(m^{\mathrm{source}}_1, m^{\mathrm{source}}_2) = (1.4M_\odot, 1.4M_\odot)$ and a luminosity distance of $D_{\mathrm{L}} = 200\,\mathrm{Mpc}$.
The tidal deformability parameters are set to $\Lambda_1 = \Lambda_2 = 3.1 \times 10^2$, consistent with predictions from the SLy \eos \cite{Douchin:2001sv}.
We generate $10^5$ simulated events with random sky locations and binary orientations, and estimate the measurement precision of $\tilde{\Lambda}$ for each event using the Fisher information matrix.

\begin{figure}[htbp]
	\centering
    \includegraphics[width=0.95\linewidth]{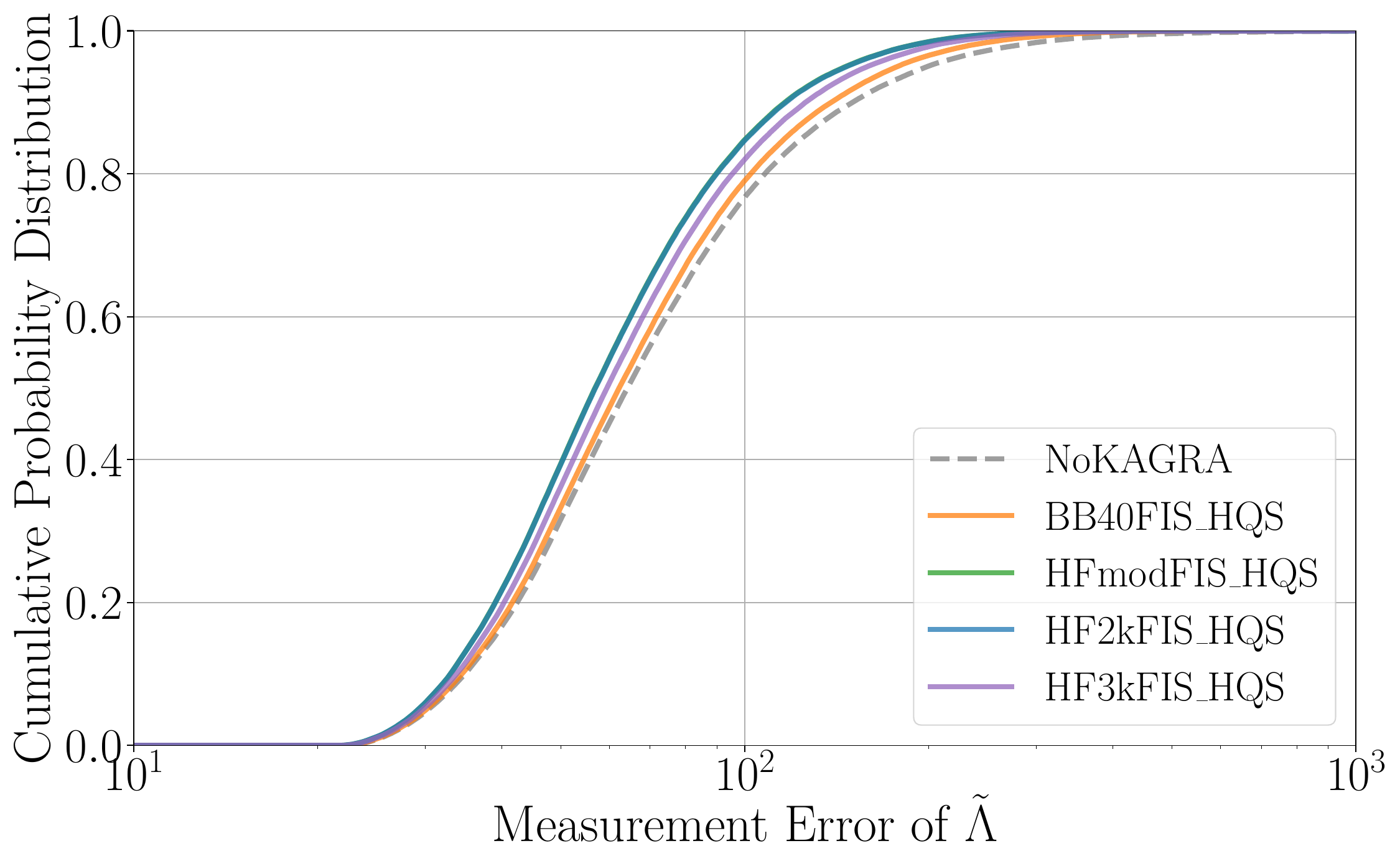}
	\caption{Cumulative distributions of measurement errors in $\tilde{\Lambda}$ for $1.4M_\odot$–$1.4M_\odot$ \bnss at a luminosity distance of $200\,\mathrm{Mpc}$, observed with the LIGO-KAGRA detector network under various KAGRA configurations. The results for HFmodFIS\_HQS and HF2kFIS\_HQS are nearly identical, and their curves overlap.} \label{fig:tidal_cumulative}
\end{figure}

\begin{table*}[htbp]
\centering
\begin{tabular}{c c}
\hline
Configuration & Median measurement error in $\tilde{\Lambda}$ \\
\hline
Without KAGRA & 64.2 \\
BB40FIS\_HQS & 62.2 \\
BB40FIS\_LB\_HQS & 62.2 \\
BB23FDS\_HQS & 62.7 \\
BB40FDS\_HQS & 62.6 \\
BB40FDS\_LB\_HQS & 62.6 \\
BB40FDS\_HQS\_BC & 62.4 \\
HFmodFIS\_HQS & 56.9 \\
HFmodFDS\_HQS & 56.7 \\
HF2kFIS\_HQS & 56.9 \\
HF2kFDS\_HQS & 56.5 \\
HF3kFIS\_HQS & 59.5 \\
HF3kFDS\_HQS & 60.9 \\
HF3k & 63.6 \\
HF3kFIS & 59.9 \\
\hline
\end{tabular}
\caption{The median measurement errors in $\tilde{\Lambda}$ for $1.4M_\odot$–$1.4M_\odot$ \bnss at a luminosity distance of $200\,\mathrm{Mpc}$, observed with the LIGO-KAGRA detector network under various KAGRA configurations.}
\label{tab:tidal_median}
\end{table*}

Figure \ref{fig:tidal_cumulative} shows the cumulative distributions of measurement errors in $\tilde{\Lambda}$ for representative KAGRA configurations.
For comparison, results from the LIGO-only network are shown as a dashed gray line.
In all cases, the inclusion of KAGRA shifts the curves leftward, indicating improved measurement precisions.
As illustrated in the figure, the HFmodFIS\_HQS and HF2kFIS\_HQS configurations yield the most significant improvements in $\tilde{\Lambda}$ precision.
This can be attributed to their enhanced sensitivity in the frequency range just before merger, where tidal effects are most prominent.
Compared to the LIGO-only case, these configurations reduce the median measurement error of $\tilde{\Lambda}$ by approximately 10\%.
The median measurement errors for other configurations are summarized in Tab. \ref{tab:tidal_median}, confirming that the HFmod and HF2k configurations generally offer the best performance.

\begin{figure}[htbp]
\centering
\includegraphics[width=0.9\linewidth]{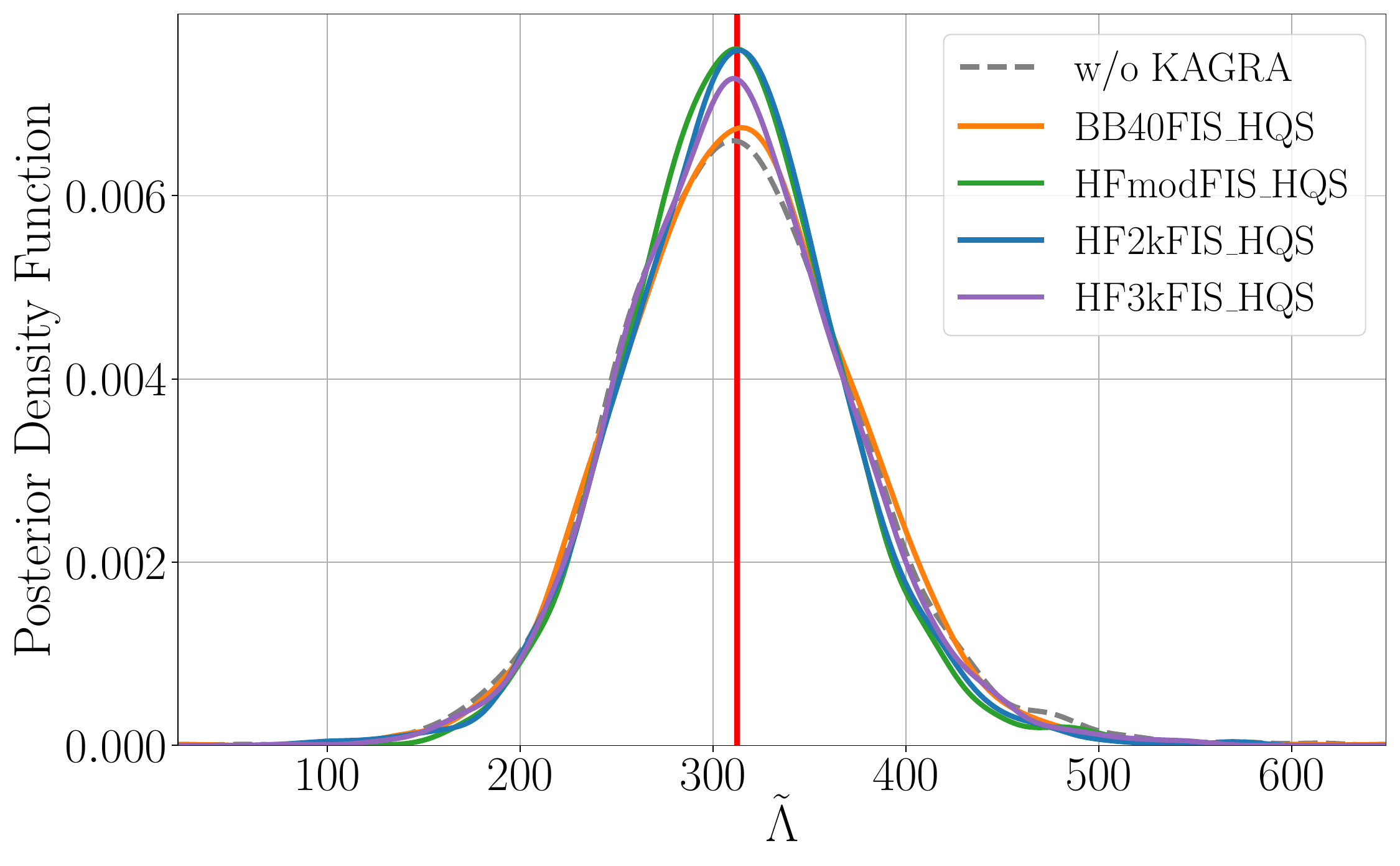} 
\caption{The posterior probability density function of $\tilde{\Lambda}$ for one of the simulated \bns signals and representative KAGRA configurations, compared to the LIGO-only result shown as a dashed gray line. The red vertical line shows the true value. The optimal \snrs are 28 and 27 for LIGO Hanford and Livingston, respectively. The optimal \snr for KAGRA is 7.6, 6.5, 5.3, and 5.1 for BB40FIS\_HQS, HFmodFIS\_HQS, HF2kFIS\_HQS, and HF3kFIS\_HQS respectively.}
\label{fig:lambda_posterior_1}
\end{figure}

\begin{figure}[htbp]
\centering
\includegraphics[width=0.9\linewidth]{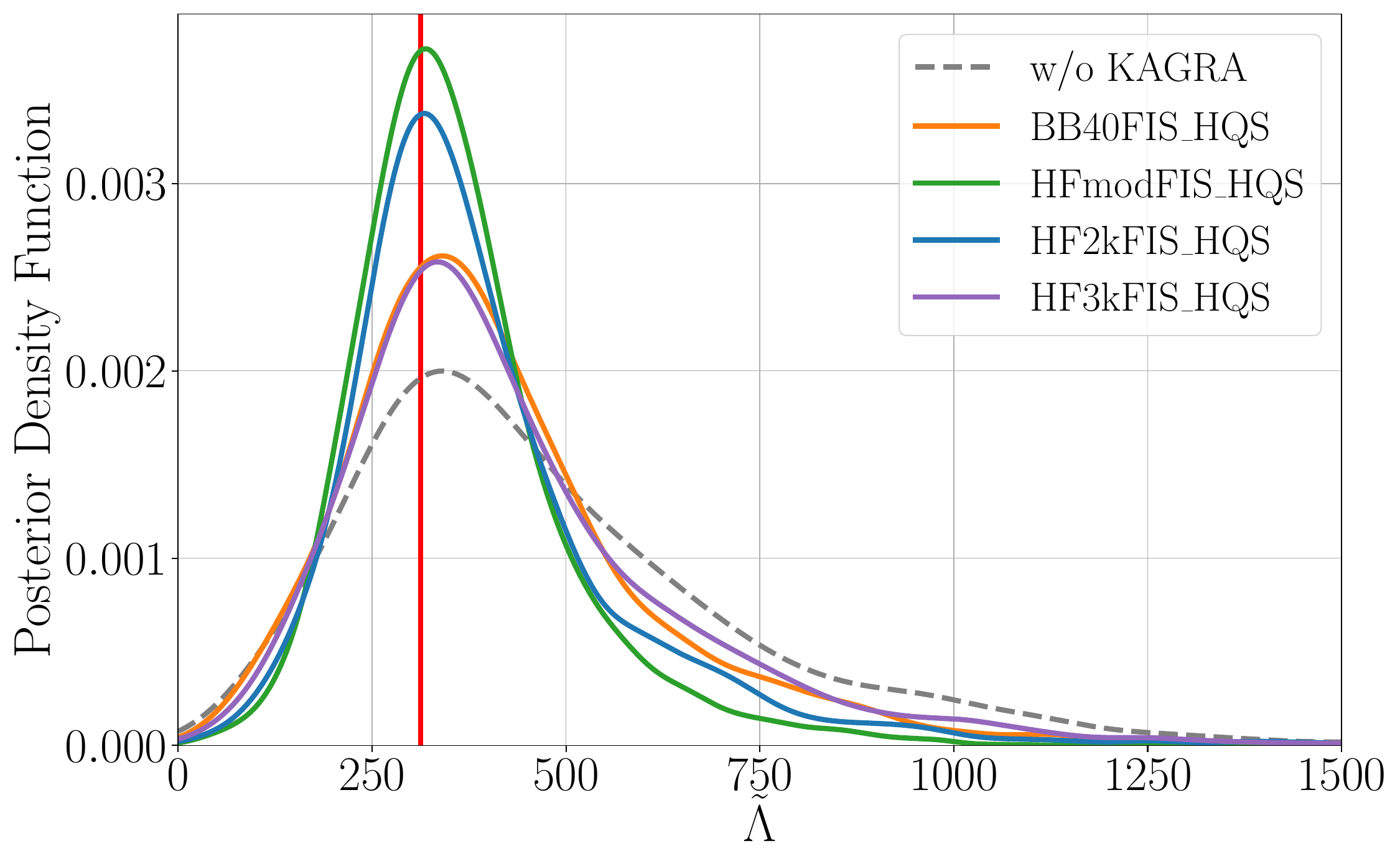} 
\caption{The posterior probability density function of $\tilde{\Lambda}$ for one of the simulated \bns signals and representative KAGRA configurations, compared to the LIGO-only result shown as a dashed gray line. The red vertical line shows the true value. The optimal \snrs are 17 and 5.2 for LIGO Hanford and Livingston, respectively. The optimal \snr for KAGRA is 7.0, 6.0, 4.9, and 4.7 for BB40FIS\_HQS, HFmodFIS\_HQS, HF2kFIS\_HQS, and HF3kFIS\_HQS respectively.}
\label{fig:lambda_posterior_2}
\end{figure}

To validate the estimates obtained from the Fisher information matrix, we also perform Bayesian parameter estimation on selected simulated signals, utilizing the Bilby software \cite{Ashton:2018jfp,Romero-Shaw:2020owr}.
For accelerating likelihood evaluations in parameter estimation, we employ the reduced order quadrature (ROQ) technique \cite{Canizares:2014fya,Smith:2016qas}, specifically utilizing the ROQ bases built in \cite{Morisaki:2023kuq}.

Figure \ref{fig:lambda_posterior_1} shows the posterior probability density function of $\tilde{\Lambda}$ for one of the simulated signals and representative KAGRA configurations, compared to the LIGO-only result shown as a dashed gray line.
The optimal \snrs are 28 and 27 for Hanford and Livingston, respectively, and range from 5.1 to 7.6 for KAGRA, depending on its configuration.
The Fisher matrix analysis predicts a reduction in measurement error by 11\% with HFmodFIS\_HQS and HF2kFIS\_HQS, by 6.9\% with HF3kFIS\_HQS, and by 2.7\% with BB40FIS\_HQS.
Consistent with this prediction, the posterior distributions are narrowest for HFmodFIS\_HQS and HF2kFIS\_HQS, as shown in the figure.
The width of the 90\% credible interval is reduced by 11\%, 6.3\%, 5.1\%, and 3.5\% with HFmodFIS\_HQS, HF2kFIS\_HQS, HF3kFIS\_HQS, and BB40FIS\_HQS respectively, roughly matching the results from the Fisher matrix.

Figure \ref{fig:lambda_posterior_2} shows the posterior probability density function of $\tilde{\Lambda}$ for another simulated signal, whose optimal \snrs are 17 and 5.2 for Hanford and Livingston, respectively, and range from 4.7 to 7.0 for KAGRA, depending on its configuration.
The Fisher matrix analysis predicts a reduction in measurement error by 29\% with HFmodFIS\_HQS and HF2kFIS\_HQS, by 21\% with HF3kFIS\_HQS, and by 9.7\% with BB40FIS\_HQS.
The actual width of the 90\% credible interval is reduced by 49\%, 35\%, 19\%, and 23\% with HFmodFIS\_HQS, HF2kFIS\_HQS, HF3kFIS\_HQS, and BB40FIS\_HQS respectively.
While the results do not exactly match the Fisher matrix predictions, HFmodFIS\_HQS and HF2kFIS\_HQS consistently yield the best performance, and the parameter estimation confirms that an improvement of over 10\% is achievable in these configurations.

\subsubsection{Binary Neutron Star Post-Merger Signals} \label{sec:cbc_post}


If a \bns merger produces a long-lived \ns remnant, \gws may be emitted in the frequency range of 1--4 kHz for several hundred milliseconds. These post-merger \gw signals contain rich information about the \ns \eos, particularly at high densities and temperatures not accessible in the inspiral phase. In the core of the remnant \ns, temperature-dependent phase transitions, such as a transition from hadronic matter to deconfined quark matter, may occur. Measuring GWs from both the inspiral and post-merger phases could therefore offer a unique opportunity to identify such phase transitions. The post-merger phases could be the only places in the Universe where such densities/temperatures occur. 

\begin{figure}[htbp]
\centering
\includegraphics[width=0.9\linewidth]{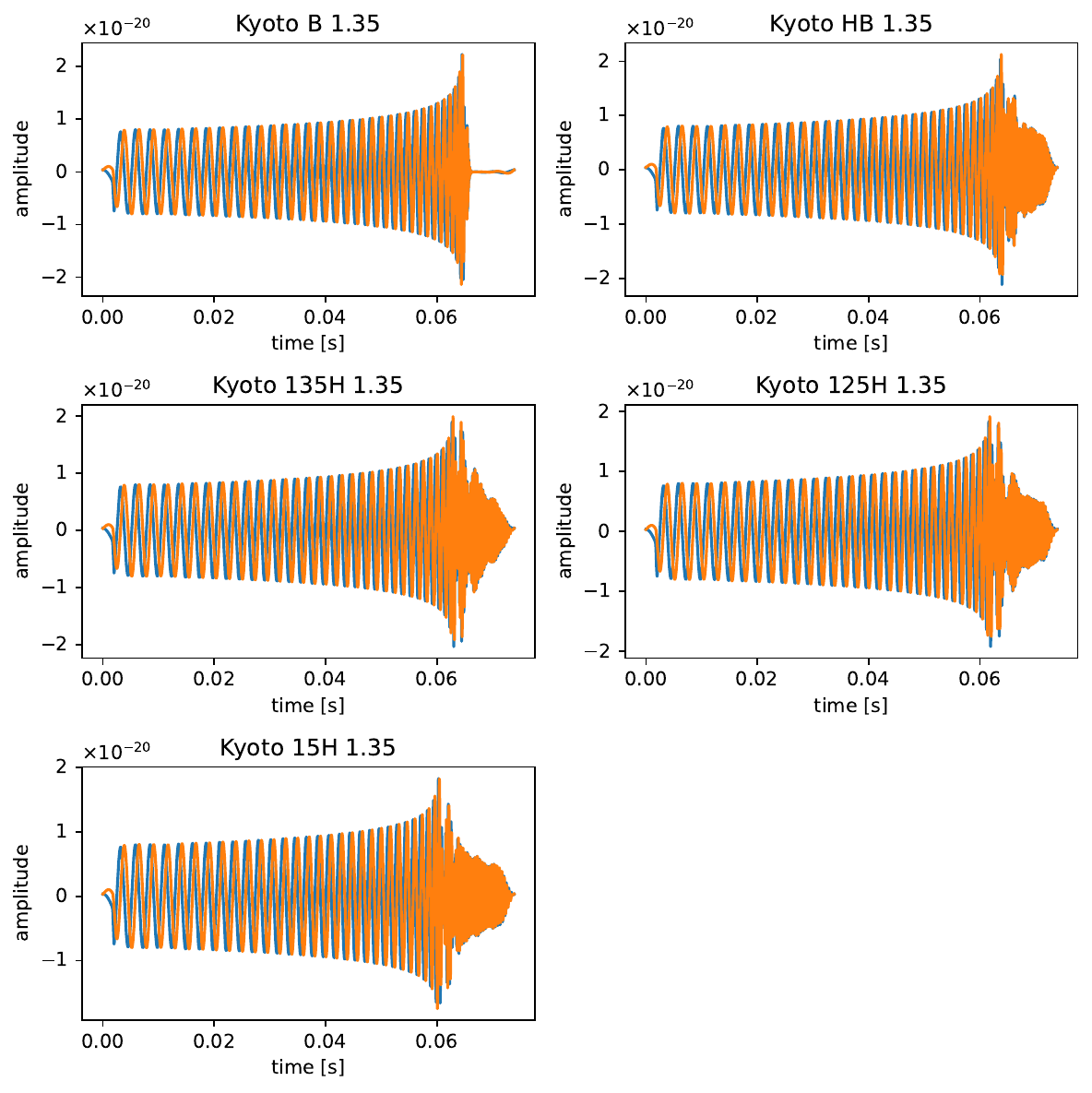} 
\caption{5 \bns waveform models by Kyoto group \cite{Kiuchi:2019kzt}. Blue and orange lines are $h_+$ and $h_{\times}$ modes respectively.}
\label{fig:KyotoGWWaveform}
\end{figure}

\begin{figure}[htbp]
\centering
\includegraphics[width=0.9\linewidth]{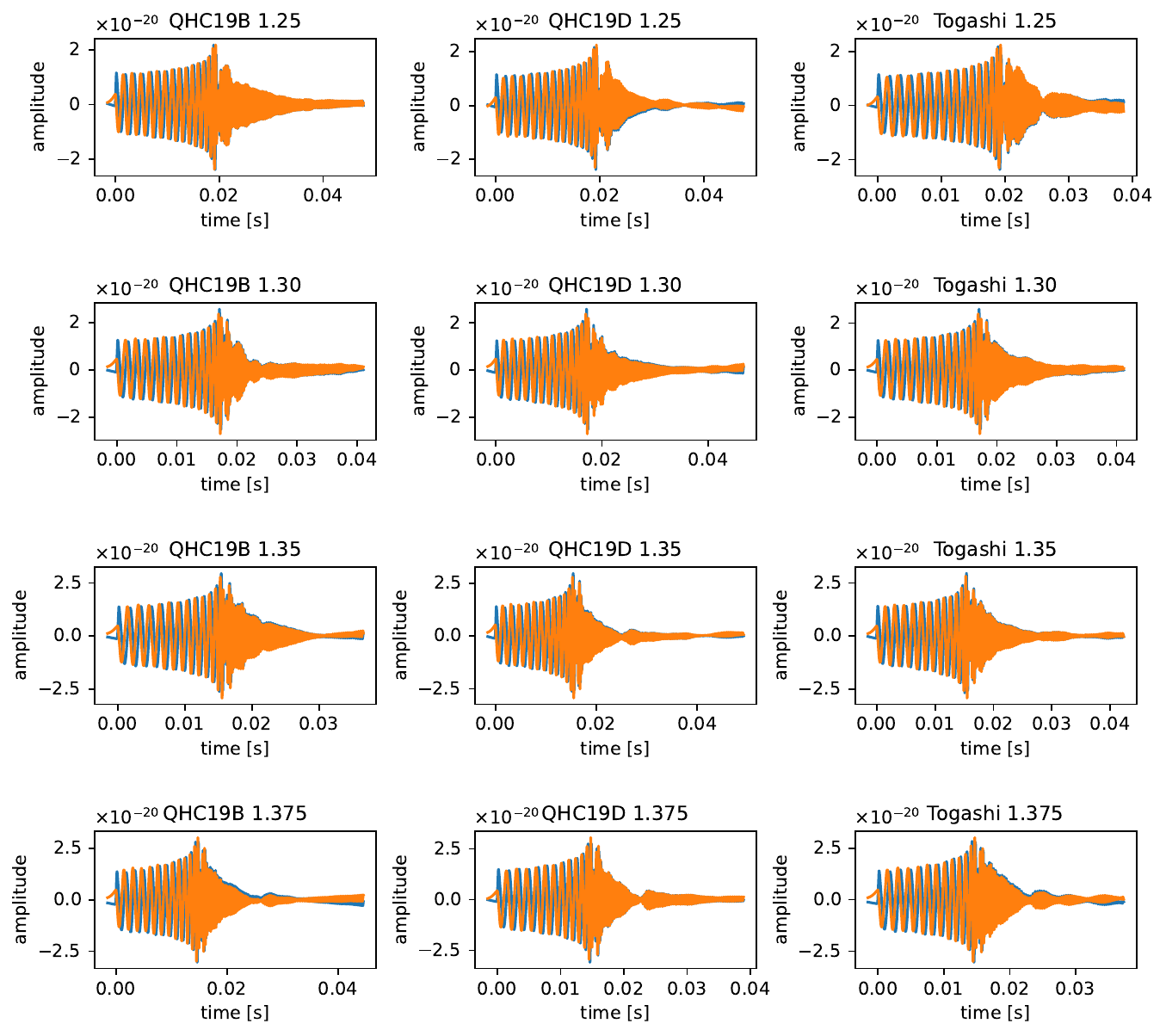} 
\caption{12 \bns waveform models by Huang et al.\cite{Huang:2022mqp}. Blue and orange lines are $h_+$ and $h_{\times}$ modes respectively.}
\label{fig:LucaGWWaveform}
\end{figure}

\begin{figure}[tbp]
\centering
\centering
\includegraphics[width=\columnwidth]{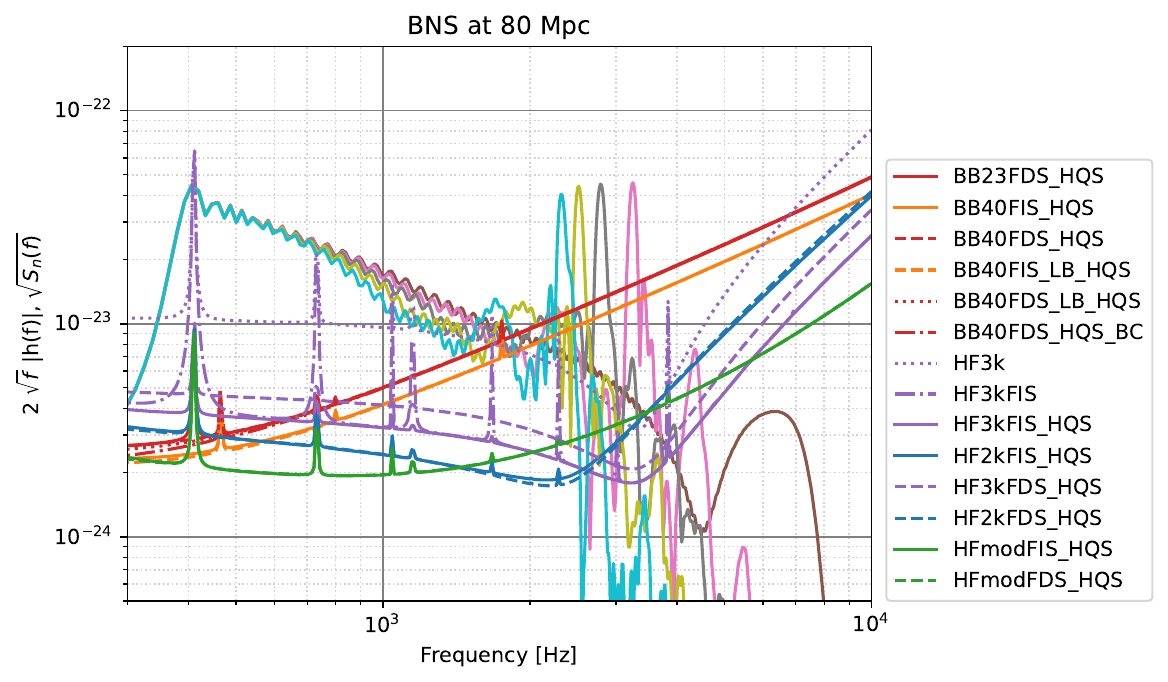}
\caption{The spectra of 5 \bns waveforms by Kyoto group \cite{Kiuchi:2019kzt}, compared with the sensitivities of proposed KAGRA configurations. Configuration names are indicated in the legend. The solid lines with blown, pink, gray, olive, cyan colors are the spectra of Kyoto B 1.35, Kyoto HB 1.35, Kyoto 135H 1.35, Kyoto 125H 1.35, and Kyoto 15H 1.35, respectively.}
\label{fig:KyotoGWandPSD}
\end{figure}
\begin{figure}[tbp]
\centering
\includegraphics[width=\columnwidth]{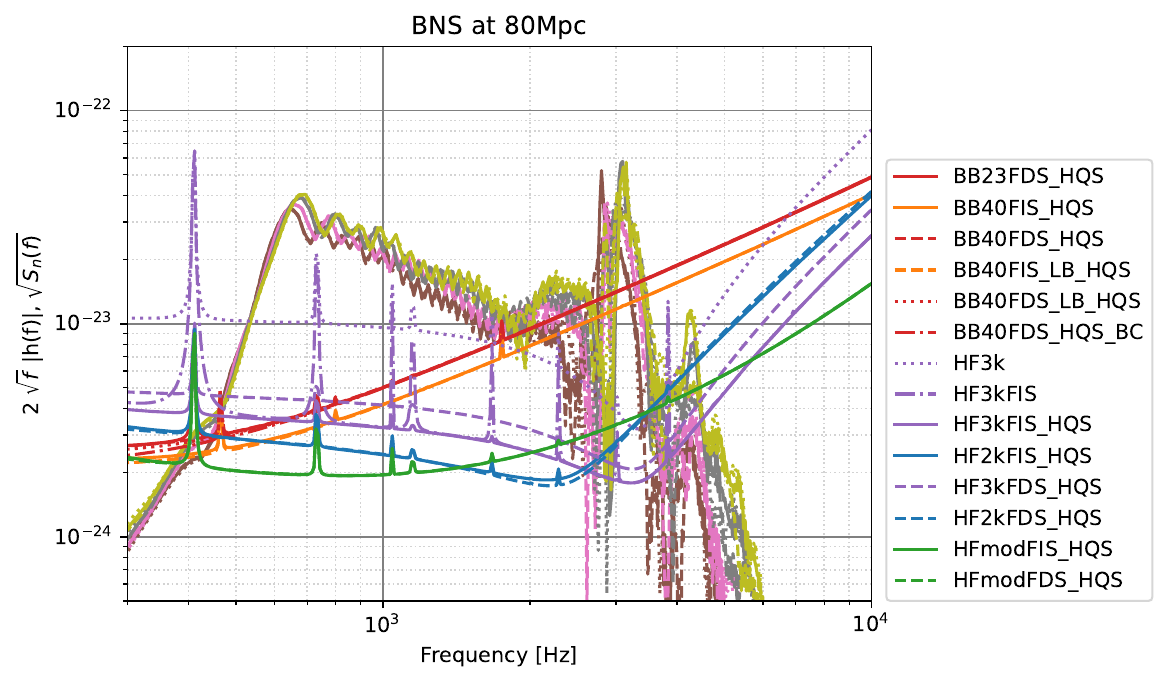}
\caption{The spectra of 12 \bns waveforms by Huang et al. \cite{Huang:2022mqp}, compared with the sensitivities of proposed KAGRA configurations. Configuration names are indicated in the legend. The solid lines with blown, pink, gray and olive colors are the spectra with masses of $1.25M_\odot$, $1.30M_\odot$, $1.35M_\odot$, $1.375M_\odot$, respectively,  and the solid, dashed, dotted lines represent QHC19B, QHC19D, and Togashi models, respectively.}
\label{fig:LucaGWandPSD}
\end{figure}


Currently, accurate modeling of post-merger GW signals relies on complex general relativistic hydrodynamics simulations.
No precise analytical waveform models exist that can be used for matched filtering or Bayesian parameter estimation methods based on templates.
Nevertheless, a signal reconstruction method called BayesWave \cite{Cornish:2014kda}, which decomposes the signal into a sum of sine-Gaussian wavelets, allow for morphology-independent recovery of post-merger waveforms.
It enables posterior estimation of key features, such as the characteristic peak frequency.
Prior studies have shown that, for signals with post-merger matched filter \snr $\rho \geq 5$, the peak frequency can be measured with an uncertainty of tens of hertz \cite{Chatziioannou:2017ixj, Torres-Rivas:2018svp}.



Following the methodology used in \cite{Ackley:2020atn}, we estimate the number of detectable post-merger signals, adopting $\rho = 5$ as the detection threshold.
The sensitive volume for this threshold is computed with a Monte Carlo method, where events are uniformly distributed within the horizon distance and isotropically over inclination angles, and the fraction with $\rho \geq 5$ is recorded.
Multiplying this sensitive volume by a \bns merger rate, $105.5^{+190.2}_{-83.9}\, \text{Gpc}^{-3}\, \text{year}^{-1}$ \cite{KAGRA:2021duu}, yields the expected detection rates. To evaluate the impact of various equations of state and simulation configurations, we employ 5 waveforms with equal-mass $(1.35, 1.35)\, M_\odot$ binaries from the Kyoto group, and 12 waveforms with varying component masses and \eoss from Huang et al., shown in Figs. \ref{fig:KyotoGWWaveform} and \ref{fig:LucaGWWaveform}. When we compute $\rho$, we only consider the frequency components above $1600\,\mathrm{Hz}$ to isolate the post-merger contribution.

\begin{table}[tbp]
\footnotesize
\begin{tabular}{|c|c|c|c|c|}
\hline
    & minimum of & maximum of & minimun of & maximum of\\
PSD & horizon (Mpc) & horizon (Mpc) & event rate (1/year) & event rate (1/year)\\ \hline
A+             & 26.1 & 39.4  & 1$\times 10^{-4}$ & 6$\times 10^{-3}$ \\ \hline
A\#            & 60.1 & 90.7  & 2$\times 10^{-3}$ & 8$\times 10^{-2}$ \\ \hline
BB23FDS HQS    & 13.9 & 21.0  & 2$\times 10^{-5}$ & 1$\times 10^{-3}$ \\ \hline
BB40FIS HQS    & 16.7 & 25.1  & 4$\times 10^{-5}$ & 2$\times 10^{-3}$ \\ \hline
BB40FDS HQS    & 13.8 & 20.9  & 2$\times 10^{-5}$ & 1$\times 10^{-3}$ \\ \hline
BB40FIS LB HQS & 16.7 & 25.2  & 3$\times 10^{-5}$ & 2$\times 10^{-3}$ \\ \hline
BB40FDS LB HQS & 13.8 & 20.9  & 2$\times 10^{-5}$ & 1$\times 10^{-3}$ \\ \hline
BB40FDS HQS BC & 13.8 & 20.9  & 2$\times 10^{-5}$ & 9$\times 10^{-4}$ \\ \hline
HF3k           & 23.9 & 63.4  & 1$\times 10^{-4}$ & 3$\times 10^{-2}$ \\ \hline
HF3kFIS        & 60.9 & 127.9 & 2$\times 10^{-3}$ & 2$\times 10^{-1}$ \\ \hline
HF3kFIS HQS    & 67.3 & 132.4 & 2$\times 10^{-3}$ & 2$\times 10^{-1}$ \\ \hline
HF2kFIS HQS    & 66.9 & 104.0 & 2$\times 10^{-3}$ & 1$\times 10^{-1}$ \\ \hline
HF3kFDS HQS    & 54.1 & 111.6 & 1$\times 10^{-3}$ & 1$\times 10^{-1}$ \\ \hline
HF2kFDS HQS    & 69.6 & 108.7 & 3$\times 10^{-3}$ & 1$\times 10^{-1}$ \\ \hline
HFmodFIS HQS   & 55.7 & 84.2  & 1$\times 10^{-3}$ & 6$\times 10^{-2}$ \\ \hline
HFmodFDS HQS   & 55.7 & 84.2  & 1$\times 10^{-3}$ & 6$\times 10^{-2}$ \\ \hline
\end{tabular}
\caption{Horizon distances and expected detection rates of \bns post-merger signals under proposed KAGRA configurations, as well as the A+ and A\# configurations of LIGO. The \snr threshold is set to 5. The ranges reflect uncertainties in the \bns merger rate, \eoss, and simulation results. } 
\label{tab:NumEventPostMerger}
\end{table}


Table \ref{tab:NumEventPostMerger} shows the horizon distances and expected detection rates for proposed KAGRA configurations, as well as the A+ and A\# configurations of LIGO.
As expected, the HF configurations yield higher detection rates compared to the BB configurations, reaching up to $\mathcal{O}(0.1)\, \mathrm{year^{-1}}$.
Notably, the HF2k and HF3k configurations can surpass the detection rates of the LIGO configurations, highlighting a unique scientific opportunity for KAGRA in the post-merger regime.

\subsection{Continuous Waves}



A continuous \gw or continuous wave (CW) is a kind of \gw
that lasts so long that we need to consider the Doppler effects of the Earth's rotation and orbital motion 
to accumulate a \snr.  Requiring that the Doppler frequency shift $f_{\GW} v_{\oplus}/c$ does not cross one 
frequency bin $1/T$ during an observation period $T$ gives us  $T\le 640$ seconds at $f_{\GW} = 1$kHz.

Possible continuous wave sources include, but are not limited to: rotating \nss with non-axisymmetric mass distribution, 
oscillations of \nss (r-mode, f-mode), precession and glitches of \nss,  
giant flares of magnetars, dark matter clouds around \bhs, 
ultra-light dark matters around the Earth, etc. Quark stars and other compact stars are not excluded. 
In any case, there has been no successful detection of continuous waves yet. 
See, e.g., \cite{2018ASSL..457..673G,2022Galax..10...72P,2023LRR....26....3R} for reviews.
For our purpose of assessing which proposed configuration in  the KAGRA upgrade plan gives us a higher chance of possible CW detection, we consider 
rotating \nss with non-axisymmetric mass distribution until the section \ref{subsubsec_unknown_pulsars}.
Prospects for detection of possible post-glitch CWs are studied in Sec. \ref{subsubsec:pulsar_glitch}.

As of writing, the ATNF Pulsar Catalog \cite{2005AJ....129.1993M} lists about 4000 pulsars, and about 1000 of them have rotation frequencies 
larger than 5 Hz. Of the pulsars with rotation frequencies above 5 Hz, about 400 pulsars are in binary systems.
Phase 1 of the Square Kilometer Array (SKA1)  is expected to find 16000 normal pulsars and 2300 millisecond pulsars.
In phase 2, it is potentially expected to find all pulsars in the Milky Way Galaxy whose beams are directed toward us \cite{2015aska.confE..40K}.

\subsubsection{Rapidly rotating non-axisymmetric neutron stars} \label{sec:atnf}

Rapidly rotating compact stars with non-axisymmetric mass distributions are typical sources of CW.
The frequency of the \gw is twice the rotation frequency, and the amplitude of the \gw is
\begin{align}
h_{q} &\simeq 
1.1\times10^{-27}\left(\frac{\epsilon}{10^{-7}}\right)\left(\frac{{\cal I}}{10^{38}{\rm kg\cdot m^2}}\right)
\left(\frac{r}{\rm 1 kpc}\right)^{-1}\left(\frac{f_{\GW}}{100{\rm Hz}}\right)^2
\label{eq:CW_hq}
\end{align}
where ${\cal I}$ is the moment of inertia around the rotation axis and $\epsilon$ is the degree of non-axisymmetry around the rotation axis.
The maximum possible value of $\epsilon$ depends on the equation of state of the compact star and possibly its internal magnetic field. 
When the magnetic field is not considered, the maximum value of $\epsilon$
is expected to be about $\epsilon\sim10^{-7}$ for an isolated \ns.
On the other hand, if the star has a strong magnetic field of the order of $10^{15}$ G like a magnetar, 
or if it is in an accreting binary system, or if it is a newly born star, the maximum value of $\epsilon$
could be about $\epsilon\sim10^{-5}$.
It has also been pointed out that observations may indicate a possible minimum value of $\epsilon$ ($\sim10^{-9}$) 
\cite{2018ApJ...863L..40W}.

Pulsars lose angular momentum by radiating electromagnetic waves, pulsar winds, and \gws.
Assuming that the time variation of rotation frequency is entirely due to \gws
then the upper limit of the \gw amplitude can be obtained. 
Assuming, e.g., the Crab pulsar, 
\begin{align}
h_{q, {\rm sd}} &\simeq 
1\times10^{-24}\left(
\frac{{\cal I}}{10^{38}{\rm kg\cdot m^2}}
\frac{|\dot f_{\rm s}|}{4\times10^{-10} {\rm Hz/s}}
\frac{30{\rm Hz}}{f_{\rm s}}
\right)^{1/2}\left(\frac{r}{2.5 \rm kpc}\right)^{-1}
\label{eq:CW_hqsd}
\end{align}
This upper limit is called the spin-down upper limit.
\footnote{To estimate $\dot f_{\rm s}$, we need to take into account the Shklovskii effect. However, in this paper we do not correct $\dot f_{\rm s}$ but requiring $\dot f_{\rm s} <0$ for simplicity, expecting that overall tendency would not change.}

The quantity to be compared with $h_{\rm q}$ and $h_{\rm q,sd}$ 
is what quantifies how large amplitude of CW can be detected by a search.
If we disregard the computational cost of data analysis, it can be written as follows.
\begin{align}
h_{\rm ul} &= C\sqrt{\frac{S_h}{T}}
\label{eq:CW_hul}
\end{align}
where $C$ is a proportionality constant that depends on the adopted analysis method and is assumed to be $C\sim10.8$ 
here (corresponding to the Bayesian 95 $\%$ strain amplitude upper limit from a time domain known pulsar search method using a coherent integration \cite{2005PhRvD..72j2002D}). 
$S_h(f)$ is the detector sensitivity (one-sided power spectrum) and $T$ is the observation time.
The latest result \cite{2025arXiv250101495T} shows that the upper limits of the CW amplitudes are below the spin-down limit for 29 pulsars.
The minimum value of the upper limit of the CW amplitude is $6.4\times10^{-27}$ for J0537-6910
(rotation frequency of 62.03 Hz). Also, the minimum value of the upper limit of $\epsilon$ is $8.8\times10^{-9}$
for J0437-4715 (rotation frequency of 173.69 Hz).

\begin{figure}[htbp]
\centering
\includegraphics[width=14cm]{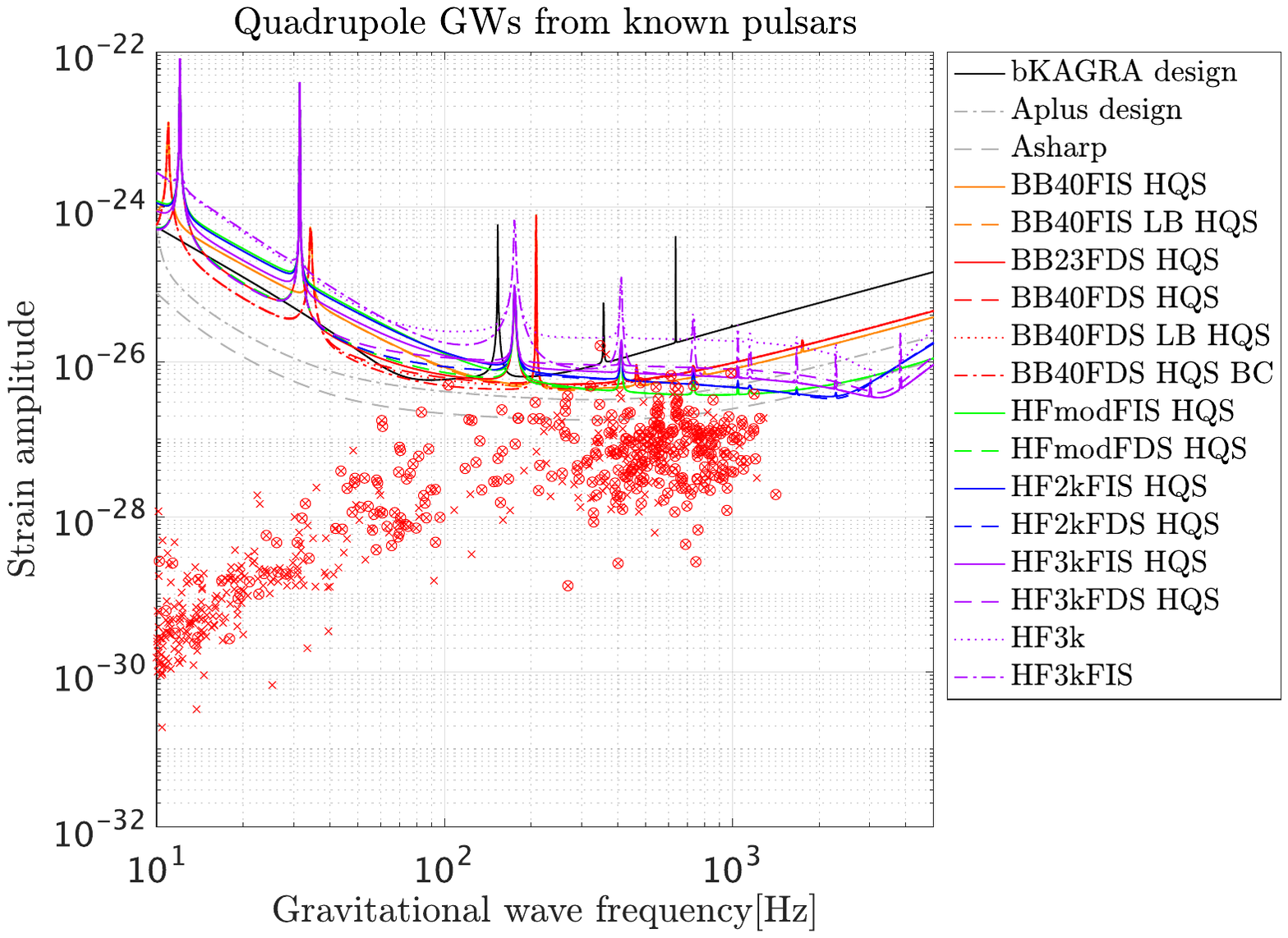} 
\caption{A comparison of the detector sensitivities of various configurations and 
quadrupole \gw amplitudes from the ATNF pulsars with known rotational frequencies, frequency derivatives ($\dot{f}_s < 0$), and distances.
One year of an optimal search is assumed.
For each pulsar, the smaller of the spin-down strain upper limit (Eq. \eqref{eq:CW_hqsd}) and the quadrupole gravitational wave strain amplitude (Eq. \eqref{eq:CW_hq}) is plotted, assuming an ellipticity of $\epsilon = 10^{-7}$ and a moment of inertia ${\cal I} = 10^{38}~\mathrm{kg \cdot m^2}$.
Each cross indicates a pulsar with 
known parameters, while crosses enclosed in circles indicate pulsars in binary systems.
}
\label{fig:CW_knownPSR_Sh}
\end{figure}

\begin{figure}[htbp]
\centering
\includegraphics[width=14cm]{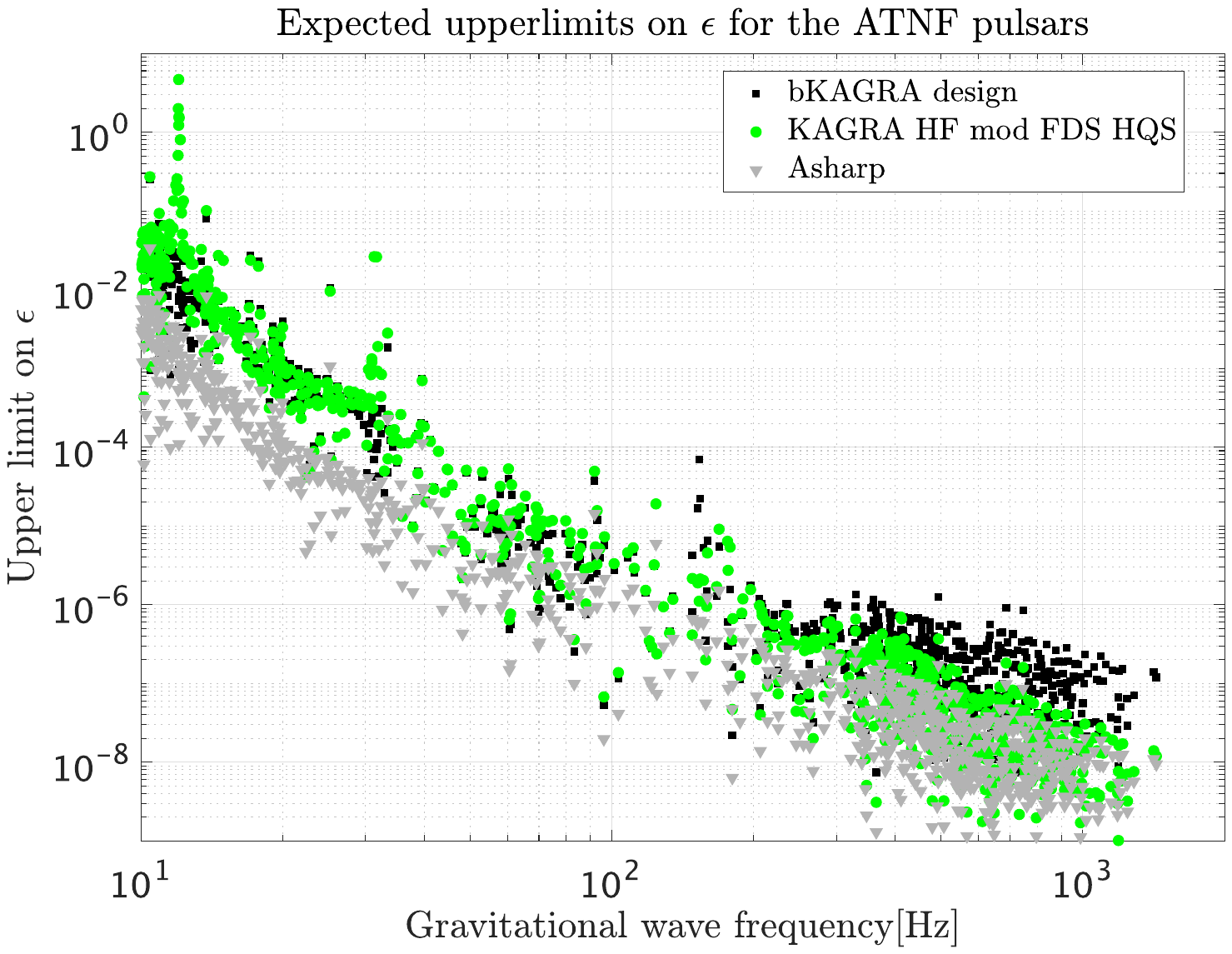} 
\caption{Possible maximum limits on $\epsilon$. 
For each pulsar, we plot the possible maximum limits on $\epsilon$ assuming three 
configurations of detectors without taking account of the spin-down upper limit. }
\label{fig:CW_knownPSR_epsilon}
\end{figure}

\begin{figure}[htbp]
\centering
\includegraphics[width=14cm]{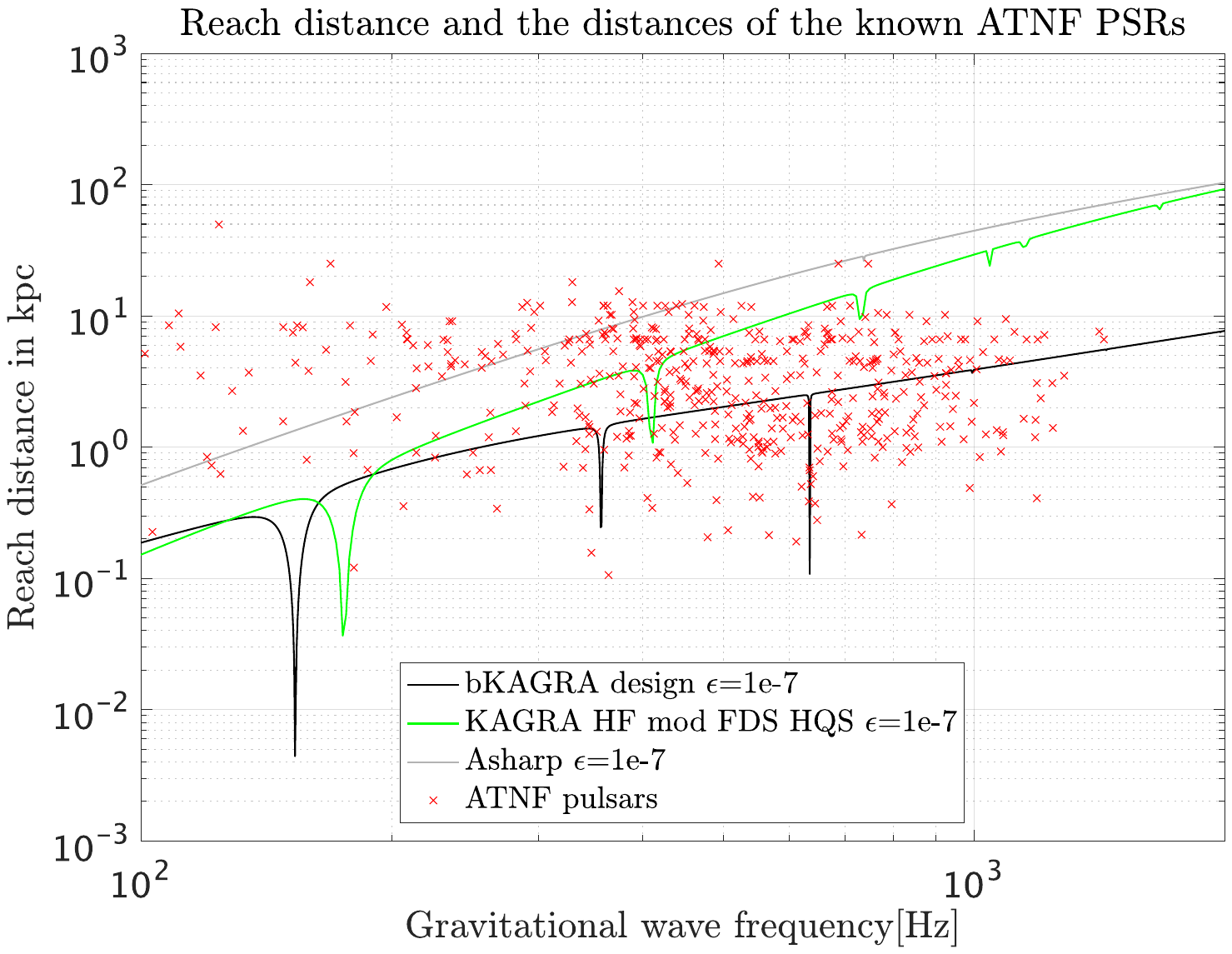} 
\caption{The maximum reach in kpc of detectors for pulsars assuming $\epsilon = 10^{-7}$.
The closest pulsar/\ns is at $\sim 130$pc. 
}
\label{fig:CW_knownPSR_distance}
\end{figure}

Assuming one year of  an optimal search, 
Fig. \ref{fig:CW_knownPSR_Sh} shows a comparison of the detector sensitivities of various configurations and 
quadrupole \gw amplitudes from known pulsars. For those pulsars and three configurations of detectors, 
Fig. \ref{fig:CW_knownPSR_epsilon} shows the possible upper limits on $\epsilon$. 
Figure \ref{fig:CW_knownPSR_distance}  shows the possible distance to which we would be able to detect 
CWs from pulsars assuming $\epsilon = 10^{-7}$ using three configurations of detectors.
All those figures suggest that middle to high frequencies ($\sim 70 - 1500$Hz)  would be promising. 

The table \ref{tbl:CW_summary_table} shows the number of ``detectable" ATNF pulsars 
for various configurations of detectors. Here we assume a CW from a pulsar is ``detectable" 
if the smaller of Eqs. \eqref{eq:CW_hq} and \eqref{eq:CW_hqsd} is larger than Eq. \eqref{eq:CW_hul} 
for that pulsar assuming one year of coherent integration. 
From this table, we conclude that 
(1) The ``KAGRA HF moderate" configuration is preferable to 
``KAGRA HF n kHz" and ``KAGRA BB" configurations because the HF-moderate detector has better sensitivity 
in $200\,\mathrm{Hz}$--$1\,\mathrm{kHz}$ where the promising sources would emit \gws, and (2)  
both FDS and FIS are fine because the difference in the sensitivity appears at lower frequencies while the strain is proportional to $f_\GW^2$.

\begin{table}[htbp]
    \centering
    \caption{Numbers of ``detectable" isolated pulsars for various configurations of detectors. Here we assume a CW from a pulsar is ``detectable" if the smaller of Eqs. \eqref{eq:CW_hq} and \eqref{eq:CW_hqsd} is larger 
    than Eq. \eqref{eq:CW_hul} 
    for that pulsar assuming one year of coherent integration. ``bKAGRA design" stands for the design sensitivity of the baseline KAGRA.}
    \label{tbl:CW_summary_table}
    \begin{tabular}{c|c||c|c}
\hline
Configuration & Number of pulsars & Configuration & Number of pulsars\\
    \hline
	A$\#$ design & 49 & 	HF3k & 0\\
	A+ design & 15 & 	HF3kFIS & 3\\
	bKAGRA design & 2 & 	HF3kFIS HQS & 3\\
	BB23FDS HQS & 2 & 	HF2kFIS HQS & 4\\
	BB40FIS HQS & 5 & 	HF3kFDS HQS & 2\\
	BB40FDS HQS & 2 & 	HF2kFDS HQS & 4\\
	BB40FIS LB HQS & 5 & 	HFmodFIS HQS & 10\\
	BB40FDS LB HQS & 2 & 	HFmodFDS HQS & 10\\	 	 	
	BB40FDS HQS BC& 5 & & \\
	\hline
    \end{tabular}
\end{table}

\subsubsection{Accreting neutron stars} 

It has long been suggested that accreting \nss would be spun up eventually to the Kepler break-up frequency. 
But none of the pulsars seem to rotate that fast. Some mechanism(s) must work and extract the angular momentum of the star 
and continuous \gws are one of the possibilities. Assuming the torque balance between the accretion and 
possible \gws, we can estimate the possible maximum \gw strain amplitude as 
\begin{align}
h_a &= 3\times 10^{-27}F_{-8}
\left(\frac{R}{10{\rm km}}\right)^{3/4}
\left(\frac{1.4 M_\odot}{M}\right)^{1/4}
\left(\frac{1{\rm kHz}}{f_{\rm s}}\right)^{1/2},
\label{eq:CW_ha}
\end{align}
where $R$ is the radius of the star and 
$F_{-8} = F/10^{-8} {\rm erg\cdot cm^{-2}\cdot s^{-1}}$\cite{2008MNRAS.389..839W} and 
$F$ is the X-ray flux. 

\begin{figure}[htbp]
\centering
\includegraphics[width=14cm]{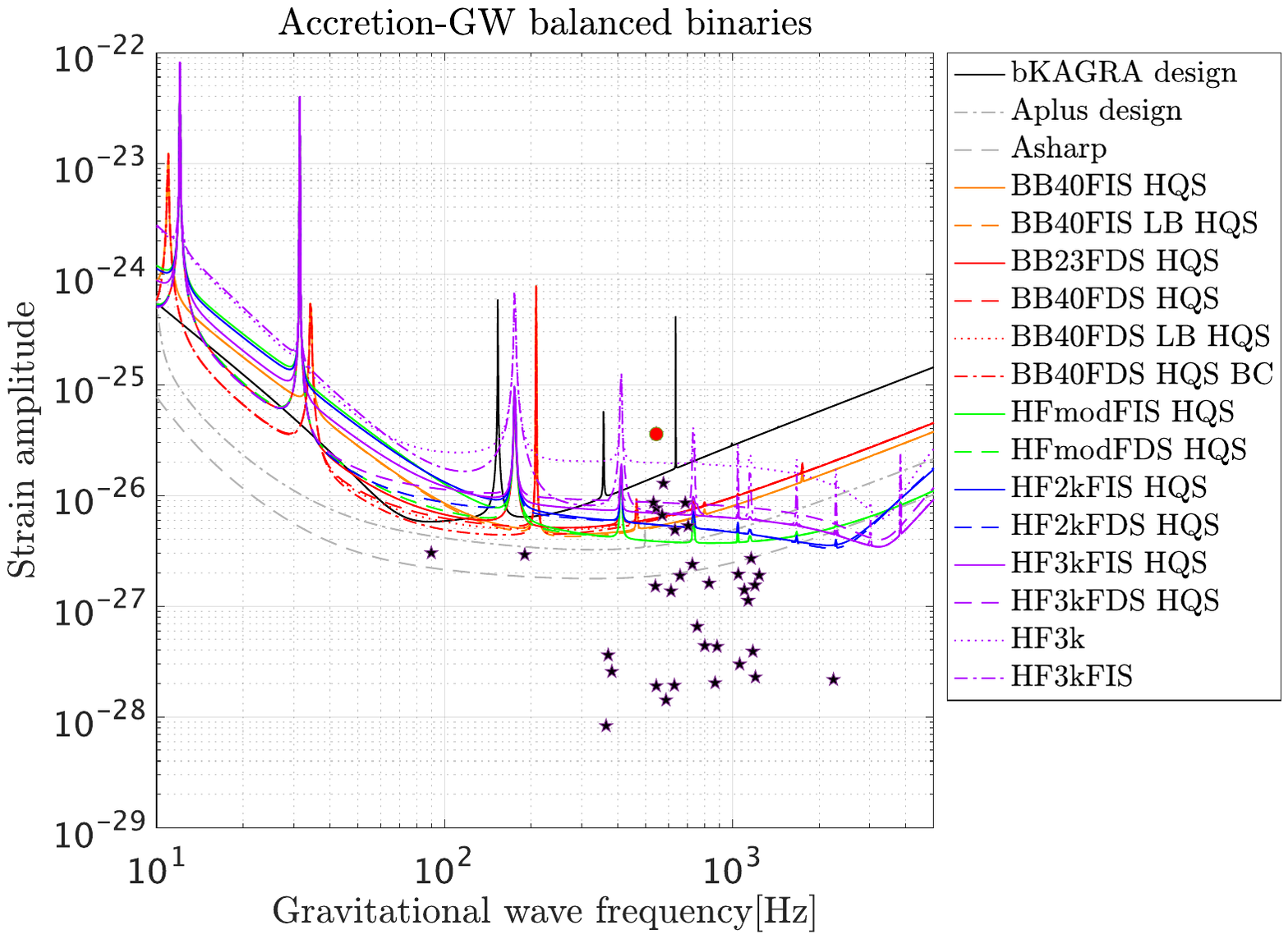} 
\caption{A comparison of the detector sensitivities of various configurations and 
quadrupole \gw amplitudes from accreting pulsars. 
One year of an optimal search is assumed. The most promising source is Scorpius X-1 (indicated by the red circle in the figure).
}
\label{fig:CW_AccretingNS_Sh}
\end{figure}

Assuming one year of  an optimal search, 
Fig. \ref{fig:CW_AccretingNS_Sh} shows a comparison of the detector sensitivities of various configurations and 
quadrupole \gw amplitudes from accreting \nss.  
There are at least four caveats: since the accretion process may be stochastic, so would the phase of the \gws 
and we are probably not able to use the optimal search. Also, the frequencies and some physical parameters of the pulsars 
needed to conduct a search are unknown, especially for Scorpius X-1. 
Thirdly and related with the above, the computational cost for this type of source is usually too high to conduct the optimal search. 
Finally, strain amplitude based on the torque balance argument may be too optimistic. 

The table \ref{tbl:CW_summary_table_accretingNS} shows the numbers of ``detectable" accreting pulsars 
for various configurations of detectors.  
From this table, we again conclude that 
``KAGRA HF moderate" is preferable to ``KAGRA HF n kHz" because the ``moderate" detector has better sensitivity 
in $200\,\mathrm{Hz}$--$1\,\mathrm{kHz}$ where the promising sources would emit \gws although the computational cost is generally higher 
at higher frequencies.

\begin{table}[htbp]
    \centering
    \caption{Numbers of ``detectable" accreting \nss for various configurations of detectors. Here 
    we assume a CW from a \ns is ``detectable" if \eqref{eq:CW_ha} is larger than \eqref{eq:CW_hul} 
    for that pulsar assuming one year of coherent integration.}
    \label{tbl:CW_summary_table_accretingNS}
    \begin{tabular}{c|c||c|c}
\hline
Configuration & Number of pulsars & Configuration & Number of pulsars\\
    \hline
	A$\#$ design & 11 & 	HF3k & 1\\
	A+ design & 8 & 	HF3kFIS & 5\\
	bKAGRA design & 1 & 	HF3kFIS HQS & 5\\
	BB23FDS HQS & 6 & 	HF2kFIS HQS & 7\\
	BB40FIS HQS & 6 & 	HF3kFDS HQS & 3\\
	BB40FDS HQS & 6 & 	HF2kFDS HQS & 7\\
	BB40FIS LB HQS & 6 & 	HFmodFIS HQS & 8\\
	BB40FDS LB HQS & 6 & 	HFmodFDS HQS & 8\\	 	 	
	BB40FDS HQS BC& 6 & & \\
	\hline
    \end{tabular}
\end{table}

\subsubsection{Unknown neutron stars}
\label{subsubsec_unknown_pulsars}

While the electromagnetic pulses from a pulsar are beamed in narrow directions, 
possible \gws from \nss are omnidirectional. Hence it may be possible that 
\nss electromagnetically undiscovered until now are just nearby and we are able to detect 
them using \gws but not electromagnetic waves. 
To assess the possibility of detecting \gws from these (electromagnetically) unknown \nss, we may need a theoretical prediction of the spatial distribution as well as the ellipticity and spin distributions of all the \nss born since the birth of our galaxy. Such a prediction in turn requires assumptions of the star formation history, stellar evolutions, spin evolutions of \nss, their orbital motions within our galaxy, and the distribution and evolutions of ellipticities of \nss. Here we briefly summarize the results of a recent study for the third generation telescopes and then comment on the KAGRA upgrade.

A recent detailed study \cite{2023ApJ...952..123P} showed that the third generation detectors 
such as Cosmic Explorer and Einstein Telescope have a better chance of detecting normal (i.e., non-recycled) \nss from the low ([5,100] Hz) to middle-frequency bands [100,500] Hz rather than in the high-frequency band [500,2500] Hz. For recycled \nss, 
the third-generation detectors have a chance of detection in the middle to high-frequency band, say, more than 40 Hz. 
We here note that the resulting numbers of detections in \cite{2023ApJ...952..123P} 
strongly depend on assumed models, especially for normal \nss. 
For example, they assumed that the maximum possible $\epsilon$ was $10^{-5}$, which might be too optimistic, 
though it is premature to conclude it was. 

\begin{figure}[htbp]
\centering
\includegraphics[width=14cm]{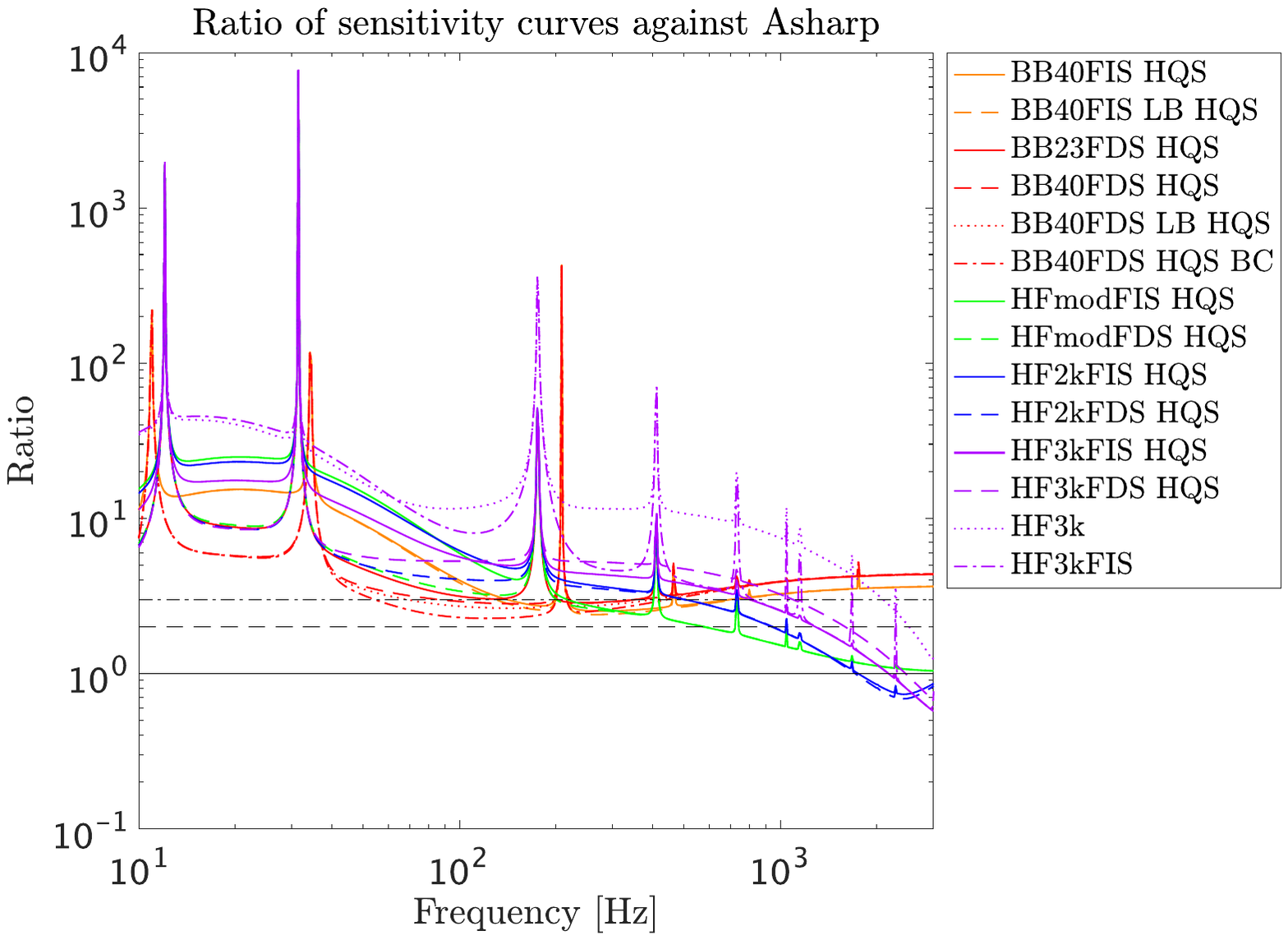} 
\caption{The ratios of the sensitivity curves of the proposed KAGRA variation against the A$\#$ detector.
The black solid, dashed, and dash-dotted lines indicate the ratio of 1, 2, and 3.
}
\label{fig:CW_sensitivity_ratio}
\end{figure}

However, whether we should improve the sensitivity of KAGRA in the era of A$\#$ at low to middle frequencies is not clear. 
Fig. \ref{fig:CW_sensitivity_ratio} shows the ratios of the sensitivity curves of the proposed KAGRA variation against the A$\#$ detector.
All the proposed sensitivity curves are above twice as bad as the A$\#$ curve except for the HFmod family above $\sim$ 500 Hz and 
the HF family above $\sim$ 1.5 kHz. 
Eq. \eqref{eq:CW_hul} tells that if the sensitivity is worse by a factor of $k$ ($\sqrt{S_h} \rightarrow k\sqrt{S_h}$), 
then we need $k^2$-times longer observation time to achieve the same \snr. 
Or if we coherently add signals from $n+1$ detectors 
where one detector has $k$ times worse sensitivity than other $n$ detectors, then the \snr would be proportional to 
$\rho_{n+1} = \rho_n\sqrt{1+1/nk}$ where $\rho_n$ is the \snr from $n$ detectors. Suppose we have  
4 A$\#$-like detectors (LIGO Hanford, Livingston, India, Virgo nEXT) and a KAGRA variation that has twice as bad sensitivity as the former four, 
then the increase 
in the \snr by adding the KAGRA variation is 6 $\%$.  
Given the expected huge computational cost for a wide-parameter-space search, the Broad-Band (BB) family may not pay.

\subsubsection{Pulsar glitch}
\label{subsubsec:pulsar_glitch}

The spins of pulsars usually slowdown due to various mechanisms e.g., the magnetodipole radiation, \ac{GW} radiation. Some pulsars sometimes experience the sudden spin-up~\footnote{The antiglitch, the sudden spindown, is a well-known phenomena for magnetors. Recently, an antiglitch of a pulsar is found~\cite{Tuo:2024pvf}. In this work, we do not consider the antiglitches.}, which is called \textit{pulsar glitches}. The mechanism of pulsar glitches is unknown.
If the glitches are the results of sudden redistributions of mass within the \ac{NS}, they might emit \acp{GW} on various time scales.
There are some models that predict the pulsar glitches can emit \acp{GW} with the duration of $\mathcal{O}(10^{3})$--$\mathcal{O}(10^{5})$ \unit{s}, depending on the relaxation time scale of the glitches.

We assess the significance of the \acp{GW} from the pulsar glitches. Our approach is based on the methods of Yim~\etal~\cite{Yim:2024eaj}, in which the radiated energy of \ac{GW} is estimated and converted into the \snr assuming the quasi-monochromatic signal model. Given the energy budget $E_\GW$ and the pulsar information, we evaluate the \snr by
\begin{equation}
  \rho^2 = \frac{5G}{2 \pi^2 c^3} \frac{A_2 E_\GW}{f_\GW^2 \dl^2 S_\un(f_\GW)}\,.
\end{equation}
Here, $f_\GW$ is the \ac{GW} frequency which is equal to twice the rotation frequency. The luminosity distance is denoted by $\dl$. $A_2$ is the function defined in Jaranowski~\etal~\cite{Jaranowski:1998qm} and depends on the pulsar position, the inclination angle, and the polarization angle. In this work, we set both the polarization angle and the inclination angle to zero. For computing $E_\GW$, we employ three \ac{GW} energy budget models: the agnostic model, the vortex unpinning model, and the transient mountain model.

We use three catalogs: ATNF Pulsar Catalog~ \cite{2005AJ....129.1993M}, JBCA Glitch Catalog~\cite{Espinoza:2011pq}, and ATNF Pulsar Glitch Table~\cite{2005AJ....129.1993M}. We use the Python library \texttt{psrqpy}~\cite{psrqpy} to query the catalogs.
We clean the catalog 
following the method of Yim~\etal~\cite{Yim:2024eaj} to get two datasets A and B. The dataset A consists of 694 glitches, while the dataset B contains 96 glitches. The moment of inertia is fixed at
$\calI = $ \qty{1e38}{kg.m^2}
for all pulsars.

\begin{table}[htbp]
  \centering
  \caption{\label{tab: pulsar glitch summary} List of the number of detectable pulsar glitches with $\rho \geq 10$.}
  \begin{tabular}{cccc}
    Configuration & Agnostic & Vortex unpinning & Transient mountain \\ \hline
    BB23FDS\_HQS & 32 & 0 & 3 \\
    BB40FIS\_HQS & 25 & 0 & 0 \\
    BB40FDS\_HQS & 50 & 0 & 6 \\
    BB40FIS\_LB\_HQS & 25 & 0 & 0 \\
    BB40FDS\_LB\_HQS & 52 & 0 & 6 \\
    BB40FDS\_HQS\_BC & 53 & 0 & 7 \\
    HF3k & 18 & 0 & 0 \\
    HF3kFIS & 18 & 0 & 0 \\
    HF3kFIS\_HQS & 23 & 0 & 0 \\
    HF2kFIS\_HQS & 22 & 0 & 0 \\
    HF3kFDS\_HQS & 31 & 0 & 2 \\
    HF2kFDS\_HQS & 32 & 0 & 3 \\
    HFmodFIS\_HQS & 22 & 0 & 0 \\
    HFmodFDS\_HQS & 32 & 0 & 3 \\
  \end{tabular}
\end{table}

We define a glitch as detectable if the associated \gw signal has a \snr greater than 10. 
The numbers of detectable glitches are summarized in Table~\ref{tab: pulsar glitch summary}. For the agnostic model, we expect to detect $\calO(10)$ \ac{GW} events from the pulsar glitches in any proposed configurations. The spectrum BB40FDS\_HQS\_BC is expected to detect the most events among the proposed configurations. For the vortex unpinning model, there are no detectable glitches. We estimate that $\calO (1)$ events can be detected for the transient mountain model. The reader should be reminded that we use catalog A for the agnostic and the vortex unpinning models, and catalog B for the transient mountain model, and they have a different number of pulsar glitches. For any model, the configurations named ``BB40FDS'' are the best, and the configurations with BB23HQS, HF with FDS are the second-best choices.


\subsection{Stochastic GW background}

The \ac{SGWB} is the persistent \acp{GW} generated by random processes. The typical sources of \ac{SGWB} are the quantum fluctuation in the primordial universe, the ensemble of compact binaries, and the ensemble of supernovae.
\acp{SGWB} from these sources can be modeled by the power law,
\begin{equation}
    \Omega_\GW (f) = \Omega_{\GW,0} \left( \frac{f}{10 \mathrm{\ Hz}} \right)^\alpha\,.
\end{equation}
within the frequency band of the ground-based detectors.
Following the \ac{LVK}'s isotropic \ac{SGWB} searches~\cite{KAGRA:2021kbb}, we use three fiducial values for $\alpha$: $\alpha=0$ for the primordial \ac{SGWB}, $\alpha=2/3$ for the ensemble of the compact binary coalescence, and $\alpha=3$ for the ensemble of various astrophysical sources such as supernovae.
Correlating two or more detectors is the standard strategy to search for \ac{SGWB}. Allen~\& Romano~\cite{Allen:1997ad} derived the \snr of the cross-correlation search for the isotropic \ac{SGWB},
\begin{equation}
    \rho_{IJ} = \frac{3H_0^2}{10\pi^2} \sqrt{2 T_\mathrm{obs}} \left[ \int^\infty_0 \dd{f} \frac{\gamma^2_{IJ}(f) \Omega^2_\GW (f)}{f^6 S_I(f) S_J(f)} \right]^{1/2}\,,
\end{equation}
where the subscripts $I,J$ indicate the interferometers, $H_0$ is the Hubble constant, and $T_\mathrm{obs}$ is the observational period. The function $\gamma_{IJ}(f)$ is called \textit{overlap reduction function}, describing the strength of the correlation between two interferometers.

We consider the cross-correlation between LIGO-Hanford, LIGO-Livingstone, and KAGRA.
We assume a one-year observational period.
We set the integration range from 3 Hz to 1000 Hz. To quantify the contribution of KAGRA joining to the LIGO's two interferometers, we define the improvement factor by
\begin{equation}
    \text{Improvement factor} = \frac{\sqrt{\rho^2_\mathrm{HL} + \rho^2_\mathrm{HK} + \rho^2_\mathrm{LK}} }{\rho_\mathrm{HL}} - 1\,.
\end{equation}
We assume that both LIGO-Hanford and LIGO-Livingstone have a PSD of A\#.
We use the library \texttt{pygwb}~\cite{Renzini:2023qtj} to calculate the overlap reduction function. The Hubble parameter is taken from Planck 18~\cite{Planck:2018vyg}.

Table~\ref{tab: improvement factor for SGWB} shows the improvement factors for the \ac{SGWB} model with $\alpha=0, 2/3$ and 3. It shows that for any models, for any proposed PSDs, the improvement factors do not exceed 2~\%. The overlap reduction functions rapidly decay to zero for the frequency higher than $\sim$ 100 Hz, leading that the high frequency sensitivity does not affect to the performance for the \ac{SGWB} search. The PSD of LIGO A\# is much better than that of KAGRA for the low frequency region. The geometrical condition of the two interferometers of LIGO is also good, i.e., the distance between the two interferometers of LIGO is shorter than the distance between LIGO and KAGRA. The arm directions of LIGO Hanford and Livingston are almost aligned. It results that the overlap reduction function of LIGO Hanford and Livingston is more advantageous than those of LIGO and KAGRA. These are the reasons why KAGRA's contribution is not significant for any proposed PSDs. We conclude that all PSDs are equivalent in the context of the isotropic \ac{SGWB} search.

\begin{table}[htbp]
    \centering
    \caption{\label{tab: improvement factor for SGWB}
    Improvement factors for SGWB models with $\alpha=0, 2/3$ and 3.}
    \begin{tabular}{cccc}
        \multirow{2}{*}{Configuration} & \multicolumn{3}{c}{Improvement factor} \\
        & $\alpha=0$ & $\alpha=2/3$ & $\alpha=3$ \\ \hline
        BB40FIS\_HQS & $5.56 \times 10^{-5}$ & $9.80 \times 10^{-5}$ & $8.05 \times 10^{-3}$ \\
        BB40FIS\_LB\_HQS & $5.64 \times 10^{-5}$ & $1.01 \times 10^{-4}$ & $8.80 \times 10^{-3}$ \\
        BB23FDS\_HQS & $1.50 \times 10^{-4}$ & $2.66 \times 10^{-4}$ & $7.83 \times 10^{-3}$ \\
        BB40FDS\_HQS & $2.85 \times 10^{-4}$ & $4.36 \times 10^{-4}$ & $9.22 \times 10^{-3}$ \\
        BB40FDS\_LB\_HQS & $2.98 \times 10^{-4}$ & $4.69 \times 10^{-4}$ & $1.02 \times 10^{-2}$ \\
        BB40FDS\_HQS\_BC & $3.31 \times 10^{-4}$ & $5.59 \times 10^{-4}$ & $1.29 \times 10^{-2}$ \\
        HFmodFIS\_HQS & $2.18 \times 10^{-5}$ & $4.01 \times 10^{-5}$ & $6.99 \times 10^{-3}$ \\
        HFmodFDS\_HQS & $1.37 \times 10^{-4}$ & $2.36 \times 10^{-4}$ & $9.41 \times 10^{-3}$ \\
        HF2kFIS\_HQS & $2.42 \times 10^{-5}$ & $4.23 \times 10^{-5}$ & $4.20 \times 10^{-3}$ \\
        HF2kFDS\_HQS & $1.39 \times 10^{-4}$ & $2.26 \times 10^{-4}$ & $5.83 \times 10^{-3}$ \\
        HF3kFIS\_HQS & $3.93 \times 10^{-5}$ & $6.53 \times 10^{-5}$ & $3.38 \times 10^{-3}$ \\
        HF3k & $1.15 \times 10^{-5}$ & $2.12 \times 10^{-5}$ & $5.31 \times 10^{-4}$ \\
        HF3kFIS & $1.07 \times 10^{-5}$ & $2.21 \times 10^{-5}$ & $1.86 \times 10^{-3}$ \\
        HF3kFDS\_HQS & $1.22 \times 10^{-4}$ & $1.80 \times 10^{-4}$ & $3.05 \times 10^{-3}$
    \end{tabular}
\end{table}

\section{Hardware upgrade\label{sec:hardware}}
The hardware upgrades and maintenance of KAGRA are critical for enhancing the sensitivity and reliability of the detector over the next decade. This section discusses potential upgrade paths, noise sources, and possible measures for improvement, as well as various upgrade options under consideration.

\subsection{General Remarks}

KAGRA has several hardware-related advantages, notably its underground location and cryogenic operation. These features lead to reduced environmental disturbances such as seismic noise and temperature fluctuations, improved interferometer stability, and lower thermal noise—provided the mechanical quality factors (Q-values) of the materials are not significantly worse than expected. These advantages offer a unique foundation for enhancing sensitivity, and we focus on upgrade plans that can effectively leverage them.

The upgrade of the KAGRA \gw detector can follow two main approaches: Broad Band (BB) and Narrow Band (NB). BB aims to improve sensitivity across a wide range of frequencies, which is scientifically preferable as it allows for the detection of various \gw sources, each with its own characteristic frequency. However, achieving improved sensitivity across a broad frequency range requires significant noise reduction across multiple sources, demanding considerable resources in terms of both funding and human resources. NB upgrades focus on enhancing sensitivity within a specific frequency range. This allows for targeted noise reduction strategies, requiring fewer resources than BB upgrades.

One notable NB option is the High Frequency (HF) upgrade, which enhances sensitivity in the high-frequency range. The sensitivity of interferometric \gw detectors at high frequencies is often limited by shot noise, which aligns well with theoretical predictions. Consequently, noise reduction strategies for shot noise are relatively well understood and effective. However, factors such as birefringence inhomogeneity could degrade common-mode noise rejection, which must be carefully considered. Given KAGRA’s limited resources, an HF-focused NB upgrade may provide a viable balance between feasibility and scientific impact. Other NB options include:

\begin{itemize}
    \item \textbf{Low Frequency (LF) Upgrades:} Improving low-frequency sensitivity can enable the detection of heavier black holes and other astrophysical phenomena. However, this approach presents significant technical challenges. Many low-frequency noise sources are not yet fully understood, making their mitigation risky.
    \item \textbf{Middle Frequency (MF) Upgrades:} Enhancing sensitivity around 100 Hz primarily requires reducing coating thermal noise. Increasing beam size can help, but this necessitates larger mirrors, which introduces both technical and cost-related challenges. Developing high-Q coatings suitable for low-temperature operation is a necessary step, requiring dedicated research and development efforts.
\end{itemize}

\subsection{Noise Sources and Possible Upgrade Measures}

This section outlines key noise sources and potential upgrade measures for KAGRA, with a focus on BB and HF improvements. It builds upon existing white papers but emphasizes more concrete upgrade strategies.

\subsubsection{Low Frequencies}

The low-frequency region of KAGRA's sensitivity curve is primarily limited by seismic, Newtonian, and suspension thermal noise, along with contributions from quantum radiation pressure effects. Effective suppression of these noise sources is crucial for improving sensitivity at frequencies below 100 Hz. Since \gw signals from massive binary systems and certain astrophysical processes appear prominently in this range, addressing low-frequency noise will expand KAGRA’s scientific reach.

\paragraph{Seismic Noise}

Ground vibrations contribute significantly to low-frequency noise. KAGRA employs sophisticated vibration isolation systems, but further improvements may be necessary, especially if the observation band is extended below 10 Hz. Enhancing vibration isolation performance can help, but for improvements above 10 Hz, refining local control strategies to reduce mirror vibration RMS is crucial for stable operation and alignment maintenance. Seismic noise originates from a variety of environmental sources, including tectonic activity, human operations, and natural seismic movements. Future strategies to mitigate this noise include better seismic sensor networks for active noise cancellation and improvements in the suspension design to reduce residual motion of the test masses.

\paragraph{Newtonian Noise}

Newtonian noise originates from fluctuating gravitational fields caused by seismic activity and atmospheric density variations. This noise is particularly challenging to mitigate as it is a direct consequence of mass density fluctuations in the environment. In underground facilities like KAGRA, Newtonian noise is lower than at surface-level observatories due to reduced seismic activity. However, residual contributions are expected to remain significant, such as water flow in the KAGRA tunnel. Future strategies to suppress Newtonian noise involve improved seismic monitoring and active cancellation techniques, where real-time environmental data is used to generate noise-subtracting signals. Advanced computational modeling of local mass movements can further refine Newtonian noise estimates and guide the design of active noise suppression systems.

\paragraph{Suspension Thermal Noise} 
Thermal noise in the suspension system arises from the internal damping of materials used in the suspension fibers and isolation stages. The mechanical loss in the sapphire fibers and blade springs mainly contributes to test mass displacement noise. Since KAGRA operates at cryogenic temperatures, thermal noise is significantly reduced compared to room-temperature detectors. However, the mechanical loss of those installed in the KAGRA suspension has been measured to be significantly lower than expected values, which will result in the suspension thermal noise limiting the KAGRA sensitivity in the near future. Better characterization of the mechanical quality factor of the suspension components will aid in refining noise estimates and reduction strategies. Potential approaches involve using high-quality bonding at the fiber connection point, or developing monolithic sapphire suspension.

\paragraph{Quantum Radiation Pressure Noise}

Quantum radiation pressure noise results from fluctuations in the momentum transfer of photons to the test masses. This effect is particularly relevant at low frequencies, where it competes with designed suspension thermal noise. The level of radiation pressure noise depends on the circulating power in the interferometer and the optical spring effect introduced by the signal recycling cavity. Future upgrades may include the implementation of frequency-dependent squeezing to manipulate quantum noise distribution, as well as increasing the circulating power while maintaining stability through improved thermal management and active control techniques. Increasing the reflectivity of the SRM can reduce the radiation pressure noise. Therefore, the optimization of the SRM reflectivity is critical for considering the upgraded plan. Investigating alternative optical configurations, such as optimized signal recycling mirror reflectivity, could also lead to further reductions in radiation pressure noise.

\paragraph{Control Noise Couplings}

Control noise arises due to feedback mechanisms in the suspension system. The local control system ensures interferometer stability but also introduces additional displacement noise into the DARM channel. The dominant sources include feedback from the main suspension control loops, particularly in the Type-A suspensions of the end test masses. Strategies for reducing control noise involve optimizing the loop gains, improving sensor readout, and refining actuation methods to minimize injected noise. Additionally, implementing advanced feedforward techniques can further suppress unwanted coupling. Control noise contributions can be further reduced by improving sensor precision, reducing ADC/DAC noise, and developing more sophisticated filtering techniques.

\subsubsection{Middle Frequencies}

The middle-frequency range of the detector, from tens to hundreds of Hz, is primarily affected by acoustic noise, thermal noise from optical coatings, and substrate thermal noise. Improvements in this range are critical for detecting compact binary coalescences and other astrophysical sources.

\paragraph{Acoustic Noise}

Environmental acoustic disturbances couple into the interferometer primarily through scattered light and mechanical vibrations. These noises originate from vacuum pumps, cooling fans, and other KAGRA site equipment. To mitigate their impact, additional acoustic shielding and improved isolation of sensitive optical components are essential. Identifying and dampening mechanical resonances within the vacuum system also play a critical role in reducing this noise. Future efforts may involve more detailed characterization of acoustic coupling paths and improved vibration isolation for critical optical components. Additionally, further development of scattered light mitigation techniques and optimization of the vacuum chamber structure could yield further improvements.

\paragraph{Coating Thermal Noise} 

Coating thermal noise is one of the dominant noise sources in the mid-frequency range of \gw detectors. It arises from mechanical dissipation in the dielectric mirror coatings used for high reflectivity. The choice of coating materials and layer structure directly impacts this noise. At cryogenic temperatures, the mechanical loss of coatings changes, and further optimization can be required to achieve minimal thermal noise. Research into alternative materials such as crystalline coatings is ongoing. The implementation of lower-loss coatings and expanding the beam diameter on the test mass by installing more massive sapphire mirrors will be crucial for reducing this noise contribution in future upgrades.

\paragraph{Substrate Thermal Noise}

Substrate thermal noise arises from mechanical losses in the bulk material of the test masses. Although sapphire provides lower thermal noise compared to fused silica at cryogenic temperatures, the substrate thermal noise still contribute partially to the sensitivity limit. The dominant sources of substrate thermal noise are thermo-elastic loss at first, and Brownian loss secondly. Ongoing studies aim to better characterize the loss mechanisms in sapphire at low temperatures. Further reduction in the Brownian thermal noise could be achieved through improved crystal growth techniques and surface polishing. Expanding the laser beam size can also reduce the substrate thermal noise.

\subsubsection{High Frequencies}

At higher frequencies, the sensitivity of KAGRA is primarily limited by quantum noise and laser technical noise. Strategies for improvement focus on reducing shot noise and stabilizing laser intensity and frequency fluctuations. Enhanced quantum noise suppression techniques are essential for optimizing high-frequency performance and extending the detector’s reach to a broader range of astrophysical events.

\paragraph{Shot Noise}

Shot noise, arising from quantum fluctuations of light, dominates at frequencies above a few hundred Hz. Increasing the circulating power in the arm cavities helps reduce shot noise, though it also necessitates improved thermal compensation and higher-quality optics to minimize associated effects. Future improvements may include implementing frequency-dependent squeezing techniques, which have already been demonstrated in other \gw detectors to redistribute quantum noise and enhance sensitivity at high frequencies. Investigations into advanced mirror coatings and optical system refinements could further reduce shot noise contributions.

\paragraph{Laser Noises}
\begin{itemize}
    \item \textbf{Frequency Noise} Frequency noise couples into DARM due to asymmetries in the interferometer optics and imperfect mode matching. Planned improvements include direct frequency excitation measurements to refine coupling estimates and optimizing stabilization feedback. 
    \item \textbf{Intensity Noise} Intensity fluctuations in the laser contribute through classical radiation pressure noise at low frequencies and sensing noise at higher frequencies. One approach is the implementation of an improved power stabilization scheme at the laser source, reducing intensity fluctuations before amplification and injection into the interferometer.
\end{itemize}

\subsection{Current status and limitations of hardware}

In this section, we discuss the current status and limitations of KAGRA's hardware, as well as the requirements for further sensitivity improvements. At present, the dominant noise sources are suspension control noise at low frequencies, suspension thermal noise and acoustic induced noises such as scattered light and input beam jitter at middle frequencies, and shot noise and frequency noise at high frequencies. The frequencies at which suspension control noise dominates are low enough that the BNS range is not significantly affected. The planned replacement of the sapphire test masses before O5 is expected to reduce both input jitter and frequency noise due to better symmetry of both detector arms. The implementation of RSE configuration with higher input power will suppress the shot noise. Consequently, the most critical current limitation is the suspension thermal noise, which originates from a mechanical Q-value of approximately $10^5$, significantly lower than the designed value of $5\times 10^6$.

\subsection{Upgrade Options}

This section presents various upgrade options by combining the aforementioned noise reduction strategies to achieve BB and HF improvements. Several sensitivity curves corresponding to different upgrade scenarios are considered.

\subsubsection{Broad Band (BB) Options}

BB upgrades aim to improve sensitivity across the entire frequency spectrum. Since these upgrades enhance sensitivity at all frequencies compared to KAGRA’s designed sensitivity, they provide a more comprehensive improvement in physical and astrophysical investigations. Although the mechanical loss of the sapphire fiber was measured to be larger than the designed one, we need to improve that to enhance the potential of the BB options enough, and here we assume the designed loss in the sensitivity estimation which we define as HQS. In the BB options, the following upgrade possibilities will be adopted alone or combinedly:

\begin{itemize}
    \item \textbf{Frequency Independent Squeezing (FIS):} Implementing FIS reduces quantum noise across all frequencies, improving broadband sensitivity.
    \begin{itemize}
        \item Input squeezing: 10 dB
        \item Total optical losses outside the interferometer: 5\%
    \end{itemize}
    \item \textbf{Frequency Dependent Squeezing (FDS):} Implementing FDS reduces quantum noise across all frequencies, improving broadband sensitivity.
    \begin{itemize}
        \item Input squeezing: 6 dB
        \item Total optical losses outside the interferometer: 10\%
        \item Filter cavity: Length 60 m, linewidth 40 Hz, round trip loss 30 ppm
    \end{itemize}
    \item \textbf{40 kg Mirrors:} Increasing the sapphire mirror mass to 40 kg reduces suspension thermal noise, further enhancing sensitivity.
    \item \textbf{Larger Beam Size:} Expanding the beam size reduces coating thermal noise. Given that the current sapphire mirror mass is 23 kg, scaling up to 40 kg suggests an approximate beam size increase of 1.2 times.
    \item \textbf{Better Coating:} Replacing the coating with AlGaAs reduces coating thermal noise significantly. This assumes a loss angle of $1 \times 10^{-5}$.
\end{itemize}

\begin{figure}
    \centering
    \begin{subfigure}{0.49\textwidth}
        \centering
        \includegraphics[width=\textwidth]{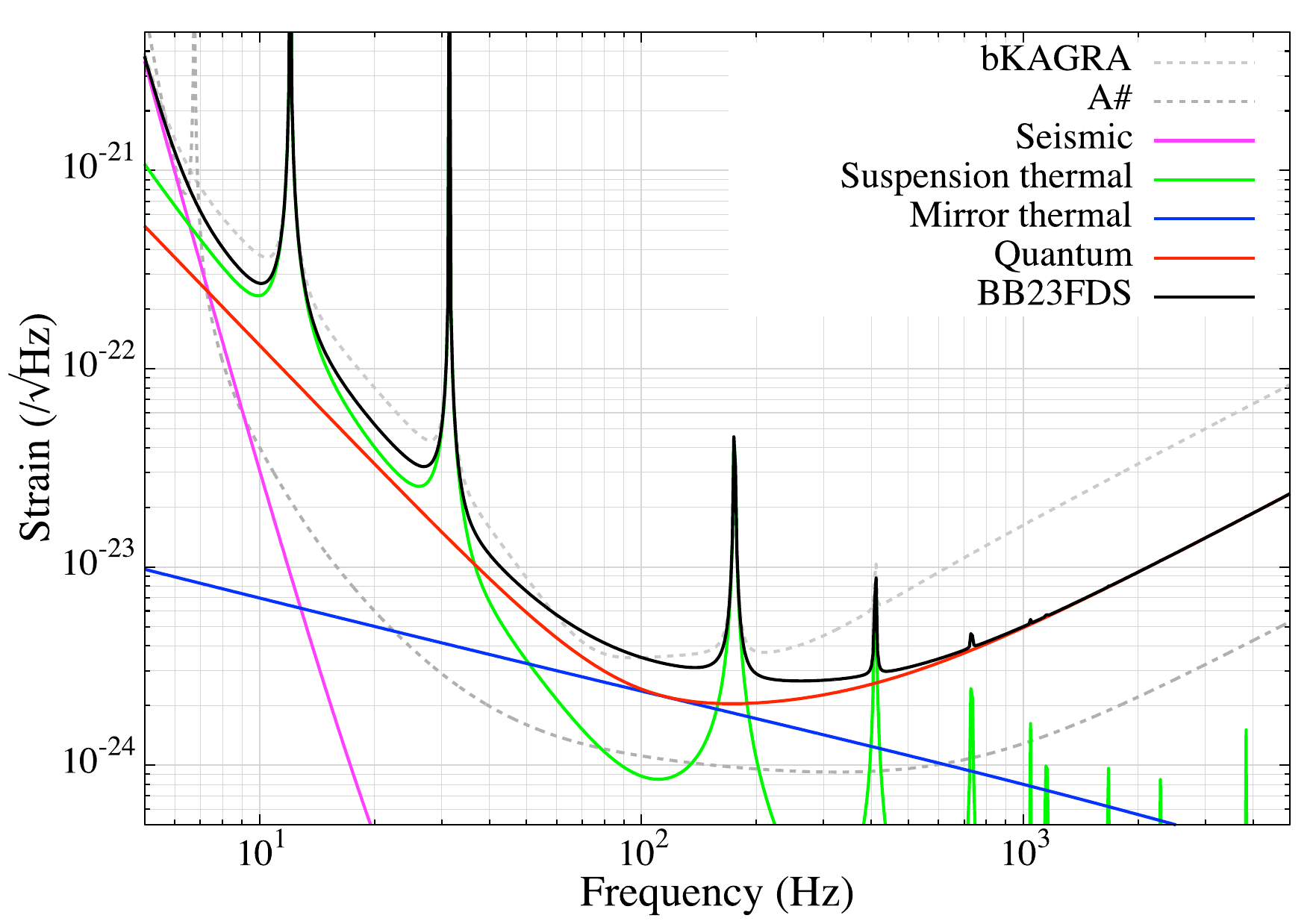}
        \caption{BB23FDS-HQS}
    \end{subfigure}
    \begin{subfigure}{0.49\textwidth}
        \centering
        \includegraphics[width=\textwidth]{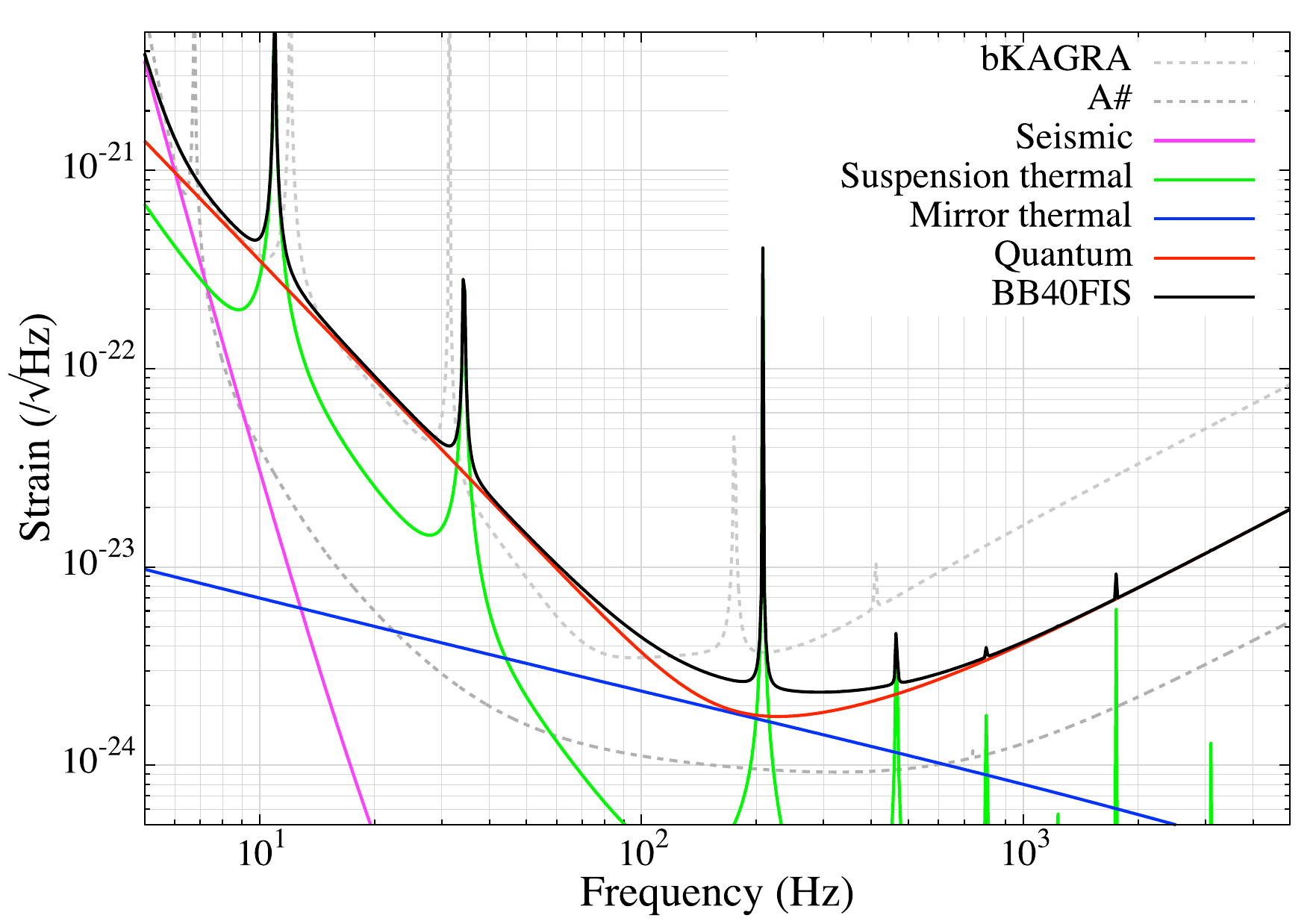}
        \caption{BB40FIS-HQS}
    \end{subfigure}
    
    \begin{subfigure}{0.49\textwidth}
        \centering
        \includegraphics[width=\textwidth]{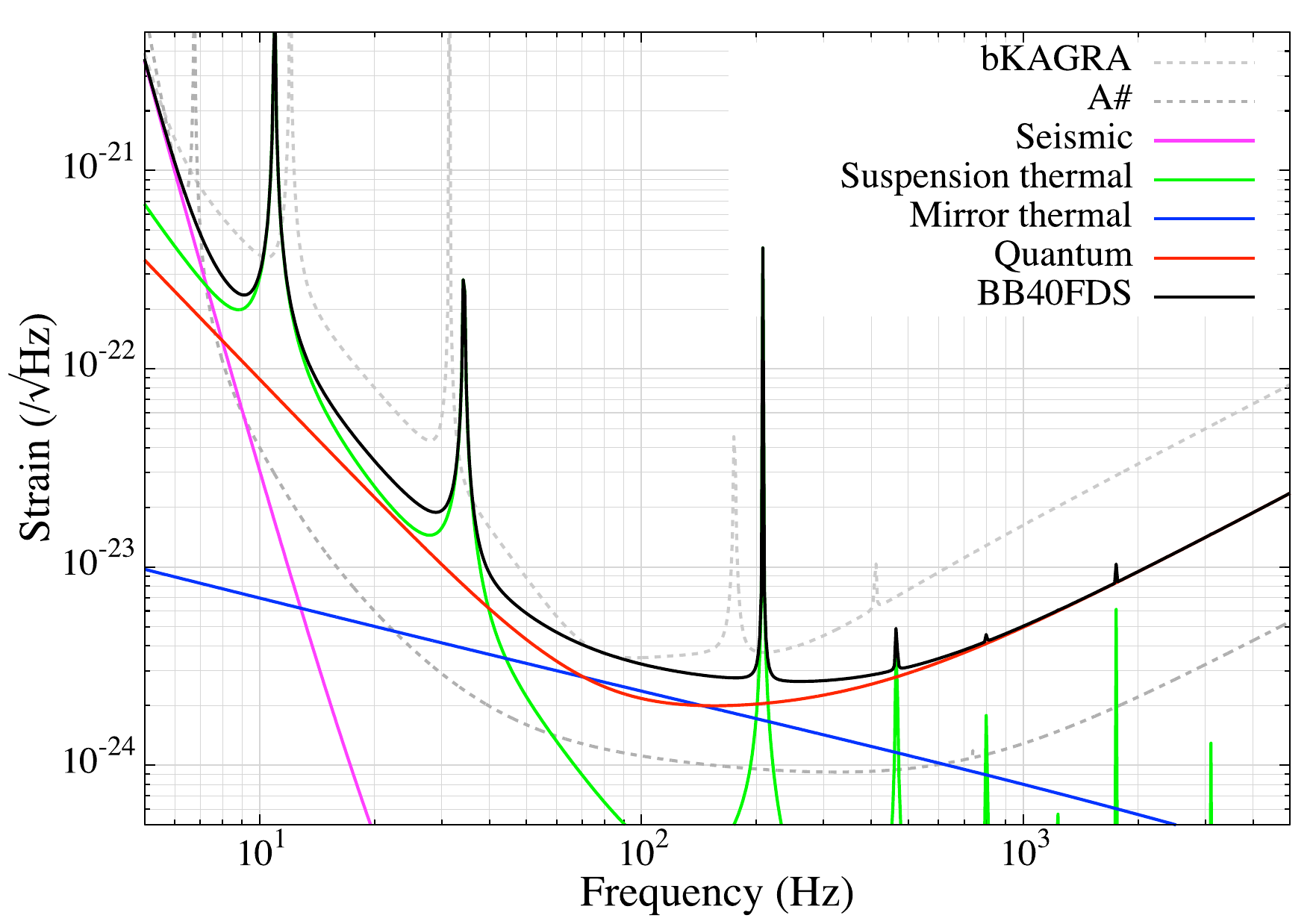}
        \caption{BB40FDS-HQS}
    \end{subfigure}
    \begin{subfigure}{0.49\textwidth}
        \centering
        \includegraphics[width=\textwidth]{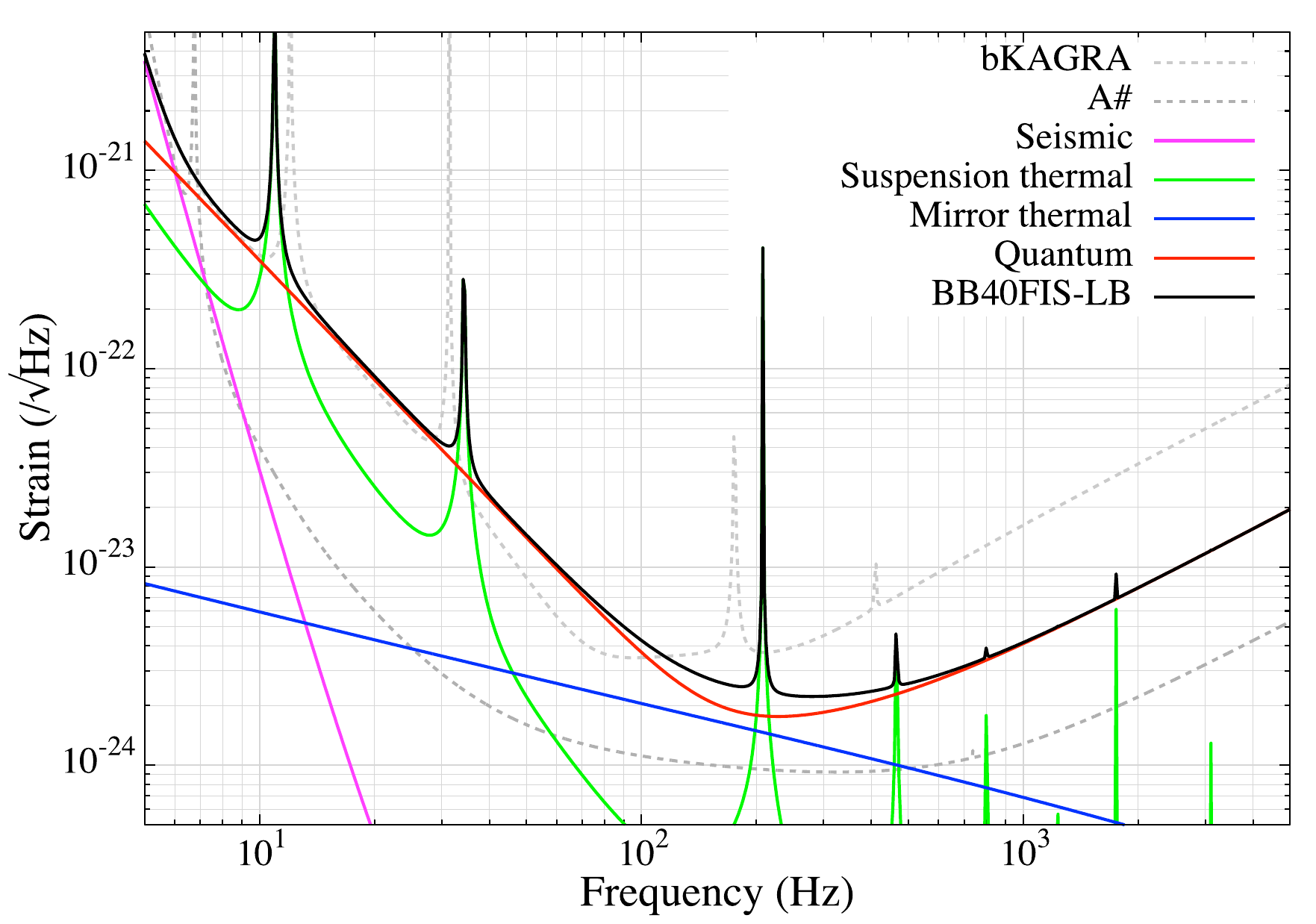}
        \caption{BB40FIS-LB-HQS}
    \end{subfigure}
    
    \begin{subfigure}{0.49\textwidth}
        \centering
        \includegraphics[width=\textwidth]{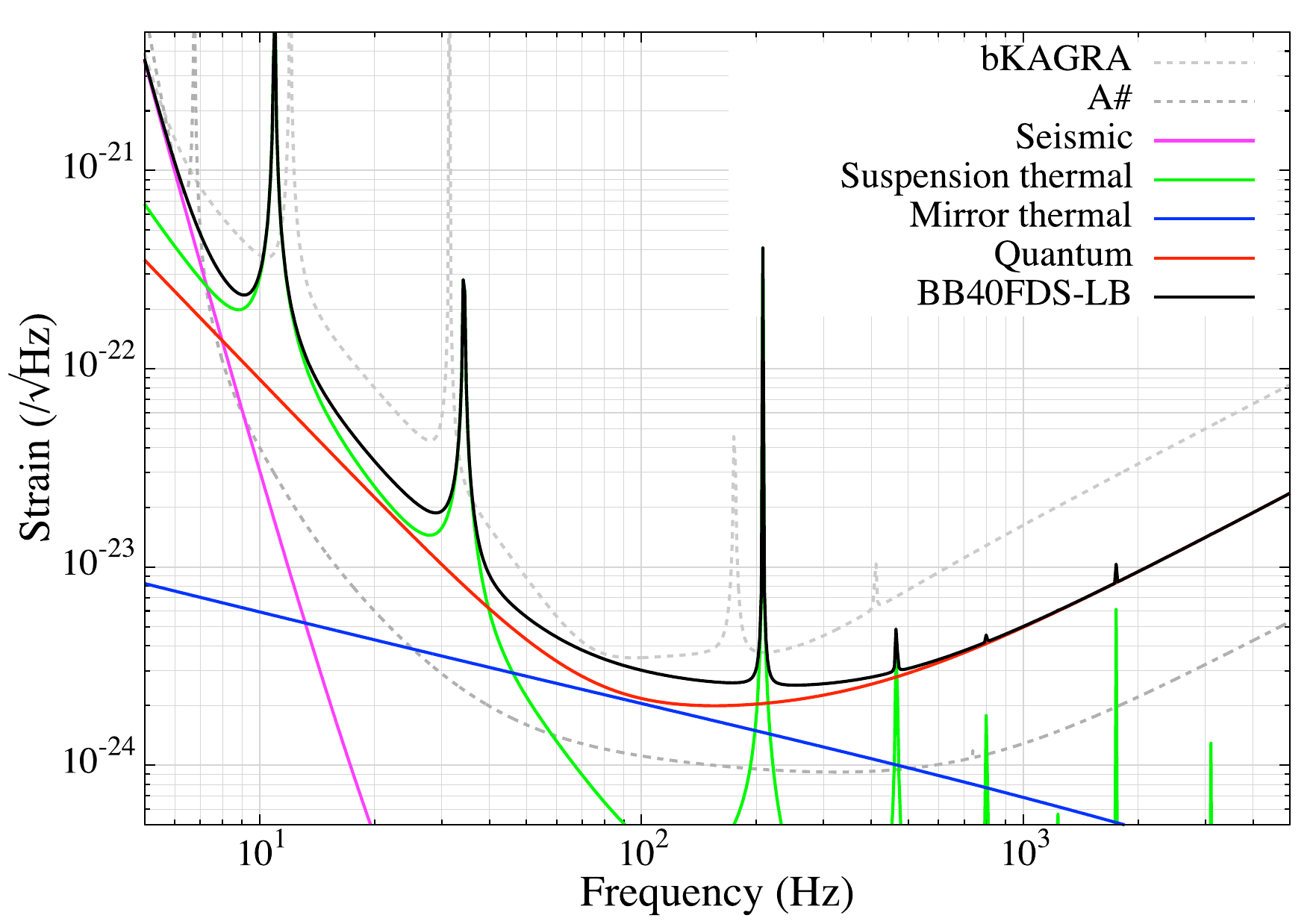}
        \caption{BB40FDS-LB-HQS}
    \end{subfigure}
    \begin{subfigure}{0.49\textwidth}
        \centering
        \includegraphics[width=\textwidth]{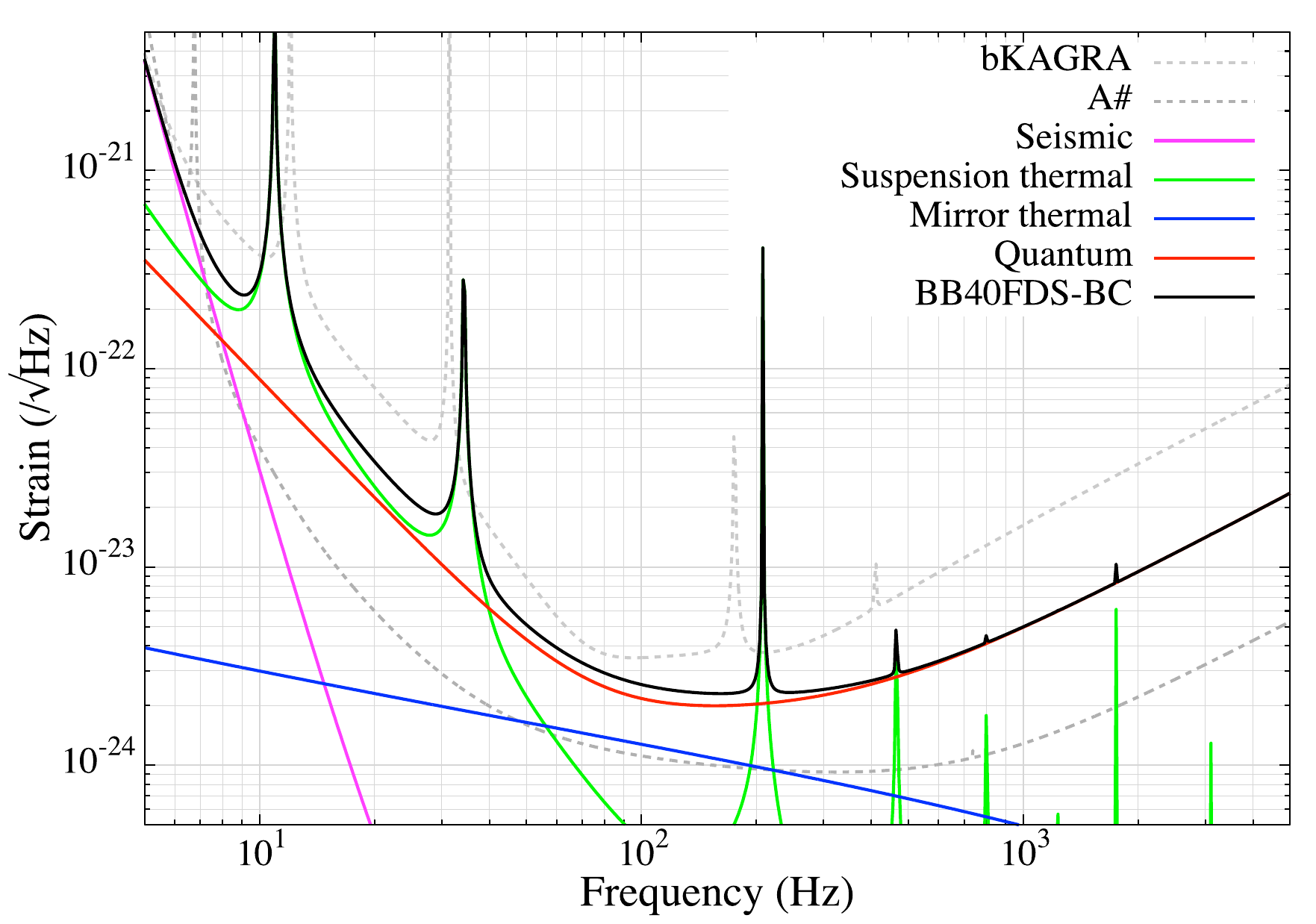}
        \caption{BB40FDS-BC-HQS}
    \end{subfigure}
    
    \caption{Broadband options. The magenta, green, blue, red solid line represents the seismic, suspension thermal, mirror thermal, and quantum noise, respectively. The sensitivity curves of the currently designed KAGRA and Asharp are plotted by the gray and black dotted lines.}
    \label{fig:BBoption}
\end{figure}

\subsubsection{High Frequency moderate (HFmod) Options}

HFmod upgrades provide a compromise between HF and BB approaches by maintaining some high-frequency enhancements while moderately improving middle-frequency sensitivity. Considered parameters include:
\begin{itemize}
    \item Input power: 150 W
    \item SRM reflectivity: 96\%
    \item Mirror temperature: 26 K
    \item 10 dB FIS or FDS
\end{itemize}

\begin{figure}
    \centering
    \begin{subfigure}{0.49\textwidth}
        \centering
        \includegraphics[width=\textwidth]{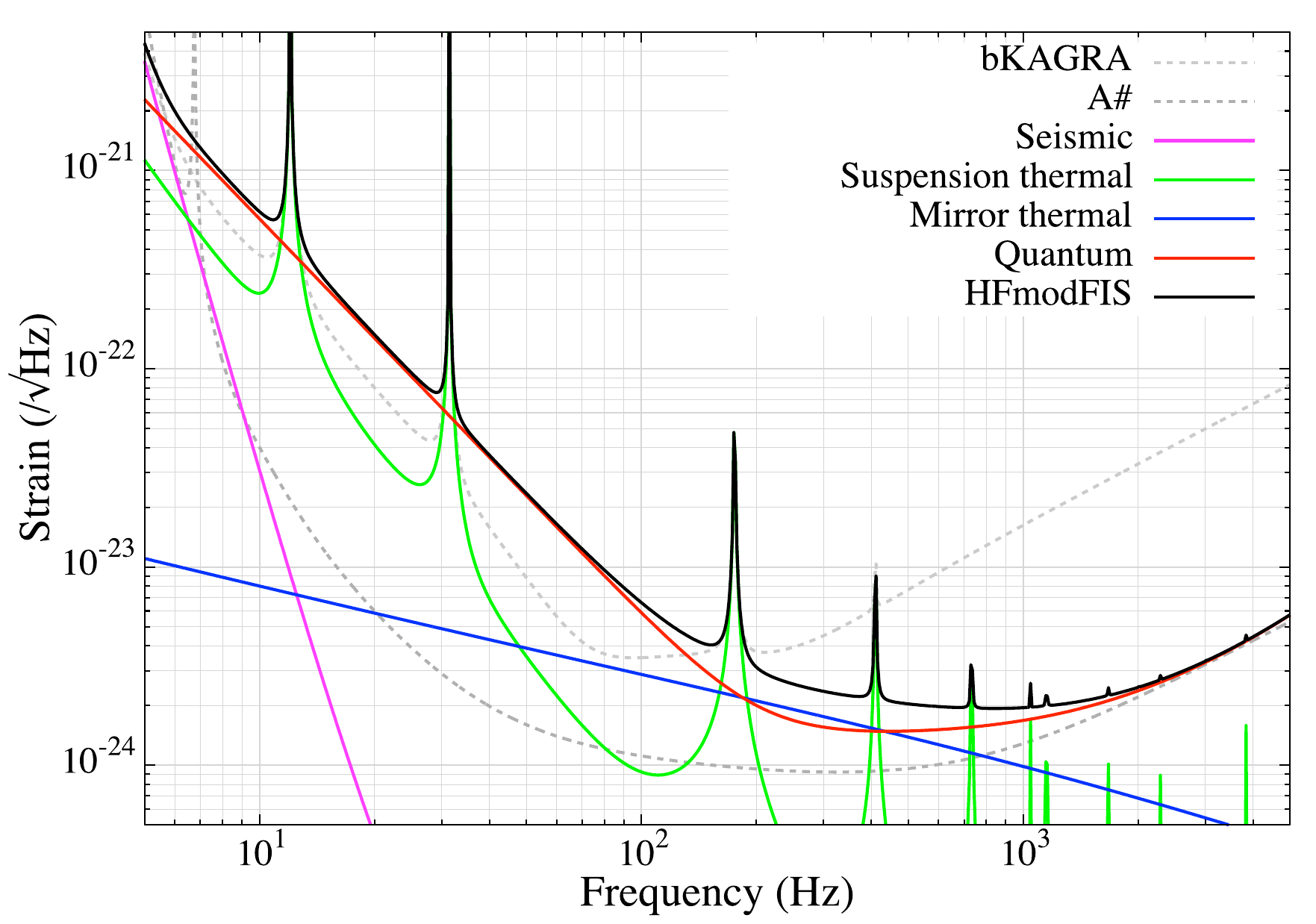}
        \caption{HFmodFIS-HQS}
    \end{subfigure}
    \begin{subfigure}{0.49\textwidth}
        \centering
        \includegraphics[width=\textwidth]{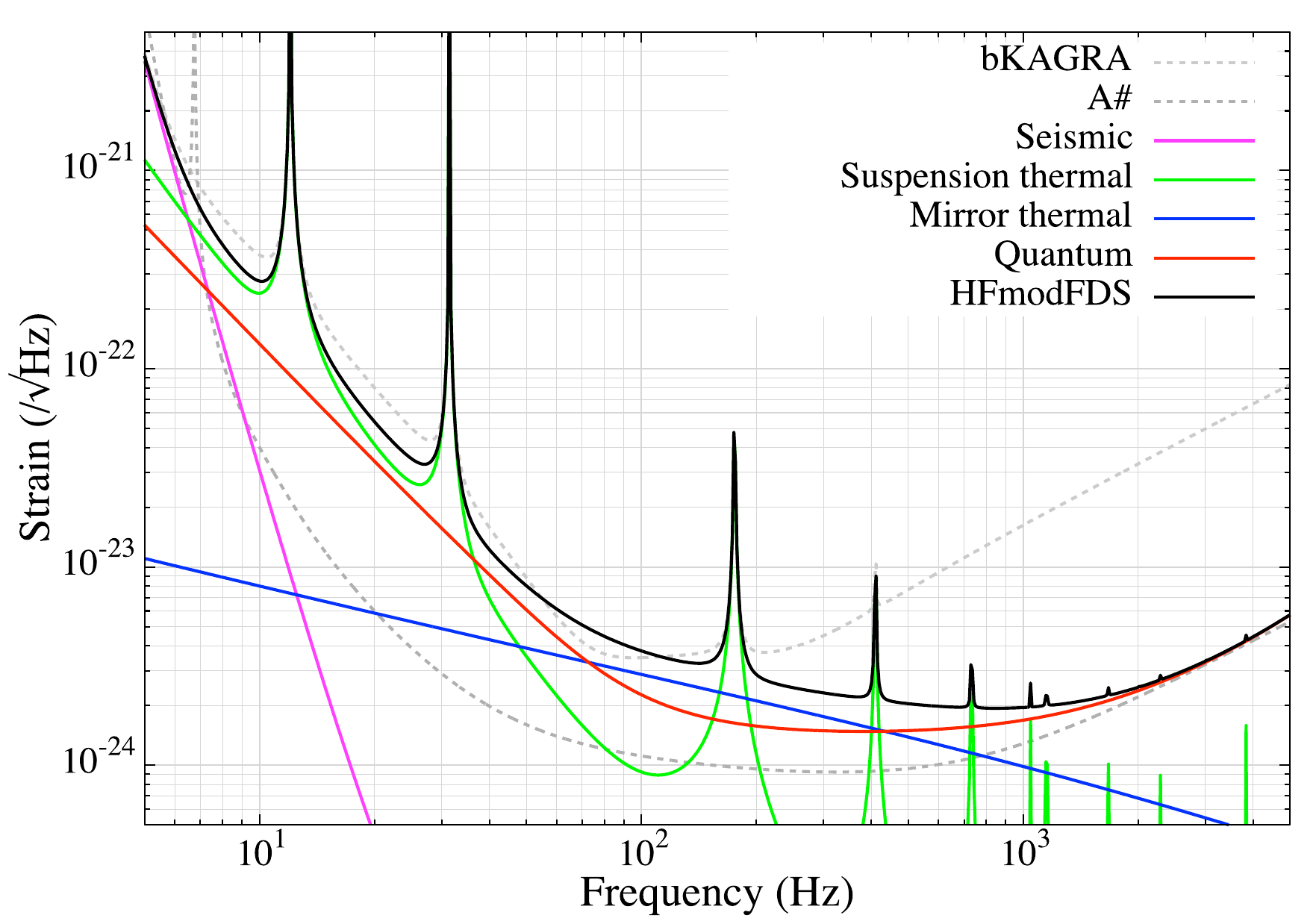}
        \caption{HFmodFDS-HQS}
    \end{subfigure}
    
    \caption{High frequency moderate options. The colors of the plots are the same as those in Fig.~\ref{fig:BBoption}.}
    \label{fig:HFmodoption}
\end{figure}

\subsubsection{High Frequency (HF) Options}

HF upgrades specifically enhance high-frequency sensitivity while sacrificing some low-frequency performance. To investigate impacts of the low-Q sapphire fiber on the HF options, some configurations do not contain the HQS options, while that is applied to others. Considered HF options include:

\begin{itemize}
    \item \textbf{3k + Frequency Independent Squeezing (FIS)}: Increasing the signal recycling mirror (SRM) reflectivity enhances \gw signal amplification at 3 kHz. Key parameters include:
    \begin{itemize}
        \item SRM reflectivity: 99.5\%
        \item Mirror temperature: 30 K
        \item Arm power: 1.3 MW
        \item Input squeezing: 10 dB
    \end{itemize}
    \item \textbf{2k + FIS}: Adjusting the input test mass (ITM) reflectivity shifts the optimal frequency to 2 kHz. Key difference:
    \begin{itemize}
        \item ITM reflectivity: 99.8\%
    \end{itemize}
    \item \textbf{3k or 2k + Frequency Dependent Squeezing (FDS)}: Combining HF optimization with FDS mitigates low-frequency degradation.
\end{itemize}

\begin{figure}
    \centering
    \begin{subfigure}{0.49\textwidth}
        \centering
        \includegraphics[width=\textwidth]{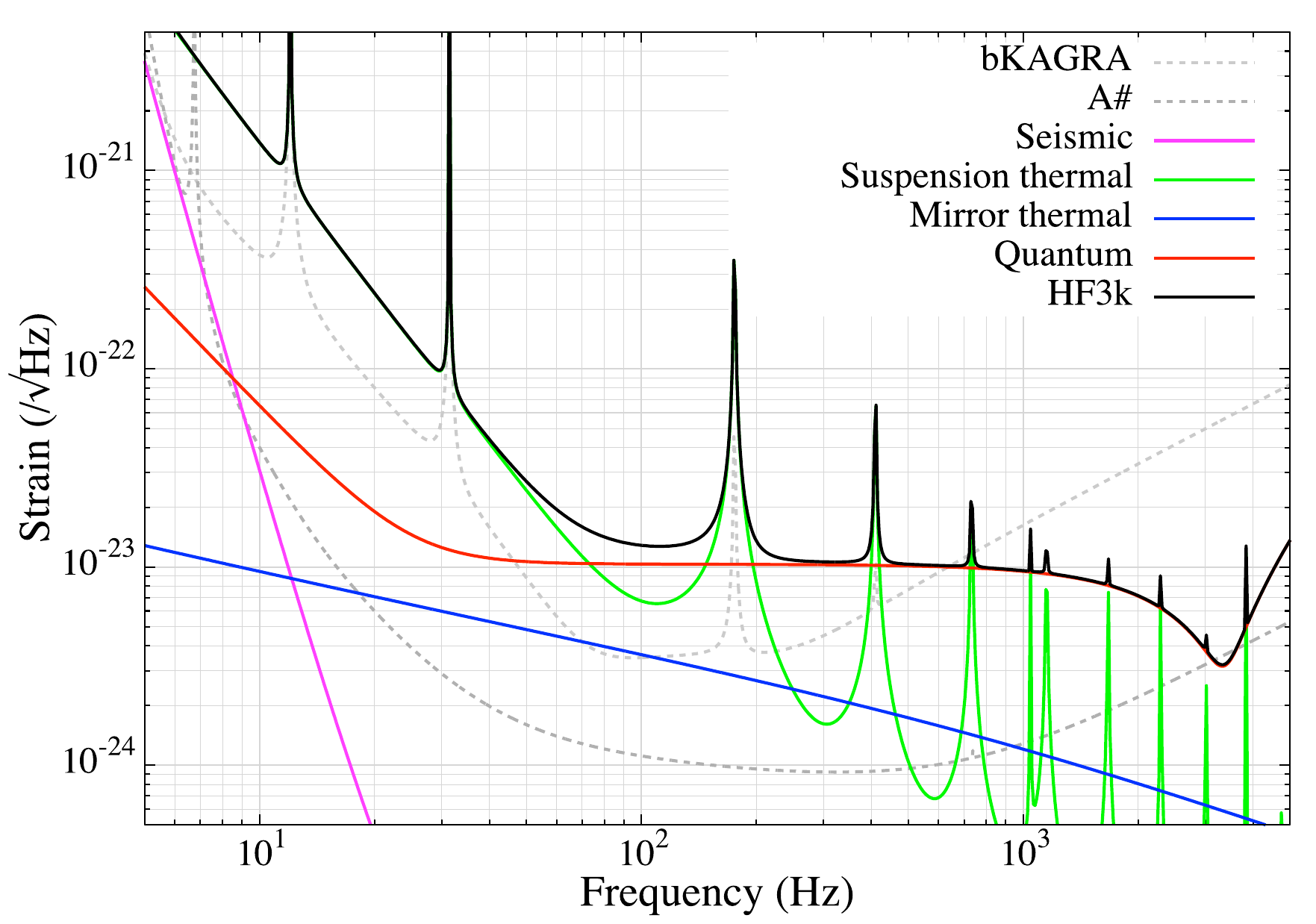}
        \caption{HF3k}
    \end{subfigure}
    \begin{subfigure}{0.49\textwidth}
        \centering
        \includegraphics[width=\textwidth]{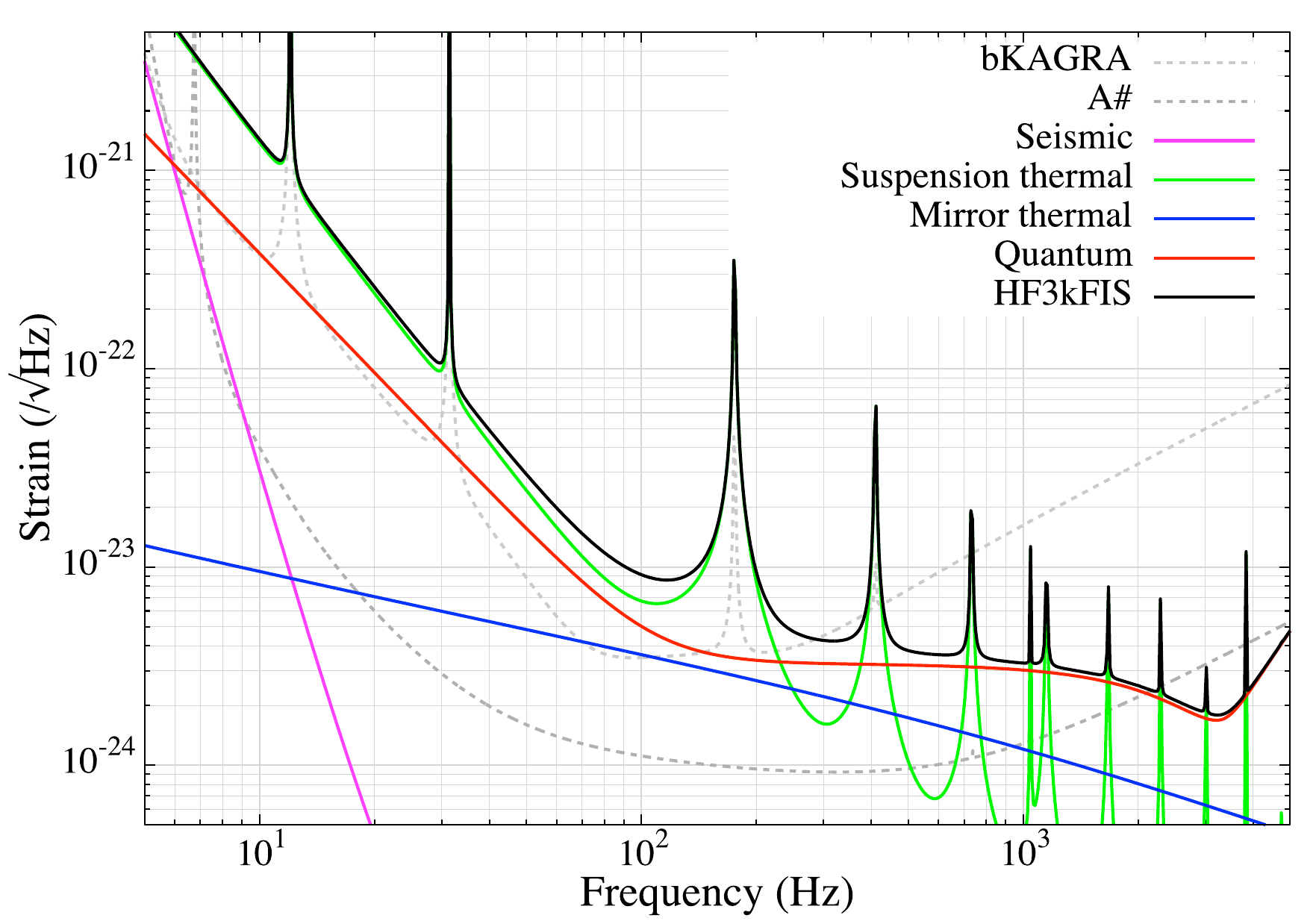}
        \caption{HF3kFIS}
    \end{subfigure}
    
    \begin{subfigure}{0.49\textwidth}
        \centering
        \includegraphics[width=\textwidth]{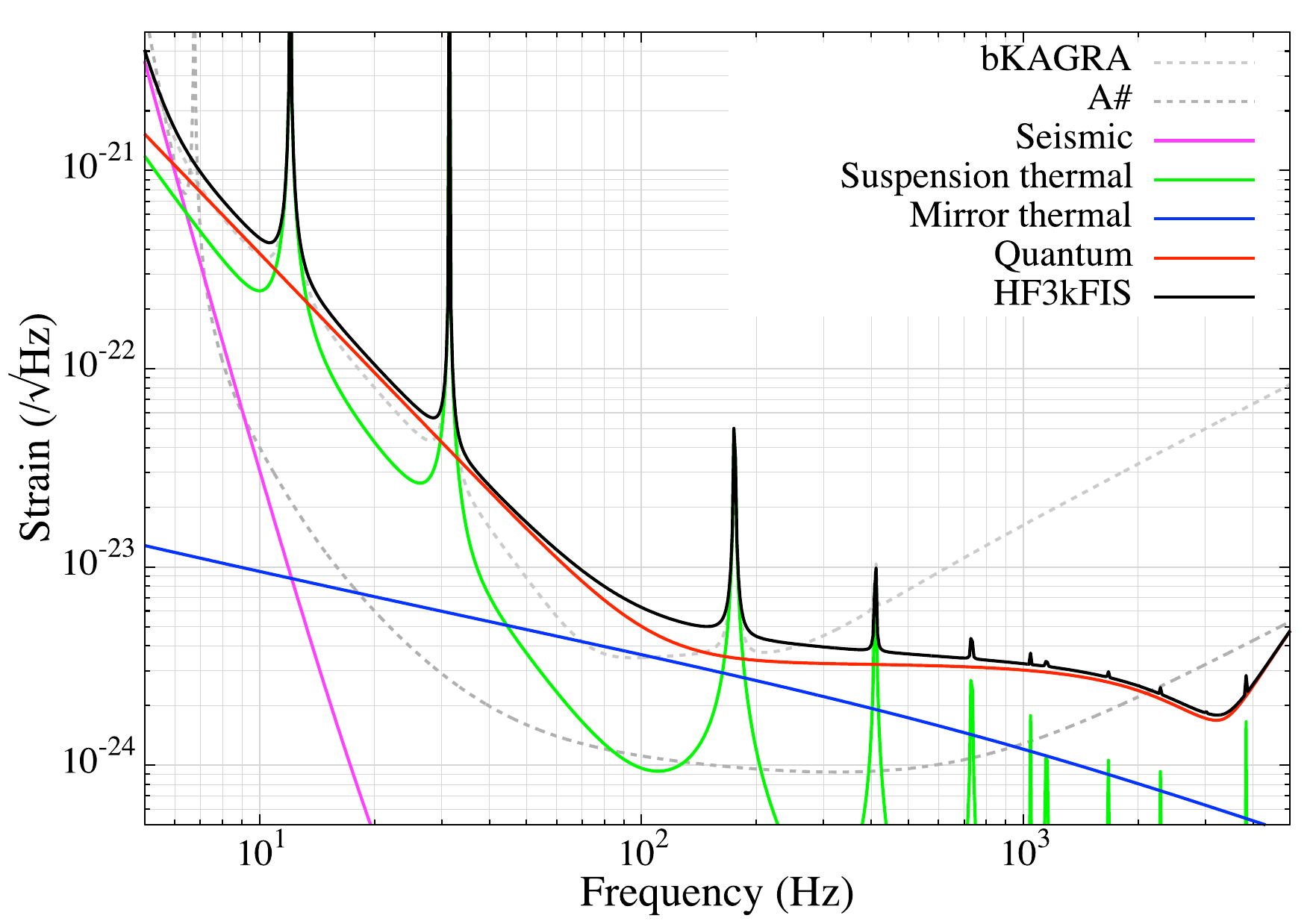}
        \caption{HF3kFIS-HQS}
    \end{subfigure}
    \begin{subfigure}{0.49\textwidth}
        \centering
        \includegraphics[width=\textwidth]{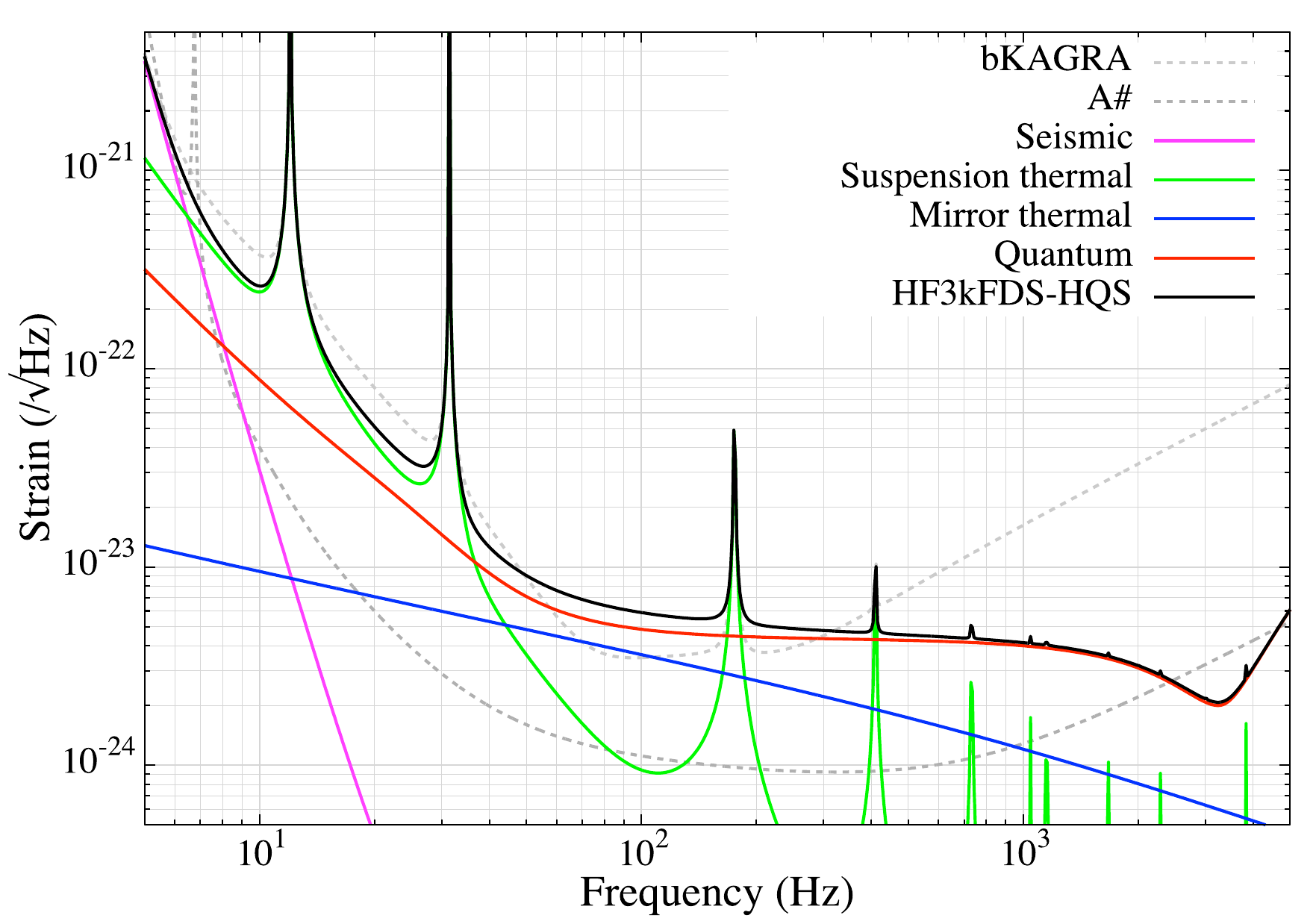}
        \caption{HF3kFDS-HQS}
    \end{subfigure}
    
    \begin{subfigure}{0.49\textwidth}
        \centering
        \includegraphics[width=\textwidth]{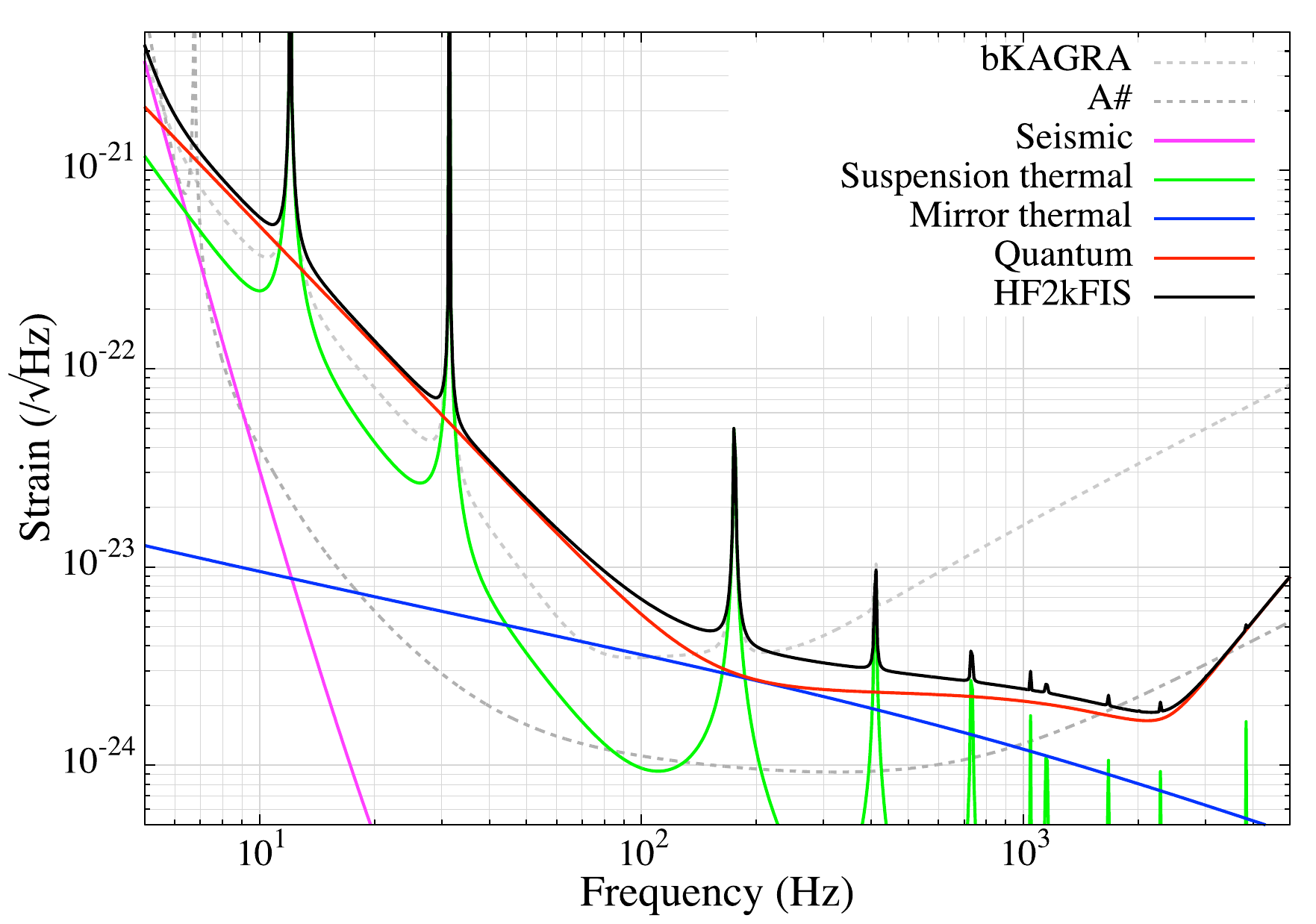}
        \caption{HF2kFIS-HQS}
    \end{subfigure}
    \begin{subfigure}{0.49\textwidth}
        \centering
        \includegraphics[width=\textwidth]{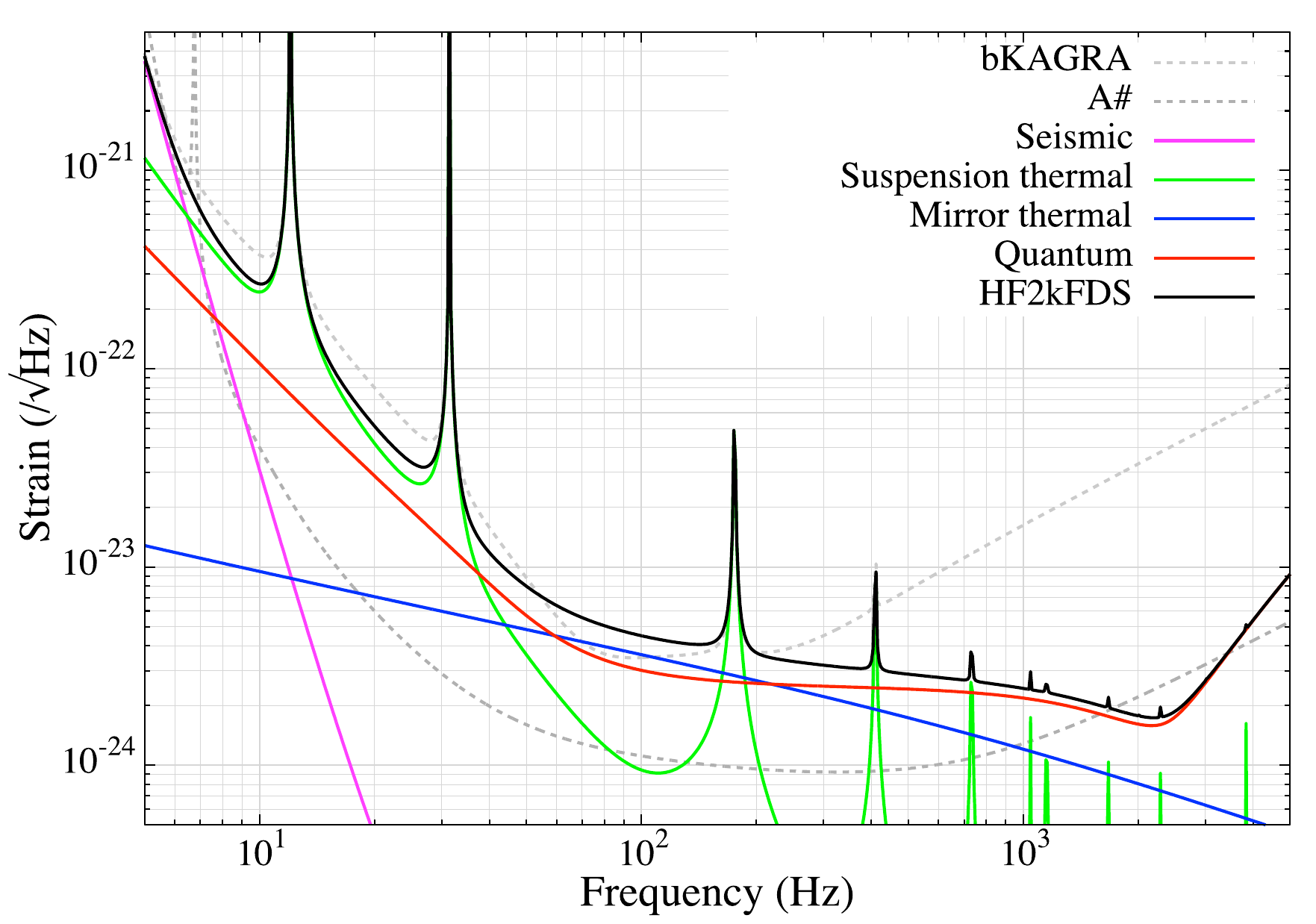}
        \caption{HF2kFDS-HQS}
    \end{subfigure}
    
    \caption{High frequency options. The colors of the plots are the same as those in Fig.~\ref{fig:BBoption}.}
    \label{fig:HFoption}
\end{figure}

\subsection{Cost and Risks}

Each upgrade approach (BB, HF, HFmod) involves specific cost and risk considerations. In this section, we evaluate the cost and risk of each upgrade option. We perform this evaluation using two tables, ``Upgrade Items" (Table\,\ref{Tbl: Upgrade items}) and ``Summary" (Table\,\ref{Tbl: Cost comparison}).




\paragraph{Terminology} In this section, \textbf{upgrade item} means a specific technology to be introduced to KAGRA, such as a high power laser or a new high reflectivity SRM. The term \textbf{upgrade option} means a collection of upgrade items to form a complete upgrade plan of KAGRA.

\subsubsection{Upgrade Items Table}

This table (Table\,\ref{Tbl: Upgrade items}) evaluates individual upgrade components using a scoring system. Each item is assessed based on several criteria:

\begin{itemize}
\item \textbf{Risk}: Technical risk. Rated from 1 to 5, with higher values indicating greater technical difficulty in implementing the upgrade.
\item \textbf{Relevance}: Reflects how critical a technology is to the overall success of the KAGRA upgrade. For instance, if a 260W laser achieves only 200W, the detector can still operate with somewhat reduced sensitivity. On the other hand, if the fabrication of SRM fails, it is impossible to operate the interferometer with the HF configuration. Therefore, the relevance of the SRM fabrication becomes high.
\item \textbf{Cost}: Estimated monetary cost to realize each upgrade item. Originally evaluated in JPY, then normalized by the average cost of all the upgrade items to make the values dimensionless.
\item \textbf{R\&D time}: Time required to develop the necessary technology and fabricate the associated equipment.
\item \textbf{R\&D FTE}: Average Full Time Equivalent human resources (FTE) necessary during the R\&D phase.
\item \textbf{Inst. time}: Time needed to install the equipment and commission it to achieve the desired performance.
\item \textbf{Inst. FTE}: Average FTE necessary during the installation and commissioning phase.
\item \textbf{Total time}: The sum of R\&D time and Inst. time.
\item \textbf{Human resources}: Expressed in person-months, the total human resources needed to complete the upgrade. This is calculated by (R\&D time)$\times$ (R\&D FTE) $+$ (Inst. time)$\times$ (Inst. FTE).
\item \textbf{Average FTE}: The average full-time equivalent personnel calculated over the entire period of upgrade.
\item \textbf{Required}: Indicates which upgrade options utilize the specific upgrade item.
\end{itemize}

Each scoring criterion is designed such that higher values represent greater cost or risk. The average value of each metric is computed, which is later used in the calculation of the \textbf{Weighted Sum} in the Summary table.

\subsubsection{Summary Table}

The \textbf{Summary} table (Table\,\ref{Tbl: Cost comparison}) presents aggregated values of risk, cost, and other metrics for each upgrade option. These aggregates are computed based on the \textbf{Required} column in the \textbf{Upgrade Items} table, which indicates which upgrade options require each upgrade item.


The row labeled \textbf{Average} at the bottom of the Summary table is a direct copy of the average values calculated in the Upgrade Items table. These averages are used to normalize each metric values in the calculation of the \textbf{Normalized Sum}.

To assess the overall technical cost and risk of each upgrade option, two indices are employed: the \textbf{Product} and the \textbf{Normalized Sum}. The Product is computed by simply taking a product of all the metric values (columns) of the upgrade option, except for the Time and Average FTE, because these factors are included in the Manpower in the form of person-month.
The Normalized Sum is calculated by normalizing each metric by its average value and then summing the results. The Product generally produces a greater spread in scores than the Normalized Sum, highlighting differences among upgrade options more prominently.

\begin{landscape}
\begin{table}
\centering
\tiny
\begin{tabular}{l|r|r|r|r|r|r|r|r|r|r|l}
\toprule
\multirow{2}{*}{Name} & \multirow{2}{*}{Risk}& \multirow{2}{*}{Relevance}& \multirow{2}{*}{Cost} & \multicolumn{1}{c|}{R\&D} &\multicolumn{1}{c|}{R\&D}  &\multicolumn{1}{c|}{Inst.}&\multicolumn{1}{c|}{Inst.} & \multicolumn{1}{c|}{Total} & \multicolumn{1}{c|}{Person-} & \multicolumn{1}{c|}{Average}& \multirow{2}{*}{Required} \\
 & & & & \multicolumn{1}{c|}{time(yr)} & \multicolumn{1}{c|}{FTE} &\multicolumn{1}{c|}{time(yr)} &\multicolumn{1}{c|}{FTE} & \multicolumn{1}{c|}{Time(yr)} & \multicolumn{1}{c|}{power(P$\cdot$M)} & \multicolumn{1}{c|}{FTE} & \\
\midrule
Higher Power Laser (260W) & 2.0 & 2 & 0.78 & 2.0 & 2.0 & 0.5 & 2.0 & 2.5 & 60.0 & 2.0 & HF3k,HF2k \\
Higher Power Laser (150W) & 1.0 & 2 & 0.2 & 1.5 & 2.0 & 0.5 & 2.0 & 2.0 & 48.0 & 2.0 & HFmod \\
High Power Compatible Cooling & 4.0 & 4 & 0.12 & 1.5 & 2.0 & 0.5 & 2.0 & 2.0 & 48.0 & 2.0 & HF3k,HF2k,HFmod \\
Thermal compensation & 2.0 & 4 & 0.39 & 2.5 & 2.0 & 0.5 & 2.0 & 3.0 & 72.0 & 2.0 & HF3k,HF2k,HFmod \\
99.5\% SRM & 1.0 & 5 & 0.12 & 1.5 & 0.2 & 0.5 & 1.0 & 2.0 & 9.6 & 0.4 & HF3k,HF2k \\
96\% SRM & 1.0 & 5 & 0.12 & 1.5 & 0.2 & 0.5 & 1.0 & 2.0 & 9.6 & 0.4 & HFmod \\
RSE lock with 99.5\% SRM & 3.0 & 5 & 0.0 & 0.0 & 0.0 & 1.0 & 2.0 & 1.0 & 24.0 & 2.0 & HF3k,HF2k \\
RSE lock with 96\% SRM & 2.0 & 5 & 0.0 & 0.0 & 0.0 & 1.0 & 2.0 & 1.0 & 24.0 & 2.0 & HFmod \\
10dB Squeezer & 3.0 & 3 & 0.39 & 2.5 & 2.0 & 0.5 & 2.0 & 3.0 & 72.0 & 2.0 & FIS,FDS \\
Filter Cavity & 3.0 & 3 & 2.35 & 2.5 & 3.0 & 0.5 & 3.0 & 3.0 & 108.0 & 3.0 & FDS \\
99.8\% ITMs & 1.0 & 2 & 1.37 & 2.5 & 0.3 & 0.5 & 2.0 & 3.0 & 21.0 & 0.58 & HF2k \\
Lock the IFO with 99.8\% ITMs & 2.0 & 5 & 0.0 & 0.0 & 0.0 & 0.5 & 2.0 & 0.5 & 12.0 & 2.0 & HF2k \\
Large TMs (40kg) & 4.0 & 5 & 3.91 & 4.5 & 2.0 & 1.0 & 4.0 & 5.5 & 156.0 & 2.36 & 40 \\
Larger Beam Size & 2.0 & 5 & 1.96 & 3.0 & 3.0 & 2.0 & 6.0 & 5.0 & 252.0 & 4.2 & LB \\
Low TN Coating (e.g. AlGaAs) & 5.0 & 3 & 3.91 & 4.0 & 1.0 & 1.0 & 4.0 & 5.0 & 96.0 & 1.6 & BC \\
Higher Q Suspension & 4.0 & 3 & 0.39 & 3.0 & 2.0 & 1.0 & 3.0 & 4.0 & 108.0 & 2.25 & HQS \\
Average & 2.5 & 4 & 1.0 & 2.0 & 1.36 & 0.75 & 2.5 & 2.78 & 70.0 & 1.92 & Average \\
\bottomrule
\end{tabular}
\caption{Upgrade items}
\label{Tbl: Upgrade items}
\end{table}

\begin{table}
\centering
\scriptsize
\begin{tabular}{l|r|r|r|r|r|r|r|r}
\toprule
Name & Risk & Relevance & Cost(MJPY) & Time(yr) & Personpower(P$\cdot$M) & Average FTE & Product & Normalized Sum \\
\midrule
BB23FDS-HQS & 10 & 9 & 3.13 & 10 & 288 & 7.25 & 8.11E+04 & 13.61 \\
BB40FIS-HQS & 11 & 11 & 4.69 & 12.5 & 336 & 6.61 & 1.91E+05 & 16.78 \\
BB40FDS-HQS & 14 & 14 & 7.04 & 15.5 & 444 & 9.61 & 6.13E+05 & 22.66 \\
BB40FIS-LB-HQS & 13 & 16 & 6.65 & 17.5 & 588 & 10.81 & 8.13E+05 & 24.45 \\
BB40FDS-LB-HQS & 16 & 19 & 9.00 & 20.5 & 696 & 13.81 & 1.90E+06 & 30.33 \\
BB40FDS-HQS-BC & 19 & 17 & 10.95 & 20.5 & 540 & 11.21 & 1.91E+06 & 30.73 \\
HFmodFIS-HQS & 17 & 26 & 1.60 & 17 & 381.6 & 12.65 & 2.71E+05 & 20.68 \\
HFmodFDS-HQS & 20 & 29 & 3.95 & 20 & 489.6 & 15.65 & 1.12E+06 & 26.56 \\
HF2kFIS-HQS & 22 & 33 & 3.56 & 21 & 426.6 & 15.23 & 1.10E+06 & 27.12 \\
HF2kFDS-HQS & 25 & 36 & 5.91 & 24 & 534.6 & 18.23 & 2.84E+06 & 32.99 \\
HF3k & 12 & 20 & 1.41 & 10.5 & 213.6 & 8.40 & 7.22E+04 & 14.51 \\
HF3kFIS & 15 & 23 & 1.80 & 13.5 & 285.6 & 10.40 & 1.77E+05 & 17.92 \\
HF3kFIS-HQS & 19 & 26 & 2.19 & 17.5 & 393.6 & 12.65 & 4.26E+05 & 22.24 \\
HF3kFDS-HQS & 22 & 29 & 4.54 & 20.5 & 501.6 & 15.65 & 1.45E+06 & 28.12 \\
Average & 2.5 & 3.81 & 1.00 & 2.78 & 70 & 1.92 &  &  \\
\bottomrule
\end{tabular}
\caption{Summary of risk and cost comparison}
\label{Tbl: Cost comparison}
\end{table}
\end{landscape}

\subsection{Necessary R\&Ds}

This section outlines the necessary research and development efforts required for the implementation of KAGRA HF and HFmode upgrades.

\subsubsection{High Power Related R\&Ds}

To achieve an arm power of 1.3 MW in KAGRA HF, an incident laser power of approximately 300 W is required. The development of a stable laser source with such high output power presents a significant technical challenge. Additionally, various issues arising from high-power laser operation must be addressed.

\vspace{3mm}
\emph{High Power Laser Development}

The current high-power laser for KAGRA uses a 70 W fiber amplifier. By cascading this system, an output of approximately 140 W can be achieved. However, the feasibility of scaling this scheme up to 300 W remains uncertain. Key questions include:
\begin{itemize}
    \item Can cascading be extended further to reach 300 W?
    \item Are there technical limitations such as Brillouin scattering with such high power operation in fiber?
    \item Can the same level of frequency and intensity stability as the current laser be achieved?
\end{itemize}

Experimental validation is required to resolve these uncertainties. If simple cascading is not feasible, an alternative approach is to coherently combine two 150 W beams. This method, however, raises concerns regarding additional noise and mode distortions, which must be evaluated.

Since 300 W-class lasers are also being considered for the next upgrades of LIGO and Virgo, establishing a collaborative development effort within the LVK community is essential.

\vspace{3mm}
\emph{Parametric Instability Study}

Parametric Instability (PI) occurs when high-order optical modes excited by mirror thermal vibrations couple with mirror mechanical modes via radiation pressure, leading to positive feedback and instability. National Astronomical Observatory of Japan (NAOJ) has initiated a simulation study to assess the risk of PI in KAGRA HF. The preliminary results suggest that PI may arise in certain modes.

Mitigation strategies must be investigated, as the resonant dampers used in LIGO may not be applicable in the cryogenic environment of KAGRA. Possible countermeasures include:
\begin{itemize}
    \item Narrow band feedback of the DARM signal to TM coil-magnet actuators at the PI frequencies.
    \item Modulating incident laser power to cancel PI effects.
    \item Using photon calibrators (PCal) to damp PI.
\end{itemize}
\vspace{3mm}
\emph{Cooling of Mirrors with High Power Operation}

High-power laser operation induces thermal gradients in the mirror substrates, causing thermal lensing. In room-temperature interferometers, this effect is significant and poses a major challenge to high-power operation. However, in cryogenic interferometers like KAGRA, the extremely high thermal conductivity of crystalline substrate reduces thermal gradients and mitigates thermal lensing effects—an advantage unique to KAGRA.

Nonetheless, effective cooling remains critical, as higher incident laser power leads to increased heat deposition in the mirrors. There are three possible solutions:

\vspace{3mm}
\emph{1. Enhancing Cooling Capacity}

Conductive cooling in KAGRA relies on sapphire fibers to transfer heat from the mirrors to the heat links. Increasing cooling capacity requires either increasing the fiber thickness or lowering the cold head temperature. Since further lowering the cold head temperature is impractical, the primary option is to increase the fiber thickness. However, this approach introduces challenges, as thicker fibers lead to an increase in suspension thermal noise and a decrease in the violin mode frequency, which could result in higher density of noise peaks at high frequencies. Therefore, it is essential to develop a balanced design that optimizes cooling efficiency while minimizing its adverse effects on noise.

\vspace{3mm}
\emph{2. Reducing Heat Generation}

Heat deposition in the mirrors arises from optical absorption in the substrate and coating. Reducing absorption will allow for higher laser power operation without excessive heating. To achieve this, it is crucial to develop ultra-high-purity sapphire substrates with lower absorption. A research effort is on going in a collaboration of NAOJ, Korea Astronomy and Space Science Institute (KASI), and institut Lumi\'ere Mati\'ere (iLM). 

For the reduction of coating absorption, cryogenic focused researches are essential. This calls for joint efforts with the ET project, which is also aiming to build a cryogenic interferometer. However, since ET operates at a different wavelength than KAGRA, the direct applicability of its coatings remains uncertain. Nevertheless, exchanging information and research findings with ET will be beneficial in advancing coating technology for KAGRA.

\vspace{3mm}
\emph{3. Allowing Higher Mirror Temperatures}

Given a constant thermal resistance, increasing the mirror temperature allows for higher permissible heat loads. Although higher temperatures lead to an increase in thermal noise, the HF upgrades have some margin to tolerate this. Furthermore, the high thermal conductivity of sapphire ensures that thermal lensing remains negligible even when the temperature is moderately elevated.

\vspace{3mm}
A combination of these approaches must be carefully implemented to optimize high-power operation strategy for KAGRA HF.

\subsubsection{Squeezing}

Squeezing is a critical technology for next-generation \gw detectors with any upgrade strategy.

\vspace{3mm}
\emph{Development of a 10 dB Squeezer}

HF upgrades aim to achieve a 10 dB squeezing level at kHz frequencies. While generating a squeezed vacuum state with over 10 dB of squeezing has already been demonstrated, integrating this into an interferometer to achieve a 10 dB quantum noise reduction remains challenging. To accomplish this, reducing interferometer losses is a prerequisite. Additionally, optimal mode-matching and alignment control between the interferometer and the squeezer must be achieved. For example, the development of wavefront sensing-based alignment control scheme for squeezer may be required.

Current squeezer development efforts are being conducted at NAOJ in collaboration with National Tsing Hua University (NTHU) in Taiwan and KASI in South Korea. There are also plans to test the developed squeezer at TAMA to refine alignment control techniques.

\vspace{3mm}
\emph{Frequency Dependent Squeezing}

HFmod upgrades aim to enhance low-frequency sensitivity through the implementation of frequency-dependent squeezing, creating a broader-band detector. A reliable method to achieve frequency-dependent squeezing is through the introduction of a filter cavity. It has been confirmed that there is sufficient space along the Y-arm to accommodate an 80 m-class filter cavity. However, constructing a filter cavity requires significant financial and human resources, making its feasibility a critical concern. Therefore, detailed studies on installation methods are necessary to assess and determine the practicality of its implementation.

An alternative approach to achieving frequency-dependent squeezing without a filter cavity is through EPR squeezing. There are ongoing discussions about the possibility of conducting large-scale cavity-based EPR squeezing experiments using the TAMA interferometer.

\subsubsection{Control of the Interferometer with a High-Reflectivity SRM}

HF upgrades require an increase in the reflectivity of the signal recycling mirror (SRM) beyond the original KAGRA design. However, it is not immediately evident whether the current interferometer control scheme will remain effective under these conditions.

Simulations are needed to evaluate the feasibility of applying existing control methods, and if necessary, new control strategies must be developed to accommodate the high-reflectivity SRM configuration.

\subsection{Summary}
From the perspective of hardware feasibility, the HF3k upgrade represents one of the least technically demanding options. It does not require major interventions on the core interferometer components. When combined with Frequency-Independent Squeezing (FIS), the HF3kFIS configuration emerges as a practical and balanced path forward, offering improved post-merger signal sensitivity with limited hardware complexity.

For enhancing the binary neutron star detection range and enabling broader-band sensitivity improvements, we can introduce Frequency-Dependent Squeezing (FDS). However, this approach demands the construction and integration of a long filter cavity, which incurs significant cost in terms of funding, human resources, and time. Therefore, FDS-based options should be weighed carefully against available resources.

Among the high-frequency upgrades, the HFmod configuration has a lower risk regarding the control of the SRC due to its moderate enhancement of signal recycling mirror reflectivity. 

In contrast, the HF2k configuration requires the replacement of the input test masses (ITMs) to modify the arm cavity finesse, which entails a substantial increase in the cost.

Within the broad-band (BB) upgrade family, BB23FDS-HQS offers a relatively low-cost entry point, though its gain in binary range remains insignificant. Other BB options, especially those involving heavier mirrors, provide greater scientific reach but at significantly higher cost and risk levels. Also we need to note that the BB options becomes almost meaningless unless we find a way to improve the suspension quality factors from the current level.

In summary, the HF3kFIS and HFmod options offer technically feasible and cost-efficient upgrade paths for KAGRA’s near-term future, while more ambitious BB and FDS-based plans may be considered in the longer term depending on scientific priorities and resource availability.


\section{Conclusions\label{sec:conclusion}}
In this study, we have evaluated 14 potential upgrade options for the KAGRA \gw detector, focusing on both scientific impact and hardware feasibility. Broadband configurations generally offer the highest detection rates for compact binary coalescence events, with the BB40FDS\_HQS\_BC option predicting $\sim 10$ binary neutron star and $\sim 10^2$ binary black hole detections per year. However, high-frequency configurations, particularly the HFmod variants, excel in other binary neutron star science goals. These include improved sky localization, with HFmod options significantly outperforming broadband configurations for GW170817-like events, as well as tighter constraints on the tidal deformability parameter, where HFmod can reduce the 90\% credible interval by $\sim 10\%$ at median and $\gtrsim 50\%$ in favorable cases where KAGRA contributes a relatively high \snr. For post-merger signals, the HF2k and HF3k configurations provide the highest expected detection rates, up to 0.1 events per year. For continuous wave signals from rotating neutron stars, HFmod configurations again show superior sensitivity in the relevant 200 Hz to 1 kHz frequency band.

From a technical perspective, the HF3kFIS configuration is among the most accessible upgrade paths, requiring minimal modifications to the core interferometer components while offering meaningful gains in post-merger signal sensitivity. The HFmod configuration, which involves only a modest adjustment to the signal recycling mirror reflectivity, also presents a low-risk and technically feasible option. In contrast, upgrades involving frequency-dependent squeezing require the construction and integration of long filter cavities, representing a substantial investment of time, funding, and effort. While broadband upgrades offer significant scientific benefits, particularly when combined with heavier mirrors and squeezing techniques, they depend critically on improvements in suspension quality factors which is a current limitation. In summary, HF3kFIS and HFmod provide realistic and cost-effective upgrade options for the near-term future of KAGRA, while more ambitious broadband and frequency-dependent squeezing configurations remain promising for the longer term, depending on future resources and scientific goals.

In conclusion, our study finds that the HFmod upgrade, which enhances sensitivity across a broad frequency range above approximately 200 Hz, offers the best balance between scientific return and technical feasibility. In particular, the HFmod configuration is well suited for improving the sky localization of binary neutron star mergers and for placing tighter constraints on the tidal deformability parameter.

\section{Acknowledgment\label{sec:acknowledgment}}
This work was supported by MEXT, JSPS Leading-edge Research Infrastructure Program, JSPS Grant-in-Aid for Specially Promoted Research 26000005, JSPS Grant-inAid for Scientific Research on Innovative Areas 2905: JP17H06358, JP17H06361 and JP17H06364, JSPS Core-to-Core Program A. Advanced Research Networks, JSPS Grantin-Aid for Scientific Research (S) 17H06133 and 20H05639 , JSPS Grant-in-Aid for Transformative Research Areas (A) 20A203: JP20H05854, the joint research program of the Institute for Cosmic Ray Research, University of Tokyo, National Research Foundation (NRF), Computing Infrastructure Project of Global Science experimental Data hub Center (GSDC) at KISTI, Korea Astronomy and Space Science Institute (KASI), and Ministry of Science and ICT (MSIT) in Korea, Academia Sinica (AS), AS Grid Center (ASGC) and the National Science and Technology Council (NSTC) in Taiwan under grants including the Science Vanguard Research Program, Advanced Technology Center (ATC) of NAOJ, and Mechanical Engineering Center of KEK.

\bibliographystyle{unsrt}
\bibliography{references}

\end{document}